\documentclass[useAMS,usenatbib,10pt]{mn2e}
\usepackage{times}  % to get nice fonts for PDF
\usepackage{listings}
\usepackage{graphics}
\usepackage{subfigure}
\usepackage{graphicx}
\usepackage{amssymb}
\usepackage{hyperref}
\usepackage{aas_macros}
\usepackage{lscape}
\usepackage{rotating}

\def\simprop{ \lower .75ex \hbox{$\sim$} \llap{\raise .27ex \hbox{$\propto$}} }
\def\eagle{{\sc eagle}}

\title[The fundamental plane of star formation in galaxies]{
The fundamental plane of star formation in galaxies revealed by the EAGLE hydrodynamical simulations}
\author[Claudia del P. Lagos et al.]{
\parbox[t]{\textwidth}{
\vspace{-1.0cm}
Claudia del P. Lagos$^{1,2}$\thanks{E-mail: claudia.lagos@icrar.org},
Tom Theuns$^{3}$, Joop Schaye$^{4}$, Michelle Furlong$^{3}$, Richard G. Bower$^{3}$, Matthieu Schaller$^{3}$, Robert A. Crain$^{5}$, James W. Trayford$^{3}$, Jorryt Matthee$^{4}$}
\vspace*{6pt} \\
$^{1}$International Centre for Radio Astronomy Research (ICRAR), M468, University of Western Australia, 35 Stirling Hwy, Crawley, WA 6009, Australia.\\
$^{2}$Australian Research Council Centre of Excellence for All-sky Astrophysics (CAASTRO), 44 Rosehill Street Redfern, NSW 2016, Australia.\\
$^{3}$Institute for Computational Cosmology, Department of Physics,
University of Durham, South Road, Durham, DH1 3LE, UK.\\
$^{4}$Leiden Observatory, Leiden University, PO Box 9513, 2300 RA Leiden, The
Netherlands.\\
$^{5}$Astrophysics Research Institute, Liverpool John Moores University, 146
Brownlow Hill, Liverpool, L3 5RF, UK.
\vspace*{-0.5cm}}

\begin{document}

%\date{Accepted ???. Received ???; in original form ???}

\pagerange{\pageref{firstpage}--\pageref{lastpage}} \pubyear{2012}

\maketitle

\label{firstpage}

\begin{abstract}
 We investigate correlations between different physical properties of star-forming galaxies in the ``Evolution and Assembly of GaLaxies and their Environments'' ({\sc eagle}) cosmological hydrodynamical simulation suite over the redshift range 
$0\le z\le 4.5$. A principal component analysis reveals that neutral gas fraction 
($f_{\rm gas, neutral}$), stellar mass ($M_{\rm stellar}$) and star formation rate (SFR) account 
for most of the variance seen in the population, with galaxies tracing a two-dimensional, nearly flat, 
surface in the three-dimensional space of $f_{\rm gas, neutral}-M_{\rm stellar}-$SFR with little scatter. 
The location of this plane varies little with redshift, whereas galaxies themselves move along the plane 
as their $f_{\rm gas, neutral}$ and SFR drop with redshift. The positions of galaxies along the plane 
are highly correlated with
gas metallicity. The metallicity can therefore be robustly predicted from $f_{\rm gas, neutral}$, or from
 the $M_{\rm stellar}$ and SFR.
We argue that the appearance of this ``fundamental plane of star formation'' is 
a consequence of self-regulation, with the plane's curvature set by the dependence of the SFR on gas density and metallicity. 
We analyse a large compilation of observations spanning the redshift range $0\lesssim \rm z\lesssim 3$, 
and find that such a plane is also present in the data. The properties of the observed fundamental plane of star formation are in 
good agreement with {\sc eagle}'s predictions.
\end{abstract}


\section{Introduction}

The star formation rate in a galaxy depends on the interplay between many physical processes, such as 
the rate at which the galaxy's halo accretes mass from the intergalactic medium (IGM), the rate of shocking 
and cooling of this gas onto the galaxy, and the details of how a multi-phase interstellar medium (ISM) converts gas into stars or launches it
into a galactic fountain or outflow (see e.g. \citealt{Benson10} and \citealt{Somerville14} for recent reviews).
The complexity and non-linearity of these processes make it difficult to understand which processes dominate, and if and how
this changes over time.

The identification of tight correlations between physical properties of galaxies (\lq scaling relations\rq) can be very 
valuable in reducing the apparent variety in galaxy properties, enabling the formulation of simple relations that 
capture the dominant paths along which galaxies evolve. Recent efforts have been devoted to studying the 
star formation rate-stellar mass relation (e.g. \citealt{Brinchmann04}; \citealt{Noeske07}), 
stellar mass-gas metallicity relation (e.g. \citealt{Tremonti04}; \citealt{Lara-Lopez10}; \citealt{Mannucci10}; \citealt{Salim14}), and 
the stellar mass-gas fraction relation (e.g. \citealt{Catinella10}; \citealt{Saintonge11}). We begin by reviewing some of these relations.

It has long been established that star-forming galaxies display a tight correlation between star formation rate (SFR) and 
stellar mass ($M_{\rm stellar}$), and that the normalisation of this relation increases with 
redshift ($z$, e.g. \citealt{Brinchmann04}; \citealt{Noeske07};   
\citealt{Daddi07}; \citealt{Rodighiero10}). This \lq main sequence\rq\ of star-forming galaxies 
has a $1~\sigma$ scatter of only $\approx 0.2$~dex, making it one of the tightest known scaling relations.

\citet{Lara-Lopez10} and \citet{Mannucci10} showed that the scatter in the 
$M_{\rm stellar}-$gas metallicity ($Z_{\rm gas}$) relation (hereafter the MZ relation) is strongly 
correlated with the SFR, and that galaxies in the redshift range $z=0$ to $z\approx 2.5$ populate 
a well-defined plane in the $3$-dimensional space of $M_{\rm stellar}-Z_{\rm gas}-$SFR. 
{\citet{Mannucci10} noted that this relation evolves, breaking down at $z\gtrsim 3$, whith  
\citet{Salim15} reporting even stronger evolution.} The 
current physical interpretation {of the MZ-SFR dependence}  
is that when galaxies accrete large quantities of gas, their SFR increases, and the
(mostly) low-metallicity accreted gas dilutes the metallicity of the ISM (e.g. \citealt{Dave12}; 
\citealt{DeRossi15}). A corollary of this interpretation is that there should be a correlation between 
the scatter in the MZ relation and the gas content of galaxies. Whether the residuals of the MZ relation 
are more strongly correlated with the gas content than with the SFR 
would depend on whether the gas metallicity is primarily set by the dilution of the ISM due to accretion, or by the enrichment 
due to recent star formation. In reality both should play an important role. 

\citet{Hughes13}, \citet{Bothwell13} and \citet{Lara-Lopez13} show that the residuals of the MZ relation 
are also correlated with the atomic hydrogen (HI) content of galaxies, and that the scatter in the correlation 
with HI is smaller than in the correlation with the SFR.
 \citet{Bothwell15} extended the latter work to include molecular hydrogen (H$_2$) and argue that the correlation between 
the residuals relative to the MZ fits are more strongly correlated with the H$_2$ content than with the SFR of galaxies.

In parallel there have been extensive studies on the scaling relations between gas content, $M_{\rm stellar}$ and SFR. 
Local surveys such as the Galex Arecibo SDSS Survey (GASS; \citealt{Catinella10}), the
CO Legacy Database for GASS (COLD GASS; \citealt{Saintonge11}), the Herschel Reference Survey (HRS;
\citealt{Boselli14c} and \citealt{Boselli14b}), the ATLAS$^{\rm 3D}$ \citep{Cappellari11} and the 
APEX Low-redshift Legacy Survey for MOlecular Gas (ALLSMOG; \citealt{Bothwell14}), have allowed the exploration 
of the gas content of galaxies selected by $M_{\rm stellar}$. Analysis of these data revealed that $M_{\rm H_2}/M_{\rm HI}$ 
correlates with $M_{\rm stellar}$, and $M_{\rm HI}/M_{\rm stellar}$ anti-correlates 
with $M_{\rm stellar}$ (e.g. \citealt{Saintonge11}; \citealt{Catinella10}). Such local surveys also allow investigating how galaxy properties 
correlate with morphology: both $M_{\rm HI}/M_{\rm stellar}$ and $M_{\rm H_2}/M_{\rm stellar}$ decrease from 
irregulars and late-type galaxies to early-type galaxies \citep{Boselli14b}. In addition, the gas fractions 
decrease with increasing stellar mass surface density (\citealt{Catinella10}; \citealt{Brown15}).  

Surveys targeting star-forming galaxies at $z>0$ allow one to investigate if $z=0$ 
scaling relations persist, and how they evolve. The ratio $M_{\rm H_2}/M_{\rm stellar}$ increases 
by a factor of $\approx 5$ from $z=0$ to $z=2.5$ at fixed 
$M_{\rm stellar}$ (e.g. \citealt{Saintonge11}; \citealt{Geach11}; \citealt{Tacconi13}; \citealt{Santini13};
\citealt{Saintonge13}; \citealt{Bothwell14}; \citealt{Dessauges-Zavadsky14}). 
\citet{Santini13} presented measurements of dust masses and gas metallicities 
for galaxies in the redshift range $0.1\lesssim z\lesssim 3$. These authors also inferred gas masses by
assuming a relationship between the dust-to-gas mass ratio and the gas metallicity. The sample is biased
to galaxies with relatively high SFRs and dust masses, and thus most of the gas derived from dust masses 
is expected to be molecular. They showed that the (inferred) gas fraction in galaxies correlates strongly 
with $M_{\rm stellar}$ and SFR, with little scatter in gas fraction at a given $M_{\rm stellar}$ and SFR. 
This behaviour is similar to that of the ISM metallicity. The correlation has not been confirmed yet with 
alternative tracers of molecular gas such as for example carbon monoxide.

More fundamental relations presumably exhibit smaller scatter. The $Z_{\rm gas}-M_{\rm stellar}$ and 
gas fraction$-M_{\rm stellar}$ correlations have a larger scatter ($1~\sigma$ scatter of $\approx 0.35$~dex, 
e.g. \citealt{Hughes13}, and $\approx 0.5$~dex; e.g. \citealt{Catinella10,Saintonge11}, respectively) 
than the SFR$-M_{\rm stellar}$ correlation ($1~\sigma$ scatter of $\approx 0.2$~dex; e.g. 
\citealt{Brinchmann04}; \citealt{Damen09}; \citealt{Santini09}; \citealt{Rodighiero10}).
However, the scatter may of course be affected by measurement errors. 

Although these relations provide valuable insight, ultimately 
they cannot by themselves 
distinguish between cause and effect. Cosmological simulations of galaxy formation are excellent testbeds since they 
allow modellers 
to examine causality directly. Provided that the simulations reproduce the observed scaling relations, 
they can be used to build understanding of how galaxies evolve, and predict how scaling relations are established, 
how they evolve, and which processes determine the scatter around the mean trends.

In this paper we explore scaling relations between galaxies from the \lq Evolution and Assembly of GaLaxies and
their Environments\rq\ (\eagle\, \citealt{Schaye14}) suite of cosmological hydrodynamical simulations.
The \eagle\ suite comprises a number of cosmological simulations performed at a range of numerical resolution, 
in periodic volumes with a range of sizes, and using a variety of subgrid implementations to model physical 
processes below the resolution limit. The subgrid parameters of the \eagle\ reference model are calibrated to the $z=0$ 
galaxy stellar mass function,  
galaxy stellar mass - black hole mass relation, and galaxy stellar mass - size 
relations (see \citealt{Crain15} for details and motivation). We use the method described 
in \citet{Lagos15} to calculate the atomic and molecular 
hydrogen contents of galaxies. The \eagle\ reference model reproduces many observed galaxy 
relations that were not part of the calibration set, such as the evolution of the galaxy stellar 
mass function \citep{Furlong14}, of galaxy sizes (\citealt{Furlong15b}), of their 
optical colours \citep{Trayford15}, and of their atomic \citep{Bahe15} and 
molecular gas content \citep{Lagos15}, amongst others.

This paper is organised as follows. In $\S$~\ref{EagleSec} we give a
brief overview of the simulation, the subgrid physics included in the \eagle\ reference model, 
and how we partition ISM gas 
into ionised, atomic and molecular fractions. We first present the evolution of gas fractions 
in the simulation and compare with observations in $\S$~\ref{EvoGasFrac}. In $\S$~\ref{localU} we describe a 
principal component analysis of \eagle\ galaxies and 
demonstrate the presence of a fundamental plane of star formation in the simulations. We characterise this plane
and how galaxies populate it as a function of redshift and metallicity. We also show that observed galaxies 
show very similar correlations. We discuss our results and present our conclusions in $\S$~\ref{ConcluSec}. 
In Appendix~\ref{ConvTests} we present `weak' and `strong' convergence tests (terms introduced by \citealt{Schaye14}), 
and in Appendix~\ref{ModelTests} 
we show how variations in the subgrid model parameters affect the fundamental plane of star formation.

\section{The EAGLE simulation}\label{EagleSec}

\begin{table}
\begin{center}
  \caption{Features of the Ref-L100N1504 simulation used in this paper. The row list:
    (1) comoving box size, (2) number
    of particles, (3) initial particle masses of gas and (4) dark
    matter, (5) comoving gravitational
    softening length, and (6) maximum proper comoving Plummer-equivalent
    gravitational softening length. Units are indicated in each row. \eagle\
    adopts (5) as the softening length at $z\ge 2.8$, and (6) at $z<2.8$. }\label{TableSimus}
\begin{tabular}{l l l l}
\\[3pt]
\hline
& Property & Units & Value \\
\hline
(1)& $L$ & $[\rm cMpc]$ & $100$\\
(2)& \# particles &  & $2\times 1504^3$ \\
(3)& gas particle mass & $[\rm M_{\odot}]$ & $1.81\times 10^6$\\
(4)& DM particle mass & $[\rm M_{\odot}]$ & $9.7\times 10^6$\\
(5)& Softening length & $[\rm ckpc]$ & $2.66$\\
(6)& max. gravitational softening & $[\rm pkpc]$& $0.7$ \\
\hline
\end{tabular}
\end{center}
\end{table}

The \eagle\ simulation suite\footnote{See {\tt
    \footnotesize http://eagle.strw.leidenuniv.nl} and {\tt
    \footnotesize http://www.eaglesim.org/} for images, movies and data
  products. A database with many of the galaxy properties in \eagle\ is publicly available and described in 
\citet{McAlpine15}.}  (described in detail by \citealt{Schaye14}, hereafter
S15, and \citealt{Crain15}, hereafter C15) consists of a large number of cosmological
hydrodynamical simulations with different resolution, volumes and physical models,
adopting the cosmological parameters of \citet{Planck14}. S15
introduced a reference model, within which the parameters of the
sub-grid models governing energy feedback from stars and accreting BHs were calibrated to ensure a
good match to the $z=0.1$ galaxy stellar mass function and 
the sizes of present-day disk galaxies.
C15 discussed in more detail the physical motivation for the 
sub-grid physics models in \eagle\ and show how the calibration of the free parameters was performed.
\citet{Furlong14}
presented the evolution of the galaxy stellar mass function 
and found that the agreement with observations extends to $z\approx 7$. The 
optical colours of the $z=0.1$ galaxy population and galaxy sizes are in reasonable agreement with observations 
 (\citealt{Trayford15}; \citealt{Furlong15b}).

In Table~\ref{TableSimus} we summarise technical details
of the simulation used in this work, including the number of
particles, volume, particle masses, and spatial resolution.  
In Table~\ref{TableSimus}, pkpc denotes proper kiloparsecs. 

A major aspect of the \eagle\ project is the use of
state-of-the-art sub-grid models that capture unresolved physics.
We briefly discuss the sub-grid physics modules adopted by \eagle\ in
$\S$~\ref{sub-gridsec}, but we refer to S15 for more details. 
In order to distinguish models with different
parameter sets, a prefix is used. For example, Ref-L100N1504
corresponds to the reference model adopted in a simulation with the
same box size and particle number as L100N1504.  
We perform convergence tests in Appendix~\ref{ConvTests}.
In Appendix~\ref{ModelTests} we 
present a comparison between model variations of \eagle\ in Appendix~\ref{ModelTests}. 

The \eagle\ simulations were performed using an extensively modified
version of the parallel $N$-body smoothed particle hydrodynamics (SPH)
code {\sc gadget-3} (\citealt{Springel08}; \citealt{Springel05b}).
Among those modifications are updates to the SPH technique, which are collectively referred to as 
`Anarchy' (see \citealt{Schaller15b} for an analysis of the impact that these changes have on 
the properties of simulated galaxies compared to standard SPH). We use {\sc SUBFIND} 
 (\citealt{Springel01}; \citealt{Dolag09}) to identify self-bound overdensities of particles within halos (i.e. substructures). 
These substructures are the galaxies in \eagle. 

Throughout the paper we make extensive comparisons between stellar
mass, SFR, HI and H$_2$ masses and gas metallicity. Following S15, all these
properties are measured in spherical apertures of $30$~pkpc.
The effect of the aperture is minimal as shown by \citet{Lagos15} and S15.

\subsection{Sub-grid physics modules}\label{sub-gridsec}

\begin{itemize}
\item {\it Radiative cooling and photoheating rates.} Cooling and
  heating rates are computed on an element-by-element basis for gas in
  ionisation equilibrium exposed to a UV and X-ray background (model from \citealt{Haardt01}) 
  and to the Cosmic Microwave Background.
  The $11$ elements that dominate the cooling rate 
  are followed individually (i.e. H,~He,~C,~N,~O,~Ne,~Mg,~S,~Fe,~Ca,~Si). 
  (See \citealt{Wiersma09b} and S15 for details).

\item {\it Star formation.} Gas particles that have cooled to reach
  densities greater than $n^{\ast}_{\rm H}$ are eligible for
  conversion to star particles, where $n^{\ast}_{\rm H}$ is a function of metallicity, as described in        
  \citet{Schaye04} and S15.
  Gas particles with $n_{\rm H}>n^{\ast}_{\rm H}$ are assigned a SFR, $\dot{m}_{\star}$ \citep{Schaye08}:
\begin{equation}
\dot{m}_{\star}=m_{\rm g}\,A\,(1\,{\rm M}_{\odot}\,{\rm pc}^{-2})^{-n}\, \left(\frac{\gamma}{G}\,f_{\rm g}\,P\right)^{(n-1)/2},
\label{SFlaw}
\end{equation}

\noindent where $m_{\rm g}$ is the mass of the gas particle,
$\gamma=5/3$ is the ratio of specific heats, $G$ is the gravitational
constant, $f_{\rm g}$ is the mass fraction in gas (which is unity for
gas particles), $P$ is the total pressure. \citet{Schaye08} demonstrate that under the assumption of vertical hydrostatic 
equilibrium, Eq.~\ref{SFlaw} is equivalent to the Kennicutt-Schmidt relation, $\dot{\Sigma}_{\star}=A(\Sigma_{\rm g}/1\,\rm M_{\odot}\,pc^{-2})^n$ \citep{Kennicutt98}, where 
$\dot{\Sigma}_{\star}$ and $\Sigma_{\rm g}$ are the surface densities of SFR and gas, and $A=1.515\times 10^{-4}\,\rm M_{\odot}\,\rm yr^{-1}\,kpc^{-2}$ and $n=1.4$ are chosen to reproduce
the observed Kennicutt-Schmidt relation, scaled to
a Chabrier initial mass function (IMF; \citealt{Chabrier03}).  In \eagle\ we adopt a stellar IMF of 
\citet{Chabrier03}, with minimum and maximum masses of $0.1\,\rm M_{\odot}$ and $100\,\rm M_{\odot}$. 
A global temperature floor,
$T_{\rm eos}(\rho)$, is imposed, corresponding to a polytropic
equation of state, 

\begin{equation}
P\propto \rho^{\gamma_{\rm eos}}_{\rm g}, 
\label{EoS}
\end{equation}

\noindent where $\gamma_{\rm eos}=4/3$. Eq~\ref{EoS} is normalised
to give a temperature $T_{\rm eos}=8\times 10^3$~K at $n_{\rm H}=10^{-1}\,\rm cm^{-3}$,
which is typical of the warm ISM (e.g. \citealt{Richings14a}).

\item {\it Stellar evolution and enrichment.}
Stars on the Asymptotic Giant Branch (AGB), massive stars (through winds) and 
supernovae (both core collapse and type Ia) lose mass and metals that are tracked 
 using the yield tables of \citet{Portinari98}, \citet{Marigo01},
and \citet{Thielemann03}. Lost mass and metals are
added to the gas particles that are within the SPH kernel of the given star particle
(see \citealt{Wiersma09} and S15 for details). 

\item {\it Stellar Feedback.}
  The method used in \eagle\ to represent energetic feedback
  associated with star formation (which we refer to as `stellar
  feedback') was motivated by \citet{DallaVecchia12}, and consists of
  a stochastic selection of neighbouring gas particles that are heated
  by a temperature of $10^{7.5}$~K. A fraction of the energy, $f_{\rm th}$ from
  core-collapse supernovae is injected into the ISM $30$~Myr after the
  star particle forms. This fraction depends on the local metallicity and gas density, as introduced by 
  S15 and C15. The calibration of \eagle\ described in C15 leads $f_{\rm th}$ to range from $0.3$ to $3$, with the median of 
  $f_{\rm th}$ $=0.7$ for the Ref-L100N1504 simulation at $z=0.1$ (see S15). 

\item {\it Black hole growth and AGN feedback.} When halos become more
  massive than $10^{10}\,h^{-1}\,{\rm M}_{\odot}$, they are seeded
  with BHs of mass $10^5\,h^{-1}\,{\rm M}_{\odot}$. 
  Subsequent gas accretion episodes and mergers make BHs grow
  at a rate that is computed following the modified Bondi-Hoyle
  accretion rate of \citet{Rosas-Guevara13} and S15. This modification considers the
  angular momentum of the gas, which reduces the accretion rate compared
  to the standard Bondi-Hoyle rate, if the tangential velocity of the
  gas is similar to, or larger than, the local sound speed. 
  The Eddington limit
  is imposed as an upper limit to the accretion rate onto BHs.
  In addition, BHs can grow by merging.

  For AGN feedback, a similar model to the stochastic model of
  \citet{DallaVecchia12} is applied. Particles surrounding the BH 
   are chosen randomly and heated by a temperature $\Delta T_{\rm
    AGN}=10^{8.5}$~K in the reference simulation {(Table~\ref{TableSimus})} and $\Delta T_{\rm
    AGN}=10^{9}$~K in the recalibrated simulation {(used in Appendix~\ref{ConvTests})}.
\end{itemize}

\subsection{Determining neutral and molecular gas fractions}\label{SecNeutralFractions}

We estimate the transitions from ionised to neutral, and from neutral to 
molecular gas following \citet{Lagos15}. Here we briefly describe how we model these transitions.

\begin{itemize}
\item {\it Transition from ionised to neutral gas.} We use the 
fitting function of \citet{Rahmati13}, who studied the  neutral gas fraction
in cosmological simulations by coupling them to a full radiative
transfer calculation with TRAPHIC \citep{Pawlik08}. This fitting function considers collisional ionisation, photo-ionisation 
by a homogeneous UV background and by recombination radiation, {and was shown to be a good approximation at $z\lesssim 5$}. 
We adopt the model of \citet{Haardt01} for the UV background. 
Note that we ignore the effect of local sources. 
\citet{Rahmati13b} showed that star-forming galaxies produce a galactic scale photoionisation
rate of $\sim 10^{-13}\rm \, s^{-1}$, which is of a similar magnitude as the UV background at $z=0$, and smaller
than it at $z>0$, favouring our approximation.
We use this function to calculate the neutral fraction on a
particle-by-particle basis from the gas temperature and density, and
the assumed UV background.
\item {\it Transition from neutral to molecular gas.} We use the model of 
\citet{Gnedin11} to calculate the fraction of molecular hydrogen on a particle-by-particle basis.
This model consists of a phenomenological model for H$_2$
formation, approximating how H$_2$ forms on the surfaces of dust grains and is destroyed by the interstellar radiation field.
 \citet{Gnedin11} produced a suite of zoom-in simulations of galaxies with a large dynamic range in metallicity and 
ionisation field in which H$_2$ formation was followed explicitly.
Based on the outcome of
these simulations, the authors parametrised 
the fraction of H$_2$-to-total neutral gas as a function of 
the dust-to-gas ratio and the interstellar radiation field. We use this parametrisation here to model the transition from HI to H$_2$.
We assume that the dust-to-gas mass ratio scales with the local
metallicity, and the radiation field with the local surface
density of star formation, which we estimate from the properties of
gas particles (see Eq.~\ref{SFlaw}). The surface densities of SFR and neutral gas were obtained using the respective volume densities and 
the local Jeans length, for which we assumed local hydrostatic equilibrium
(\citealt{Schaye01}; \citealt{Schaye08}). Regarding the assumption of the constant dust-to-metal ratio, recent work, for example by 
\citet{Herrera-Camus12}, has shown that deviations from this relation arise at dwarf galaxies with low metallicity ($\lesssim 0.2\,z_{\odot}$). 
In our analysis, we include galaxies that are well resolved in \eagle, i.e. $M_{\rm stellar}>10^9\,\rm M_{\odot}$ (see S15 for details), 
and therefore 
we expect our assumption of a constant dust-to-metal ratio to be a good approximation.
\end{itemize}

\citet{Lagos15} also used the models of \citet{Krumholz13} and \citet{Gnedin14} to calculate the H$_2$ fraction for individual particles, 
finding similar results. 
We therefore focus here on one model only. Throughout the paper we make use of the Ref-L100N1504 simulation and we simply refer to 
it as the \eagle\ simulation. If any other simulation is used we mention it explicitly. 
{We also limit our galaxy sample to $z<4.5$, the redshift regime in which the fitting function of 
\citet{Rahmati13} provides a good approximation to the neutral gas fraction.}

\section{The evolution of gas fractions in EAGLE}\label{EvoGasFrac}

In \citet{Lagos15} we analysed the $z=0$ H$_2$ mass scaling relations and \citet{Bahe15} analysed HI mass scaling relations. Here 
we show how these scaling relations evolve and compare with observations. 
We define the neutral and molecular gas fractions as
\begin{eqnarray}
 f_{\rm gas,neutral}&\equiv&\frac{(M_{\rm HI}+M_{\rm H_2})}{(M_{\rm HI}+M_{\rm H_2}+M_{\rm stellar})},\label{fgasneutral}\\
 f_{\rm gas,mol}&\equiv&\frac{(M_{\rm H_2})}{(M_{\rm H_2}+M_{\rm stellar})}.\label{fgasmol} 
\end{eqnarray}

\noindent Note that we do not include the mass of ionised hydrogen in Eqs.~\ref{fgasneutral}~and~\ref{fgasmol} 
because it is hard to estimate {observationally}, which would make the task of comparing simulation with observations difficult. 
Similarly, in Eq.~\ref{fgasmol} we do not include HI in the denominator 
because for observations at $z>0$ there is no HI information.
\begin{figure}
\begin{center}
\includegraphics[width=0.49\textwidth]{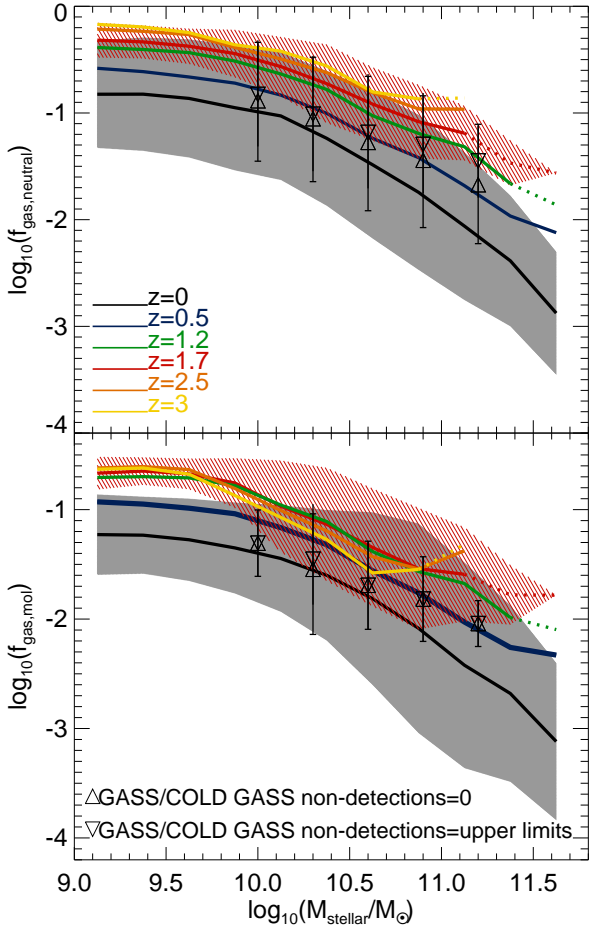}
\caption{The neutral (Eq.~\ref{fgasneutral}; top panel) and molecular (Eq.~\ref{fgasmol}; bottom panel) gas fractions 
 as a function 
of stellar mass at $z=0$, $z=0.5$, $z=1.2$, $z=1.7$, $z=2.5$, and $z=3$, as labelled, for the \eagle\ simulation. 
Lines show the median relations, and the hatched regions show the $16^{\rm th}$ to $84^{\rm th}$ percentiles. 
For clarity, the latter are shown only for $z=0$ and $z=1.7$ galaxies. Solid lines show bins with $>10$ galaxies, while dotted 
lines show bins where the number of galaxies drops below $10$. Observations at $z=0$ from GASS and COLD GASS 
 are shown using two symbols: 
upside down triangles show the medians if upper limits are taken for the non detections, and 
triangles show the median when we set HI and H$_2$ masses to zero for the non detections. The true median is bracketed by 
these two values. Errorbars show the $1~\sigma$ scatter. \eagle\ and the observations agree within $0.5$~dex.}
\label{ScalingsofH}
\end{center}
\end{figure}

Fig.~\ref{ScalingsofH} shows $f_{\rm gas,neutral}$ and 
$f_{\rm gas,mol}$ as a function of stellar mass, for 
all galaxies with $M_{\rm stellar}>10^9\,\rm M_{\odot}$ at redshifts $0\le z\le 3$ in EAGLE, {to 
match the observed redshift range.}
Both gas fractions increase with redshift at fixed stellar mass and decrease with stellar mass at a given redshift.
The slopes of the relations $f_{\rm gas,neutral}$-stellar mass and 
$f_{\rm gas,mol}$-stellar mass do not change significantly with redshift, but the normalisations evolve rapidly.
The increase of $f_{\rm gas,neutral}$ at fixed stellar mass from $z=0$ to $z\approx 2.5$ is $\approx 0.6$~dex. 
At $2.5<z<3$, $f_{\rm gas,neutral}$ shows a very weak or no evolution. 
The molecular gas fraction increases by $\approx 0.6$~dex at fixed stellar mass from $z=0$ to $z\approx 1.2$, which is faster than 
the evolution of $f_{\rm gas,neutral}$. At $1.7\lesssim z\lesssim 3$, 
 $f_{\rm gas,mol}$ shows little evolution in the stellar mass range 
$10^{9}\,\rm M_{\odot}\lesssim M_{\rm stellar} \lesssim 5\times 10^{9}\,\rm M_{\odot}$, and 
  a weak decrease with redshift for $M_{\rm stellar}\gtrsim 5\times 10^{9}\,\rm M_{\odot}$.
The increase in the neutral and molecular gas fractions from $z=0$ to $z\approx 2$ is due to the increasing accretion rate 
onto galaxies in the same redshift range.
The weak decrease in $f_{\rm gas,mol}$ at $z\gtrsim 2$ 
is due to galaxies at those redshifts having much higher interstellar radiation fields and lower gas metallicities 
than galaxies at $z<2$, conditions that hamper the formation of H$_2$ by dissociating H$_2$ and 
 reducing the amount of dust available to act as catalyst for H$_2$, respectively. 
 A significant amount of the gas with densities $>0.1\,\rm cm^{-3}$ 
remains atomic under these harsh ISM conditions, causing $f_{\rm gas,neutral}$ to continue increase with increasing redshift 
at fixed stellar mass (at least up to $z\approx 5$), 
whereas $f_{\rm gas,mol}$ decreases. On average, both $f_{\rm gas,neutral}$ and $f_{\rm gas,mol}$ increase 
by $\approx 0.6-0.7$~dex from $z= 0$ to $z\approx 2$. In the same redshift range, the specific SFR, $\rm sSFR=SFR/M_{\rm stellar}$ 
increases by a factor of $\approx 15$ \citep{Furlong14} in \eagle. 
This difference between the increase in gas fraction and SFR is a consequence of the super-linear power-law index, $n=1.4$, of the
observed star formation law, which is adopted in \eagle\ (Eq.~\ref{SFlaw}; see also discussion in 
$\S\,5.4$ in \citealt{Lagos15}). 

In Fig.~\ref{ScalingsofH} we also compare the $z=0$ \eagle\ result with the observations of GASS and COLD GASS at $z=0$. 
 The observational strategy in GASS and COLD GASS 
was to select all galaxies with $M_{\rm stellar}>10^{10}\,\rm M_{\odot}$ at $z<0.05$ from the Sloan 
Digital Sky Survey Data Release 4 and 
image a subsample of those in HI and CO(1-0). 
\citet{Catinella10} and \citet{Saintonge11} integrated sufficiently long to enable the detection of HI and H$_2$ of $>0.015\times M_{\rm stellar}$ 
 at stellar masses $M_{\rm stellar}>10^{10.6}\,\rm M_{\odot}$, or HI and 
H$_2$ masses $>10^{8.8}\,\rm M_{\odot}$ in galaxies with 
 $10^{10}\,\rm M_{\odot}<M_{\rm stellar}<10^{10.6}\,\rm M_{\odot}$.
In the case of CO observations (for COLD GASS and those discussed below), 
we adopted a conversion factor $X=2\times 10^{-20}\,\rm cm^{-2}\,(K\, km\, s^{-1})^{-1}$ (Milky-Way like; \citealt{Bolatto13}), where X is defined as

\begin{equation}
\frac{N_{\rm H_2}}{\rm cm^{-2}}=X \,\left(\frac{I_{\rm CO(1-0)}}{\rm K\, km\,s^{-1}}\right),
\end{equation}

\noindent where $N_{\rm H_2}$ is the H$_2$ column density and $I_{\rm CO(1-0)}$ is the velocity-integrated $\rm CO(1-0)$ brightness 
temperature (in traditional radio astronomy observational units). 
We show the observational results treating non detections in two different ways: by using the upper limits (upside down triangles), 
and by setting the HI and H$_2$ masses to zero. \eagle\ results are in qualitative agreement with the observations. 
The median relations of \eagle\ and 
GASS plus COLD GASS are at most $0.3$~dex from each other at $M_{\rm stellar}<10^{10}\,\rm M_{\odot}$, 
while the $1~\sigma$ scatter is $\approx 0.5$~dex.
There is some tension at $M_{\rm stellar}\gtrsim 10^{11}\,\rm M_{\odot}$, but we show later that this tension is diminished if we study the 
gas fraction-stellar mass relations in bins of SFR.
\citet{Lagos15} and \citet{Bahe15} analysed in detail how \eagle\ compares with GASS and COLD GASS, 
and we point to those papers for more comparisons (e.g. radial profiles, 
stellar concentrations, SFR efficiencies, etc.).

\begin{figure*}
\begin{center}
\includegraphics[width=0.76\textwidth]{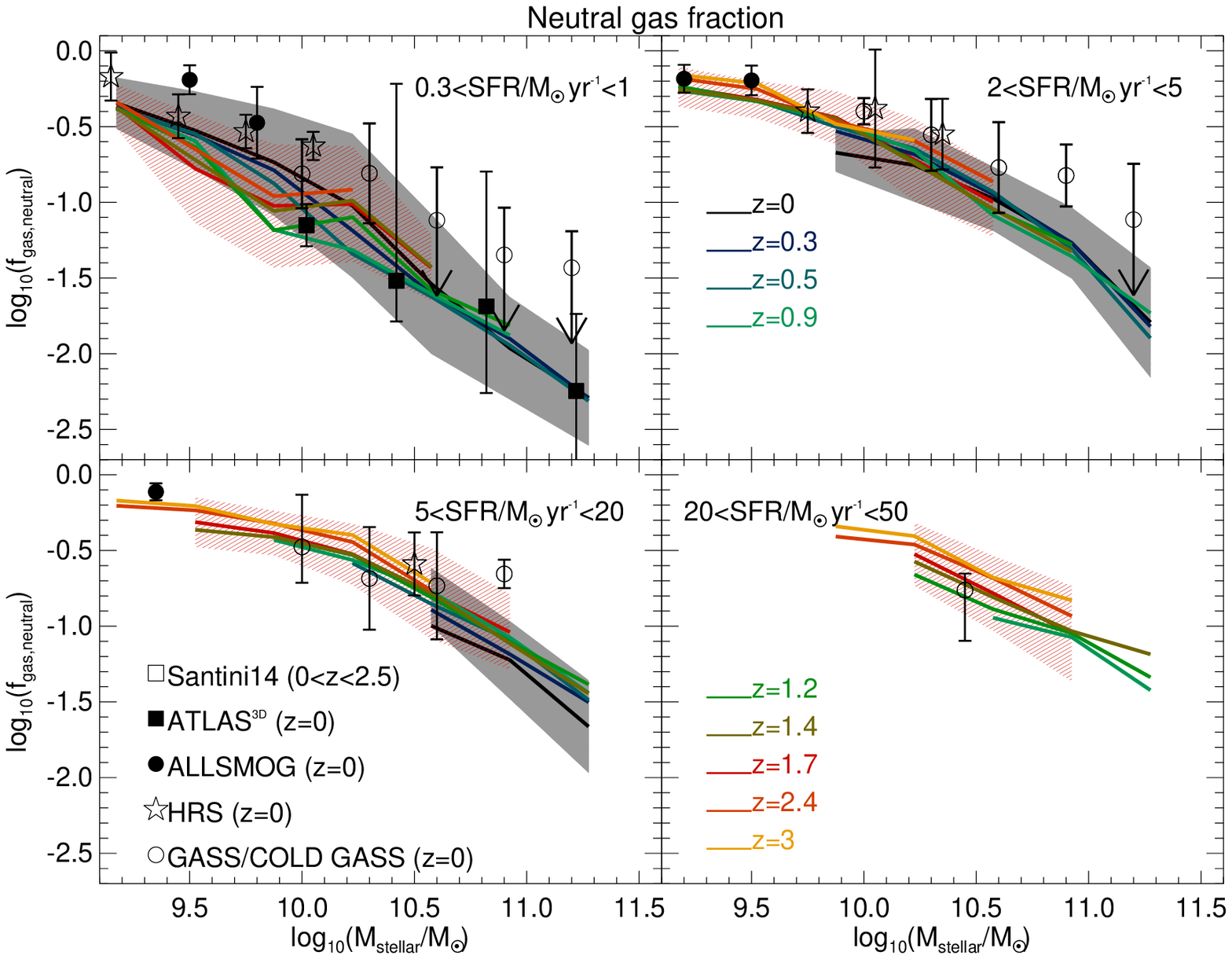}
\includegraphics[width=0.76\textwidth]{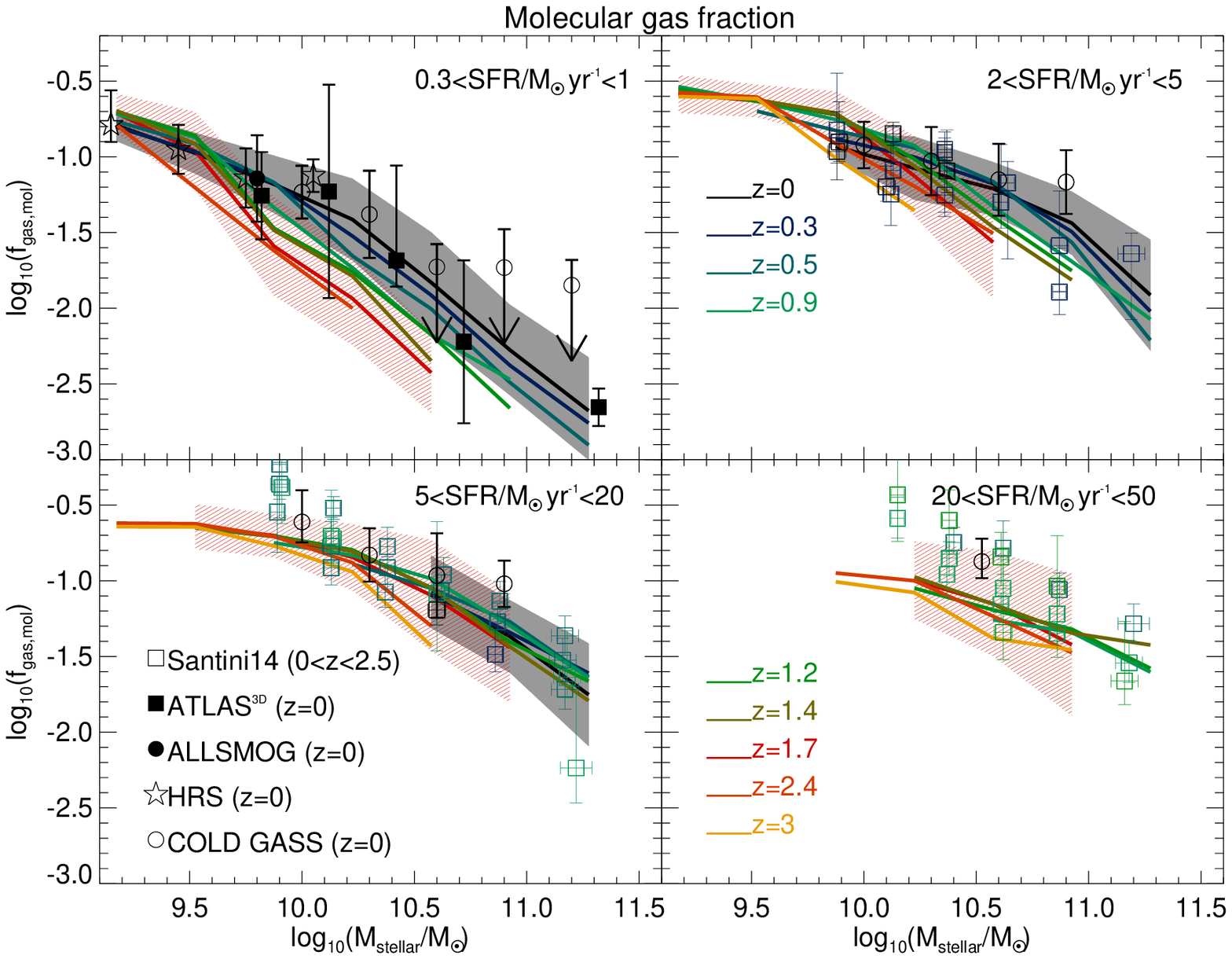}
\caption{The neutral (Eq.~\ref{fgasneutral}; top panels) and molecular (Eq.~\ref{fgasmol}; bottom panels) gas fractions as a function of stellar mass 
in bins of SFR, as labelled in each panel. 
For \eagle\ galaxies, lines show the medians, while the $16^{\rm th}$ to $84^{\rm th}$ percentiles are shown as shaded regions 
(but only for $z=0$ and $z=1.7$ galaxies).
We only show bins that have $>10$ galaxies.
Symbols show the observational result of GASS and COLD GASS  (\citealt{Catinella10} and \citealt{Saintonge11}; open circles), 
 HRS (\citealt{Boselli14c}; stars), ATLAS$^{\rm 3D}$ (\citealt{Cappellari11}; \citealt{Young11}; 
\citealt{Serra12}; \citealt{Davis14}; filled squares), ALLSMOG (\citealt{Bothwell14}; filled circles) and \citet{Santini13} (open squares).
Observations have been coloured according 
to their redshift following the same colour code we used for \eagle\ galaxies (labelled in the right panels). 
We see only weak evolution once 
the gas fraction-stellar mass relation is investigated in bins of SFR, with the remaining evolution being mostly due to 
evolution of the median SFR within each SFR bin. Overall, \eagle\ agrees well with the observations within $0.3$~dex (with the scatter 
on the observations being of a similar magnitude).}
\label{HNeutralMFEagleSF2}
\end{center}
\end{figure*}

In Fig.~\ref{HNeutralMFEagleSF2} we show the dependence of $f_{\rm gas,neutral}$ and $f_{\rm gas,mol}$ on stellar mass in four bins 
of SFR. In \eagle, both $f_{\rm gas,neutral}$ and $f_{\rm gas,mol}$ show very weak or no evolution at fixed stellar mass and SFR.
 Thus, the evolution seen in Fig.~\ref{ScalingsofH} is related to the increase of the median SFR with redshift at fixed stellar mass. 
Note that in the top-left panel of Fig.~\ref{HNeutralMFEagleSF2} there is a weak evolution of $f_{\rm gas,neutral}$ and $f_{\rm gas,mol}$ with redshift, but this is mostly due to 
the SFR slightly changing at fixed stellar mass 
within the allowed range ($0.3\,\rm M_{\odot}\,yr^{-1}<SFR<1\,\rm M_{\odot}\,yr^{-1}$). 
Galaxies with SFRs closer to $1\,\rm M_{\odot}\,yr^{-1}$ have higher $f_{\rm gas,neutral}$ and $f_{\rm gas,mol}$ than 
those galaxies having SFRs closer to $0.3\,\rm M_{\odot}\,yr^{-1}$. This means that the weak evolution displayed by \eagle\ in the scaling relations 
shown in Fig.~\ref{HNeutralMFEagleSF2} are simply due to the strong correlation between gas fraction (either neutral or molecular) and SFR. 
 Since the SFR is more strongly correlated with H$_2$ 
than with total neutral gas in \eagle\ \citep{Lagos15}, 
we see more variations in the $f_{\rm gas,mol}$-stellar mass relation than in the 
$f_{\rm gas,neutral}$-stellar mass even if we select narrow ranges of SFR (see for example the SFR bin $0.3\rm M_{\odot}\,yr^{-1}<\rm SFR<1\rm M_{\odot}\,yr^{-1}$ 
in Fig.~\ref{HNeutralMFEagleSF2}).
We come back to this 
in $\S$~\ref{localU}.

In Fig.~\ref{HNeutralMFEagleSF2} we also show observations from GASS and COLD GASS (\citealt{Catinella10}; \citealt{Saintonge11}), 
HRS (\citealt{Boselli14c} and \citealt{Boselli14b}), the ALLSMOG (\citealt{Bothwell14}), 
ATLAS$^{\rm 3D}$ (\citealt{Cappellari11}; \citealt{Young11};
\citealt{Serra12}; \citealt{Davis14}) and from \citet{Santini13}. HRS is 
a volume-limited survey, containing $323$ galaxies at distances between $15$ and $25$~Mpc, and stellar masses 
$M_{\rm stellar}\gtrsim 10^{9}\,\rm M_{\odot}$. HRS galaxies were followed-up to image CO(1-0), while HI data 
was obtained from \citet{Giovanelli05} and \citet{Springob05} (see \citealt{Boselli14c} for details). 
ALLSMOG is a survey designed to obtain H$_2$ masses for galaxies with $3\times 10^{8}\,\rm M_{\odot}\lesssim 
M_{\rm stellar}\lesssim 10^{10}\,\rm M_{\odot}$, at distances between $40$ and $110$~Mpc. 
HI data for ALLSMOG was obtained from \citet{Meyer04}, 
\citet{Springob05} and \citet{Haynes11}.
We use here the first release of \citet{Bothwell14} of $42$ galaxies.
The ATLAS$^{\rm 3D}$ survey is a volume-limited survey of $260$ early-type galaxies with 
resolved kinematics of the stellar component and ionised gas (\citealt{Cappellari11}). 
\citet{Young11} and \citet{Serra12} presented 
measurements of CO(1-0) and HI masses for ATLAS$^{\rm 3D}$ galaxies, respectively, while stellar masses and SFRs for these galaxies were 
presented in \citet{Cappellari13} and \citet{Davis14}, respectively. 
\citet{Santini13} presented measurements of $f_{\rm gas,mol}$ as a function of stellar mass
in the redshift range $0.1\lesssim z \lesssim 3$. Santini et al. measured dust masses from Herschel photometry, and
inferred a gas mass by using measured gas metallicities and a dust-to-gas mass ratio that is metallicity dependent.
 Since all their sampled galaxies have relatively high SFRs and dust masses,
 most of the gas mass derived from dust masses is expected to be molecular.
We show the observations in bins of SFR, as we did for \eagle. Some of the results from 
GASS and COLD GASS surveys are upper limits due to 
non-detections of HI and/or CO(1-0).  
From the observational side, we find broad agreement between the different surveys, even though they cover different stellar 
mass ranges and redshifts. 
We emphasise that this is the first demonstration of the stellar mass-SFR-gas fraction connection across redshifts in observational data.  
This 3-parameter relation is thus a property of real galaxies and hence is a significant observational result.

Fig.~\ref{HNeutralMFEagleSF2} shows that EAGLE's  
predictions are in good agreement with the observations, within the dispersion of the data 
and the scatter of the simulation, for all the SFR bins. The median relation of \eagle\ is usually $\lesssim 0.1-0.2$~dex from the median 
relation in the observations, but this offset of much smaller than the observed scatter ($\approx 0.3-0.5$~dex).  
For the highest SFR bin ($20\,\rm M_{\odot}\, yr^{-1}< SFR<50\,\rm M_{\odot}\, yr^{-1}$) 
there is only one observational data point for $f_{\rm gas,neutral}$ due to the lack of HI information. This data point 
corresponds to the median of $4$ galaxies belonging to GASS and COLD GASS. 
In the simulation there are no galaxies with those SFRs at $z=0$, which is due to its limited volume.
GASS and COLD GASS are based on SDSS, which has a volume at $z<0.1$ 
that is $\approx 10$ times larger than the volume of the Ref-L100N1504 simulation. Thus, the non existence of such galaxies 
at $z=0$ in \eagle\ is not unexpected.  

From Fig.~\ref{HNeutralMFEagleSF2} one concludes that there is a relation between 
 $f_{\rm gas,neutral}$, stellar mass and SFR, and between $f_{\rm gas,mol}$, stellar mass and SFR. 
These planes exist in both the simulation and the observations, which is a significant result for \eagle\ and observations.
This motivates us to analyse more in detail how fundamental these 
correlations are compared to the more widely-known scaling relations introduced in $\S$~$1$. 
With this in mind we perform a principal component analysis in the next section.

\section{The fundamental plane of star formation}\label{localU}

\subsection{A principal component analysis}\label{PCASec}

With the aim of exploring which galaxy correlations are most fundamental and how the gas fraction-SFR-stellar mass relations fit into that picture, 
we perform a principal component analysis (PCA) over $7$ properties of galaxies in the Ref-L100N1504 simulation. 
{We do not include redshift in the list of properties because we decide to only include properties of galaxies to make the interpretation of 
PCA more straightforward. However, 
we do analyse possible redshift trends in $\S$~\ref{FPGas}.}
We include 
all galaxies in \eagle\ with $M_{\rm stellar}>10^9\rm M_{\odot}$, SFR$>0.01\,\rm M_{\odot}\,yr^{-1}$, 
$M_{\rm neutral}>10^7\rm\,M_{\odot}$ and at $0\le z\le 4.5$ in the PCA. 
Here $M_{\rm neutral}$ is the HI plus H$_2$ mass.
The PCA uses orthogonal transformations to find linear combinations of variables.

  PCA is designed to return as the first principal component the combination of variables that contains the largest possible variance 
of the sample, with each subsequent component having the largest possible variance under the constraint that it is orthogonal to the 
previous components. 
In order to perform the PCA, we renormalise galaxy properties in logarithmic space by 
subtracting the mean and dividing by the standard deviation of each galaxy property. 
Table~\ref{TablePCA} shows the variables that
were included in the PCA and shows the first three principal components. We apply equal weights to the galaxies in the PCA, {which 
is justified 
by the fact that the redshift distribution of galaxies with $M_{\rm stellar}>10^9\,\rm M_{\odot}$ is close to flat 
(see bottom panel of Fig.~\ref{residuals}).} 

\begin{table*}
\begin{center}
  \caption{Principal component analysis (PCA) of galaxies in the Ref-L100N1504 simulation. Galaxies
with $M_{\rm stellar}>10^9\,\rm M_{\odot}$, $\rm SFR>0.01\,M_{\odot}\,yr^{-1}$, 
$M_{\rm H_2}/(M_{\rm H_2}+M_{\rm stellar})>0.01$ and $0\le z \le 4.5$ were included in the analysis.
The PCA was conducted with the variables: stellar mass, star formation rate, 
metallicity of the star-forming gas ($Z_{\rm SF,gas}$), {molecular, atomic and neutral gas masses} and the half-mass stellar
radius $r_{\rm 50,st}$. 
We adopt $Z_{\odot}=0.0127$.
 Before performing the PCA, we renormalise all the components by subtracting the mean and dividing by
the standard deviation (all in logarithm). In the table we show the property each component relates to, but we remind 
the reader that we renormalise them before performing the PCA.  
{The three first principal components account for $55$\%  $24$\% and
$14$\%, respectively, of the total variance, and therefore account together for $93$\% of the total variance}. The first three
PCA vectors are shown here.}
\label{TablePCA}
\begin{tabular}{l c c c c c c c}
\\[3pt]
\hline
 & (1) & (2) & (3) & (4) & (5) & (6) & (7) \\
\hline
comp. & $\hat{\rm x}_1$ & $\hat{\rm x}_2$ &$\hat{\rm x}_3$ &$\hat{\rm x}_4$ &$\hat{\rm x}_5$ &$\hat{\rm x}_6$ &$\hat{\rm x}_7$ \\
\hline 
Prop. &  $\rm log_{10}\left(\frac{M_{\rm stellar}}{\rm M_{\odot}}\right)$ & $\rm log_{10}\left(\frac{SFR}{M_{\odot}\,yr^{-1}}\right)$ & $\rm log_{10}\left(\frac{Z_{\rm SF,gas}}{Z_{\odot}}\right)$ & $\rm log_{10}\left(\frac{M_{\rm H_2}}{\rm M_{\odot}}\right)$ & $\rm log_{10}\left(\frac{M_{\rm HI}}{\rm M_{\odot}}\right)$& $\rm log_{10}\left(\frac{M_{\rm neutral}}{\rm M_{\odot}}\right)$ & $\rm log_{10}\left(\frac{\rm r_{\rm 50,\star}}{\rm kpc}\right)$\\
\hline
PC1 & $0.31$  & $-0.57$ & $-0.19$ & $-0.15$ & $0.4$ & $0.6$ & $0.06$\\
PC2 & $0.46$ & $0.04$ & $-0.31$ & $-0.51$ & $0.22$ & $-0.61$  & $0.09$\\
PC3 & $-0.19$  & $-0.68$  & $-0.14$ & $0.33$ & $-0.33$  & $-0.51$  & $0.002$\\
\hline
\end{tabular}
\end{center}
\end{table*}

We find that the first principal component is dominated by the stellar mass, SFR and the {neutral gas mass (and secondarily 
by the atomic gas mass),}
 with weaker dependencies on the molecular gas mass and the gas metallicity. 
This component accounts for {55\%} of the variance 
of the galaxy population.
The relation between the neutral gas fraction, SFR and stellar mass of galaxies define a plane in the 3-dimensional space, which we refer to as 
``the fundamental plane of star formation'', that we 
explore in detail in $\S$~\ref{FPGas}. Since this plane 
accounts for most of the variance, it is one of the most fundamental relations of galaxies. This is an important 
prediction of \eagle.

The second principal component is dominated by the stellar mass, metallicity of the star-forming gas, and molecular and neutral gas masses.
{This component is responsible for $24$\% of the variance} of the galaxy population in \eagle, 
and can be connected with the mass-metallicity relation and how its scatter is correlated with the molecular and neutral gas content. 
Note that molecular gas plays a secondary role compared to the neutral gas fraction.
This will be discussed in $\S$~\ref{MMrelation}.

The third principal component shows a correlation between all the gas components 
(molecular, atomic and neutral), {SFR and secondarily on stellar mass and gas metallicity}. This principal component shows that 
galaxies tend to be simultaneously rich (or poor) {in atomic and neutral (molecular plus atomic) hydrogen}. Note that the half-mass 
radius does not strongly appear in the first three principal components. We find that $r_{\rm 50,\star}$ appears in the 
fourth and fifth principal components, with dependencies on the stellar mass and molecular gas mass (no dependence of 
$r_{\rm 50,\star}$ on gas metallicity is seen in our analysis). 

We test how the PCA is affected by selecting subsamples of galaxies. Selecting galaxies with
$M_{\rm stellar}>10^{10}\,\rm M_{\odot}$ {has the effect of increasing the importance of 
the H$_2$ mass and metallicity on the first principal component, while 
in the second principal component we see very little difference. However, we still see that the main 
properties defining the first principal component are the stellar mass, SFR and neutral gas mass.}
If instead, we select galaxies with $M_{\rm stellar}>10^9\,\rm M_{\odot}$ that are mostly 
passive (those with $\rm 0.001\,M_{\odot}\,yr^{-1}\le SFR\le 0.1\,M_{\odot}\,yr^{-1}$), we find that the 
first principal component changes very little, while 
in the second principal component $M_{\rm H_2}$ becomes as important as 
$M_{\rm neutral}$. A selection of galaxies with 
$M_{\rm stellar}>10^{10}\,\rm M_{\odot}$ and $\rm 0.001\,M_{\odot}\,yr^{-1}\le SFR\le 0.1\,M_{\odot}\,yr^{-1}$ 
(which again correspond to mostly passive galaxies), {produces the PCA to give more weight to the gas metallicity and the H$_2$ mass in 
the first principal component, becoming more dominated 
by the stellar mass, SFR, $Z_{\rm SF,gas}$ and H$_2$ and HI masses.}
These tests show that the first principal component is always related to the fundamental plane of 
star formation that we introduce in $\S$~\ref{FPGas} regardless of whether we select massive galaxies only, passive galaxies 
or the entire galaxy population. {For galaxies with SFRs$\lesssim 0.1\,\rm M_{\odot}\,yr^{-1}$, we see that the metallicity becomes 
more prominent in the first principal component.} 
The second principal component in all the tests we did has the gas metallicity playing an important 
role and therefore is always related to the MZ relation.

{As an additional test to determine which gas phase is more important (neutral, atomic or molecular),
we present in Appendix~\ref{PCATest} 
three principal component analyses, 
in which we include stellar mass, SFR, gas metallicity and HI, H$_2$ or neutral gas mass. We find that the highest variance 
is obtained in the first principal component of the PCA that includes the neutral gas mass. If instead we include the HI or 
H$_2$ masses, we obtain a smaller variance on the first principal component. In addition, we find that the contribution of 
the metallicity of the star-forming gas in the first principal components of the PCA performed using the neutral or HI gas masses is negligible, 
while it only appears to be important if we use the H$_2$ mass instead. 
This supports our interpretation that most of the variance in the galaxy population is enclosed in the 
``the fundamental plane of star formation'' of galaxies, and that the neutral gas mass is more important than the 
HI or H$_2$ masses alone.}
In the rest of this section we analyse in detail the physical 
implications of the first two principal components presented in Table~\ref{TablePCA}, {which together 
account for $79$\% of the variance seen in the EAGLE galaxy population}.

\subsection{The fundamental plane of star formation}\label{FPGas}
\begin{figure*}
\begin{center}
\includegraphics[width=0.49\textwidth]{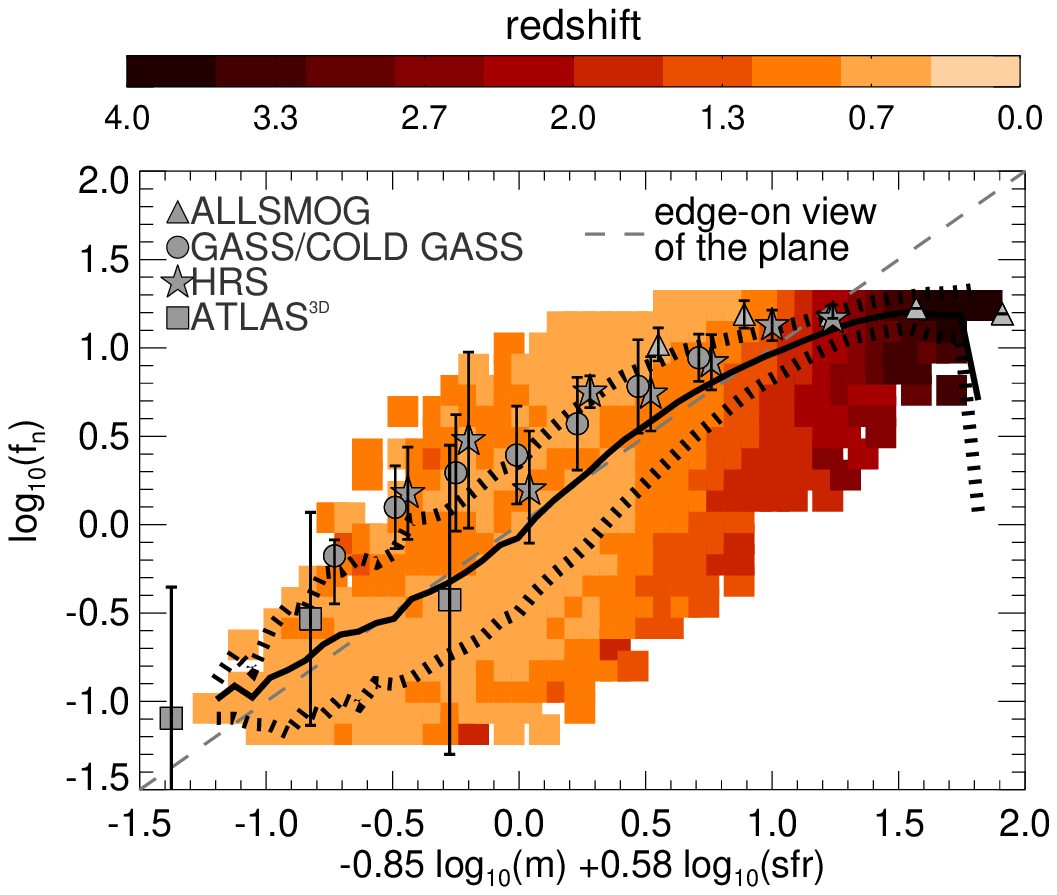}
\includegraphics[width=0.49\textwidth]{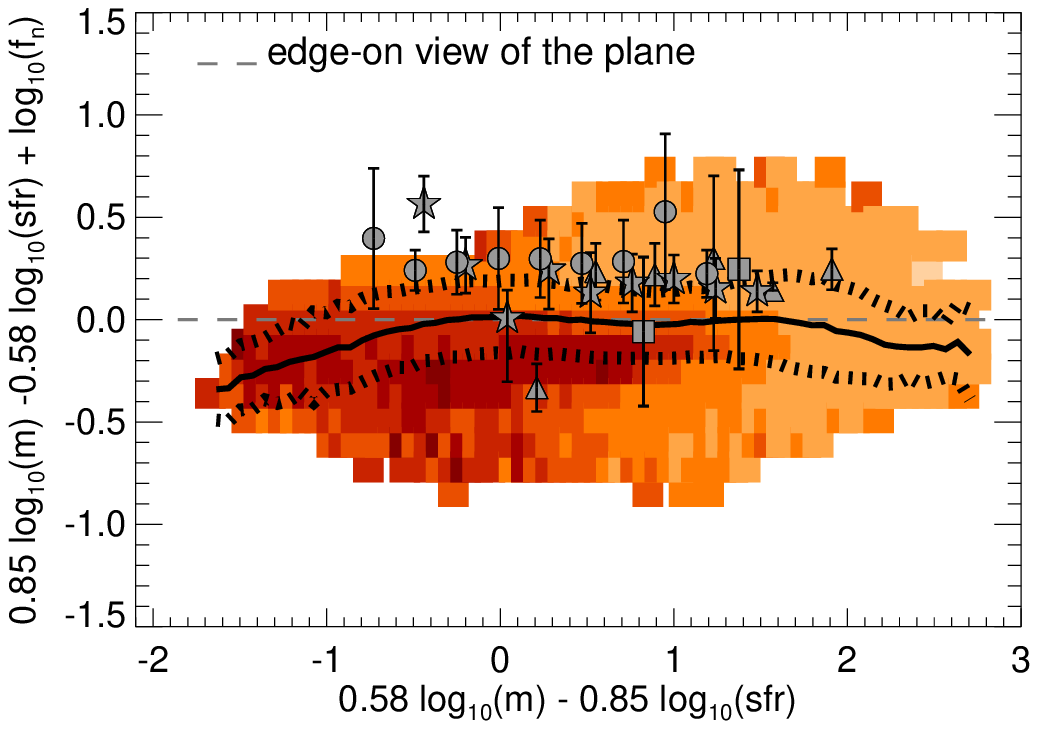}
\includegraphics[width=0.49\textwidth]{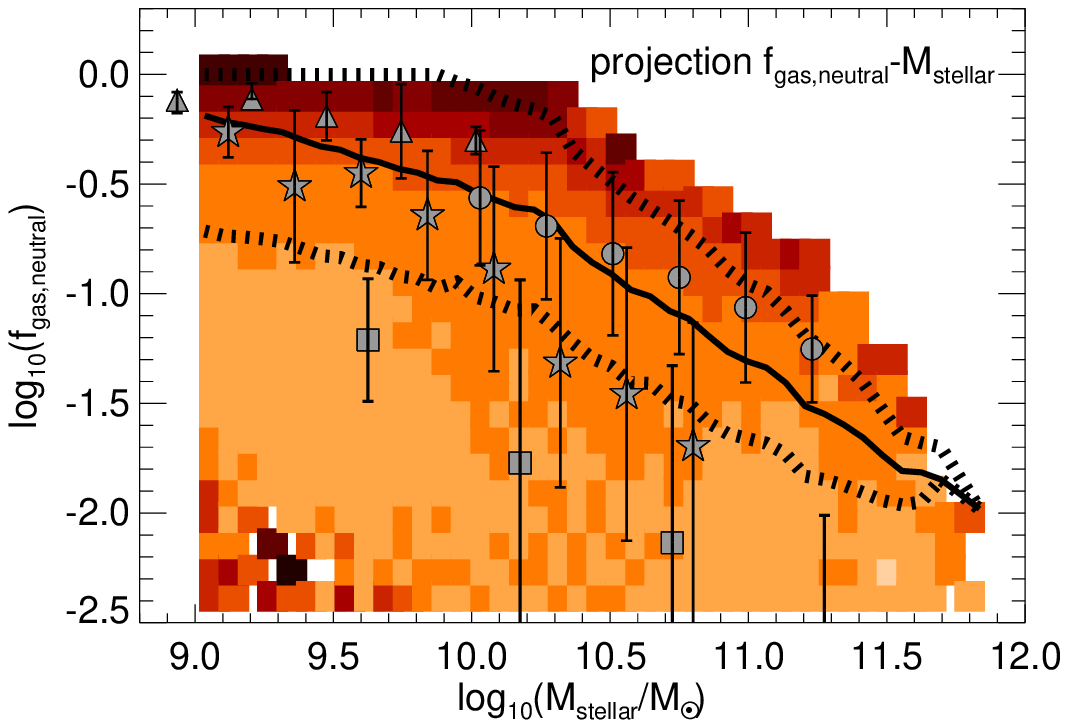}
\includegraphics[width=0.49\textwidth]{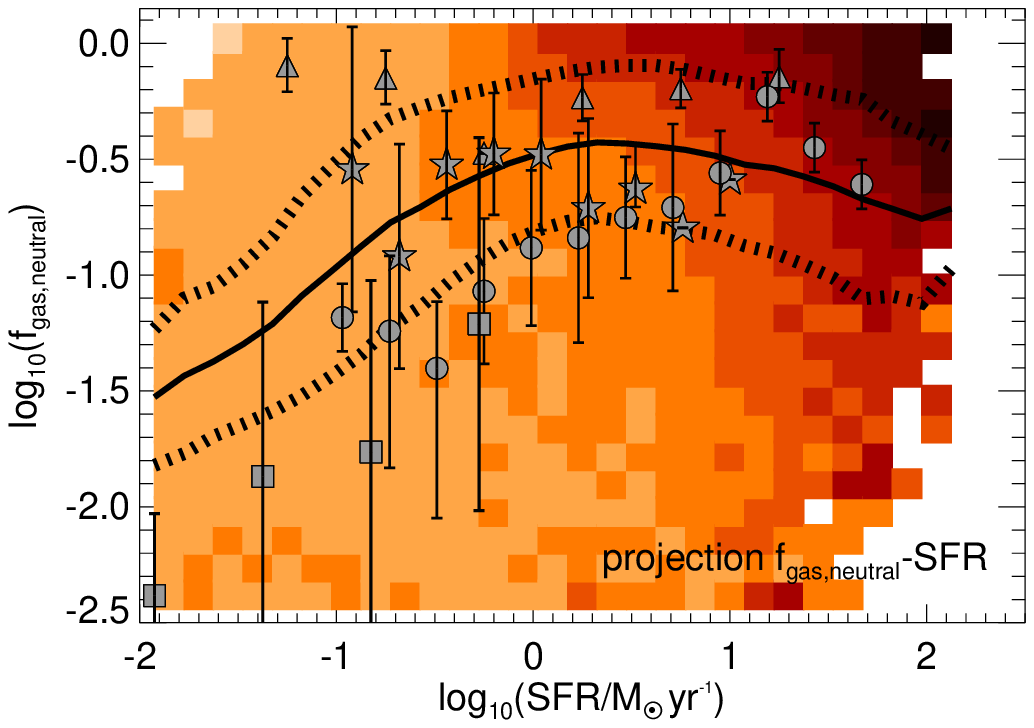}
\caption{Four views of the distribution of galaxies in the 3-dimensional space of neutral fraction, stellar mass and SFR. 
We include all \eagle\ galaxies with $M_{\rm stellar}>10^9\,\rm M_{\odot}$, in the redshift range $0\le z\le 4.5$. 
The median and $16^{\rm th}$ and $84^{\rm th}$ percentiles are shown as solid and dotted lines, respectively,  
and are shown in all the panels. 
Filled squares are coloured according to the median redshift of galaxies in bins of the horizontal and vertical axis, as indicated in the colour bar.
The top panels show edge-on views of the fitted plane of Eq.~\ref{Eqfneutral}, 
with the top-left panel showing normalised gas fraction as a function of the combination of SFR and stellar mass 
of Eq.~\ref{Eqfneutral} (see also Eq.~\ref{defvariables} for the definitions of m, sfr and f$_{\rm n}$), 
while the top-right panel shows the vector perpendicular to the {plane, $\rm \vec{v}_{\perp}=(a,b,c)$, as a function 
of a vector parallel to the plane, $\rm \vec{v}_{\parallel}=(-b, -a, 0)$, where the plane is defined 
as $\rm ax+by+cz=0$ (see Eq.~\ref{Eqfneutral})}.
The bottom panels show two projections along the axes of the 3-dimensional space that are nearly face-on views of the plane:
 $f_{\rm gas,neutral}$ vs. stellar mass (left panel) and $f_{\rm gas,neutral}$ vs. SFR (right panel).
{The dashed line in the top panels show edge-on views of the plane}.
Symbols show observations: squares correspond to GASS and COLD GASS,
 circles to HRS, squares to ATLAS$^{\rm 3D}$, and triangles to the ALLSMOG survey, as labelled in the top-left panel. 
Observations follow a plane in the 3-dimensional space of $f_{\rm gas,neutral}$, stellar mass and SFR that is very similar to the one predicted by \eagle. For a movie rotating over the 3-dimensional space please see \protect\small\tt{www.clagos.com/movies.php}.}
\label{HIFP}
\end{center}
\end{figure*}

{Here we investigate the dependence of the neutral and molecular gas fraction on stellar mass and SFR. We change from using 
gas masses in $\S$~\ref{PCASec} to gas fractions. The reason for this is that the scatter in the 3-dimensional space 
of stellar mass, SFR and neutral gas fraction or molecular gas fraction is the least compared to what it is obtained if we instead 
use gas masses or simply neutral or molecular gass mass to stellar mass ratios. We come back to this when discussing 
Eqs.~\ref{Eqfneutral} and \ref{Eqfmol}.}
 
In order to visualise a flat plane in a three-dimensional space, it helps to define vectors that are 
perpendicular and parallel to the plane, and plot them against each other in order to reveal edge-on and face-on orientations 
of the plane. This is what we do in this section. If we define a plane 
as $\rm ax+by+cz=0$, {vector perpendiculars and parallel to the plane would be 
$\rm \vec{v}_{\perp}=(a,b,c)$ and $\rm \vec{v}_{\parallel}=(-b, -a, 0)$, respectively.}
We use these vectors later to show edge-on orientations of the 
fundamental plane of star formation, which we introduce in Eqs.~\ref{Eqfneutral} and \ref{Eqfmol}. 

Fig.~\ref{HIFP} shows four views of the 3-dimensional space of neutral gas fraction, 
stellar mass and SFR. In this figure we include all galaxies in \eagle\ with $M_{\rm stellar}>10^{9}\,\rm M_{\odot}$, 
SFR$\ge 0.01 \rm \,M_{\odot}\,yr^{-1}$, and that are in the redshift range $0\le z\le 4.5$.
We show the underlying redshift distribution of the galaxies by binning each plane and colouring bins according to the 
median redshift of the galaxies.
Two of the views show edge-on orientations of the plane (i.e. with respect to the best-fit plane of Eq.~\ref{Eqfneutral} below), 
 and the other two are projections along the axes of the 3-dimensional space. 
One edge on view (top-left panel) shows the neutral gas fraction as a function of the combination of SFR and stellar mass 
of Eq.~\ref{Eqfneutral}. For the second edge-on view (top-right panel),  
 we use the perpendicular and parallel vectors defined above, with the plane being defined in Eq.~\ref{Eqfneutral}. 

\begin{figure*}
\begin{center}
\includegraphics[width=0.49\textwidth]{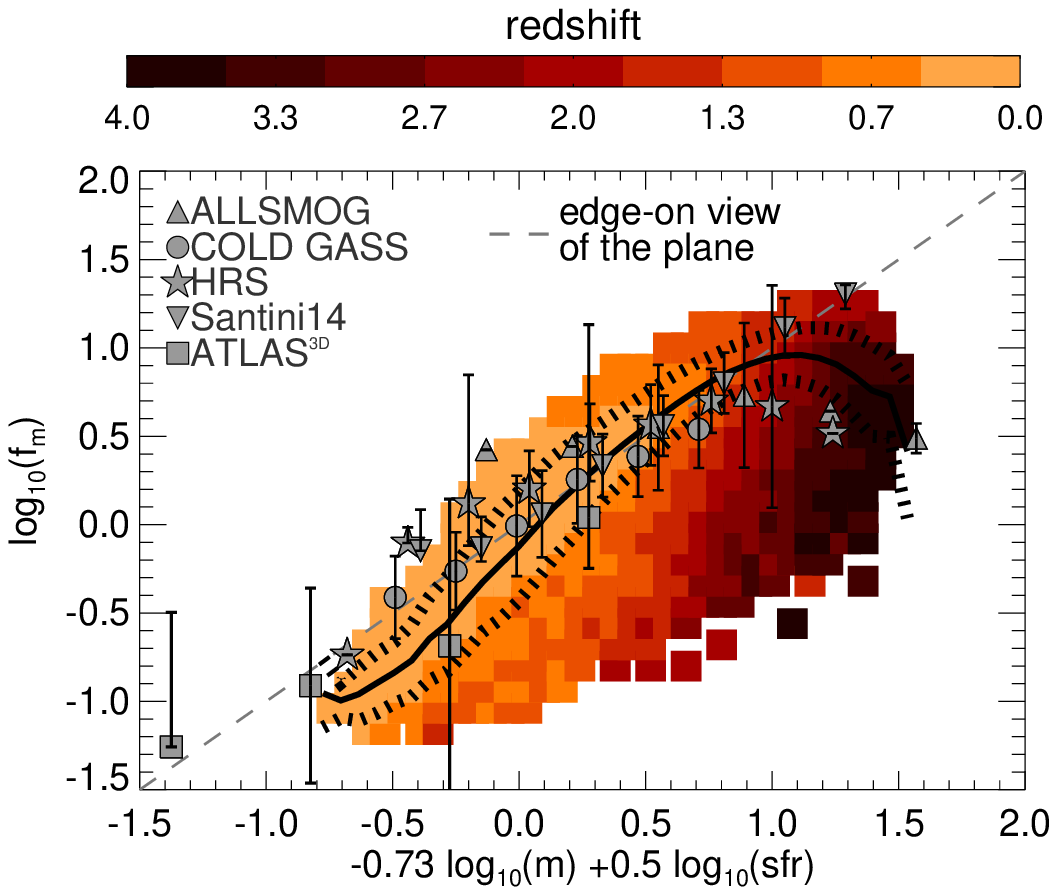}
\includegraphics[width=0.49\textwidth]{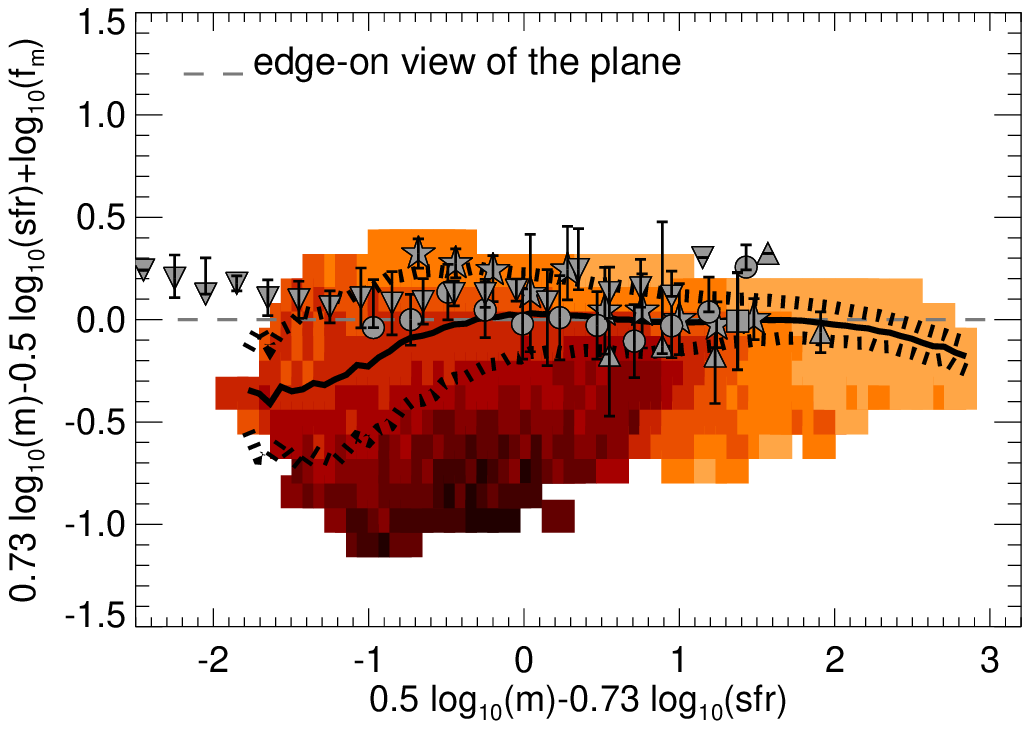}
\includegraphics[width=0.49\textwidth]{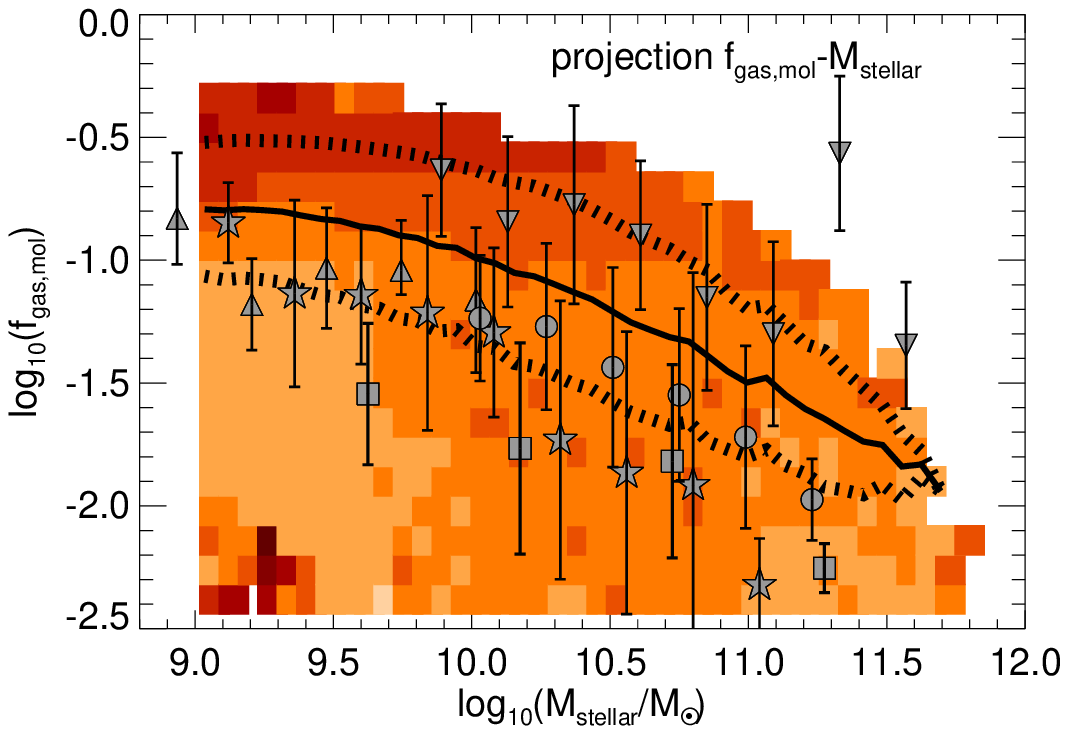}
\includegraphics[width=0.49\textwidth]{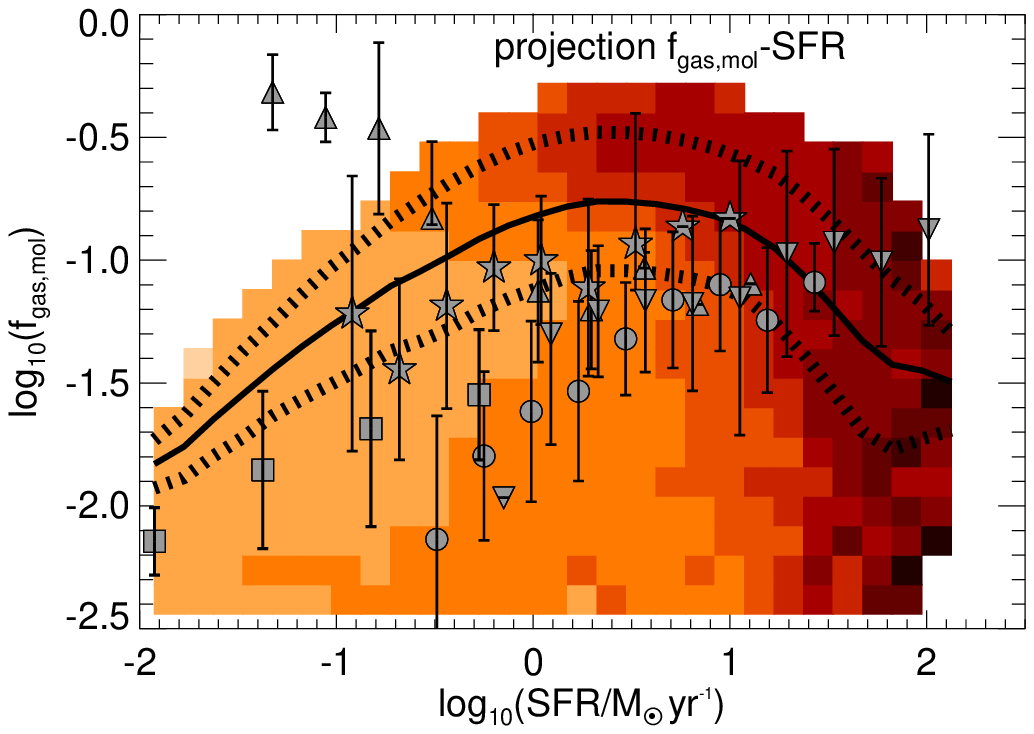}
\caption{As in Fig.~\ref{HIFP} but for the molecular gas fraction. 
For the two edge-on views of the top panels we use the plane definition of Eq.~\ref{Eqfmol} 
(see also Eq.~\ref{defvariables} for the definitions of m, sfr and f$_{\rm m}$). 
Here we also show the observational results from \citet{Santini13}, which correspond to star-forming galaxies
at $z\lesssim 3$. 
 For a movie rotating over the 3-dimensional space please see {\small \tt www.clagos.com/movies.php}.}
\label{H2FP}
\end{center}
\end{figure*}

Galaxies populate a well-defined plane, which 
shows little evolution. Galaxies evolve along this plane with redshift, in such a way that they 
are on average more gas rich and more highly star-forming at higher redshift. When we consider the molecular gas fraction instead of the 
neutral gas fraction, the situation is the same: galaxies populate a well-defined plane in the 
3-dimensional space of $f_{\rm gas,mol}$, stellar mass and SFR (shown in Fig.~\ref{H2FP}).
This means that at fixed SFR and stellar mass, there is very little evolution in $f_{\rm gas,neutral}$ and $f_{\rm gas,mol}$.  
Hence, most of the observed trend of an increasing molecular fraction with redshift 
(e.g. \citealt{Geach11}; \citealt{Saintonge13})  
is related to the median SFR at fixed stellar mass increasing with redshift (e.g. \citealt{Noeske07}; \citealt{Sobral14}).
We argue later that both the SFR and gas fraction are a consequence of the self-regulation of star formation 
in galaxies. 

For both $f_{\rm gas,neutral}$ and $f_{\rm gas,mol}$ the relation is best described by a curved surface in 3-dimensional space.
Here we provide fits of the flat plane tangential to this 2-dimensional surface at $M_{\rm stellar}=5\times 10^{10}\,\rm M_{\odot}$ and 
$\rm SFR=2\, M_{\odot}\, yr^{-1}$, which we compute using the {\tt HYPER-FIT} R package\footnote{{\tt hyperfit.icrar.org/}} of \citet{Robotham15}.
We refer to the tangential plane fitted to the $\rm f_{\rm gas,neutral}-SFR-M_{\rm stellar}$ relation 
 as ``the fundamental plane of star formation''. 
For the fitting, we weigh each galaxy by the inverse of the number density in logarithmic mass interval 
in order to prevent the fit from being biased towards the more numerous small galaxies.
The best fit planes are: 

\begin{eqnarray}
0.85\,{\rm log_{10}(m)}-0.58\,{\rm log_{10}(sfr)}+{\rm log_{\rm 10}(f_{\rm n})}&=&0,\label{Eqfneutral}\\
0.73\,{\rm log_{10}(m)}-0.50\,{\rm log_{10}(sfr)}+{\rm log_{\rm 10}(f_{\rm m})}&=&0,\label{Eqfmol}
\end{eqnarray}
 
\noindent where,

\begin{eqnarray}
{\rm m} &=&\frac{M_{\rm stellar}}{5\times 10^{10}\,\rm M_{\odot}},\,\, {\rm sfr} =\frac{{\rm SFR}}{\rm 2\, M_{\odot}\, yr^{-1}},\,\nonumber\\
f_{\rm n} &=&\frac{f_{\rm gas,neutral}}{0.046},\,\,\,\,\,\,\,f_{\rm m} =\frac{f_{\rm gas,mol}}{0.026}.\label{defvariables}
\end{eqnarray}
The fits above are designed to minimise the scatter.
The best fits of Eqs.~\ref{Eqfneutral} and \ref{Eqfmol} are shown as dashed lines 
in the top-left panels of Fig.~\ref{HIFP} and Fig.~\ref{H2FP}, respectively.
The standard deviations perpendicular to the planes calculated by {\tt HYPER-FIT} are 
$0.17$~dex for Eq.~\ref{Eqfneutral} and $0.15$~dex for Eq.~\ref{Eqfmol}, while the 
standard deviations parallel to the gas fraction axis are $0.24$~dex for Eq.~\ref{Eqfneutral} and $0.2$~dex for Eq.~\ref{Eqfmol}.
Although the scatter seen for the molecular gas fraction is slightly smaller than for the neutral gas fraction, the PCA 
points to the latter as capturing most of the variance of the galaxy population. This is because the neutral gas fraction 
is more directly connected to the process of gas accretion than the molecular gas fraction, and we discuss later that 
accretion is one of the key processes determining the existence of the fundamental planes. 
In addition, because SFR and the molecular gas mass are 
strongly correlated, only one of these properties is needed to describe most of the variance among galaxy properties. 
We also analysed the correlation between $f_{\rm gas,neutral}$ ($f_{\rm gas,mol}$) and specific SFR, and found that the scatter increases 
by $\approx 20$\% ($\approx 25$\%) relative to the scatter characterising  
Eq.~\ref{Eqfneutral}. {We find that fitting planes to the three-dimensional dependency 
of gas mass-SFR-stellar mass or gas-to-stellar mass ratio-SFR-stellar mass 
(instead of gas fraction-SFR-stellar mass, as presented in Eqs.~\ref{Eqfneutral} and \ref{Eqfmol}) 
lead to an increase in the scatter relative to was it is obtained around Eqs.~\ref{Eqfneutral} and \ref{Eqfmol} of 
 $\approx 20-30$\%. We therefore conclude that the tightest correlations (i.e. least scatter) in \eagle\ are those 
between gas fraction, stellar mass and SFR.}

Note that there is a clear turnover at $f_{\rm gas,mol}\approx 0.3$ (very clear at 
a y-axis value $\approx 0.7$ in the top-left of Fig.~\ref{H2FP}), which is produced by 
galaxies with $\rm SFR\gtrsim 15\,\rm M_{\odot}\,yr^{-1}$.
Most of the galaxies that produce this turnover are forming stars in an ISM with a very high median pressure 
(SFR-weighted pressures of $\rm log_{10}(\langle P\rangle\, k^{-1}_B/\, \rm cm^{-3}\,K)\approx 6-7$). 
The turnover is less pronounced in the neutral gas fraction relation (top left panel in Fig.~\ref{HIFP}).
Most galaxies that lie around the turn-over are at $z\gtrsim 2$. 
The fact that we do not see such strong turn-over in the neutral gas fraction is because galaxies with high SFRs 
 have an intense radiation field 
that destroys H$_2$ more effectively, moving the HI to H$_2$ transition towards higher gas pressures. 
Thus, a significant fraction of the gas with densities $n_{\rm H}\gtrsim 1\,\rm cm^{-3}$ remains atomic at high-redshift. The effect of this on the 
H$_2$ fraction is important, introducing the turn-over at high H$_2$ fractions seen in Fig.~\ref{H2FP}. 

For the neutral gas fraction we find that the fitted plane of Eq.~\ref{Eqfneutral} is a good description of the neutral gas fractions 
of galaxies in \eagle\ (note that this is also true for the higher resolution simulations shown in Appendix~\ref{ConvTests}) at
$f_{\rm gas,neutral}\lesssim 0.5$ (y-axis value $\approx 1$ in the top-left of Fig.~\ref{HIFP}). 
However, at higher neutral gas fractions, the fit tends to overshoot the gas fraction by $\approx 0.1-0.2$~dex. 
{The latter is not because the gas fraction saturates at $\approx 1$, but because there is a physical change in the ratio 
of SFR to neutral gas mass from $z=0$ towards high redshift, due to the super-linear star formation law adopted in \eagle\ and the 
ISM gas density evolution. 
We come back to this point in $\S$~\ref{PhysicalInterpretation}.} 
For the molecular gas fraction we find that the fit of Eq.~\ref{Eqfmol} describes the molecular 
gas fractions of \eagle\ galaxies well in the regime $0.02\lesssim \rm f_{\rm gas,mol} \lesssim 0.3$
 ($-0.2\lesssim \rm log_{10}(f_{m})\lesssim 1$), while at lower and at higher $f_{\rm gas,mol}$ 
the fit overshoots the true values of the gas fraction.
At the high molecular gas fractions this is due to 
 galaxies populating the turnover discussed above, that deviates from the main 
plane (which corresponds to galaxies with SFR$\gtrsim \rm 15\,M_{\odot}\,yr^{-1}$ and $f_{\rm gas,mol}\lesssim 0.3$). 

We also investigated the distribution of \eagle\ galaxies in the 3-dimensional space of star-forming gas mass, 
$M_{\rm stellar}$ and SFR at higher redshifts, $5\le z\le 7$.
We used star-forming gas mass rather than neutral or molecular gas mass, because our approximations for calculating the 
latter two may not be accurate
at these higher redshifts (see e.g. the discussion in \citealt{Rahmati13b}). 
 We find that $5\le z\le 7$ \eagle\ galaxies trace a 2-dimensional curved surface in this 3-dimensional space with little scatter.
This leads us to suggest that the process that induces the strong correlation that gives rise to the fundamental plane of 
star formation at $z\le 4.5$, is already operating
at $5\le z\le 7$. 

\begin{figure}
\begin{center}
\includegraphics[width=0.49\textwidth]{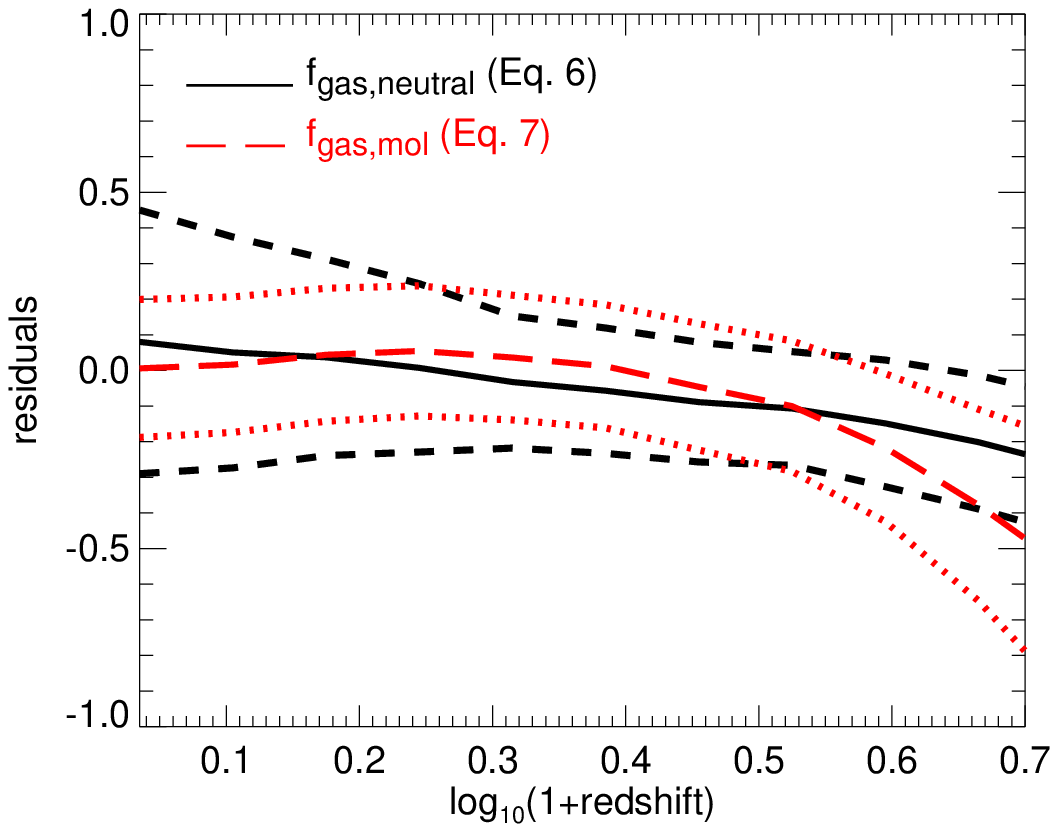}
\includegraphics[trim = 0mm 0mm 0mm 3.5mm,clip,width=0.49\textwidth]{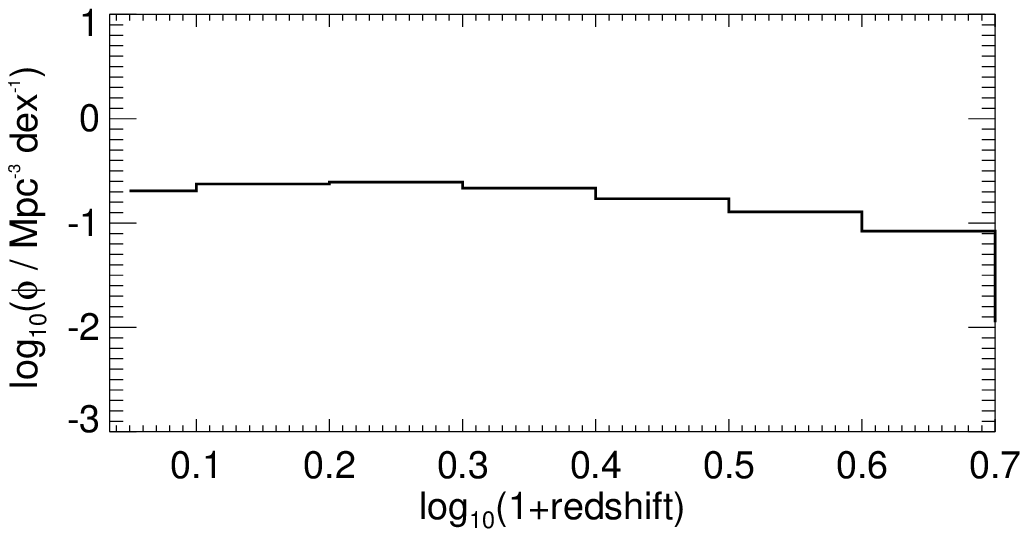}
\caption{{{\it Top panel:} Residuals of simulated galaxies from Eqs.~\ref{Eqfneutral} and \ref{Eqfmol} as a function of 
$\rm log_{\rm 10}(1+redshift)$. 
Here residuals are defined as $\rm ax+by+cz$, where $a$, $b$ and $c$ are defined in Eqs.~\ref{Eqfneutral} and \ref{Eqfmol}. 
The solid black line is the mean residual of galaxies with $M_{\rm stellar}>10^9\,\rm  M_{\odot}$ 
from the fit of Eq.~\ref{Eqfneutral} to the fundamental plane, with the dashed lines indicating the
$16^{\rm th}$ and $84^{\rm th}$ percentiles. 
The red long dashed line and red dotted lines, are the corresponding median and percentiles residuals from the fit of Eq.~\ref{Eqfmol}.
Note that the redshift at which the medians cross zero is set by the choice of normalisation, and thus it has no physical meaning. 
{\it Bottom panel:} Redshift distribution of the galaxies with $M_{\rm stellar}>10^9\,\rm M_{\odot}$i, shown at the top panel.}}
\label{residuals}
\end{center}
\end{figure}

{We show in Fig.~\ref{residuals} the residuals of the galaxies from the fits of Eqs.~\ref{Eqfneutral} and \ref{Eqfmol} as a function of 
redshift. In the case of Eq.~\ref{Eqfneutral}, we see that residuals depend very weakly on redshift, 
with the median slightly decreasing with increasing redshift. Including redshift in {\tt HYPER-FIT} leads to an increase in the scatter 
of $\approx 50$\%, indicating that including redshift does not improve the fit provided in Eq.~\ref{Eqfneutral}. 
For the molecular gas fraction fit of Eq.~\ref{Eqfmol}, we find the residuals show no dependence with 
redshift at $z<2$ ($\rm log_{\rm 10}(1+redshift)\approx0.5$), and the trend seen at higher 
redshifts is due to the turnover discussed above. Again, we observe an increase in the scatter 
of the fit if we include redshift, showing that there is no improvement by adding redshift (unless we ignore galaxies at $z<2$).}  

In Figs.~\ref{HIFP}~and~\ref{H2FP} we also 
investigate whether observed galaxies populate a similar plane in the gas
fraction, stellar mass and SFR space, as the one \eagle\ predicts. The observational datasets, which were introduced in $\S$~\ref{EvoGasFrac}, correspond to GASS,
COLD GASS, HRS, ALLSMOG, ATLAS$^{\rm 3D}$ and
 \citet{Santini13}.

We show the observations in Figs.~\ref{HIFP}~and~\ref{H2FP} in the same way as we show \eagle\ results: we 
calculate the median neutral and molecular gas fraction and the 1$\sigma$ scatter around those values 
 in the two edge-on views with respect to the best fits of Eqs.~\ref{Eqfneutral}~and~\ref{Eqfmol}, and the two projections over the axis of the 3-dimensional space. 
We find that observed galaxies follow a similar plane as galaxies in \eagle,
albeit with some surveys having neutral gas fractions $\approx 0.1-0.2$~dex higher than those found for \eagle\ galaxies at fixed stellar mass and SFR.  
For example if we compare \eagle\ with GASS plus COLD GASS, we find such an offset in the neutral gas fractions, but compared to HRS and ATLAS$^{\rm 3D}$ we find very good agreement. 
Regarding molecular fractions, we find that the observations follow a plane that is very similar to the one described by the \eagle\ galaxies,  
{as shown in Fig.~\ref{H2FP}}.
 Interestingly, the observations suggest a turn-over at high $f_{\rm gas,mol}$ similar to the one displayed by \eagle\ (see 
top-left panel of Fig.~\ref{H2FP}). This could point to real galaxies forming stars in intense UV radiation fields, as we find for \eagle\ galaxies. 

Overall, we find that the agreement with the observations is well within the scatter of both the simulation and 
observations. Note that galaxies in the observational sets used here were selected very differently and in some cases 
using complex criteria, which is easy to see in the nearly face-on views of the bottom panels of Fig.~\ref{HIFP} and Fig.~\ref{H2FP}. 
For example, ATLAS$^{\rm 3D}$ and ALLSMOG differ by $\gtrsim 1.5$~dex in the nearly face-on views. However, when the plane is seen 
edge-on, both observational datasets follow 
the same relations. 
This means that even though some samples are clearly very biased, like \citet{Santini13} towards gas-rich galaxies, 
when we place them in the 3-dimensional space of gas fraction, SFR and stellar mass, they lie on the same plane. 
The fact that observations follow a very similar plane in the 3-dimensional space of gas fraction, SFR and stellar mass 
as \eagle\ is remarkable.

\subsubsection{Physical interpretation of the fundamental plane of star formation}\label{PhysicalInterpretation}

We argue that the existence of the 2-dimensional surfaces in the 3-dimensional 
space of stellar mass, SFR and neutral or molecular gas fractions in \eagle\ is due to the 
self-regulation of star formation in galaxies. The rate of star formation is controlled by the balance between 
gas cooling and accretion, which increase the gas content of galaxies, and stellar and BH-driven outflows, that remove 
gas out of galaxies (see \citealt{Schaye10}, \citealt{Lagos10}, \citealt{Booth10}, \citealt{Haas13a} 
for numerical experiments supporting this views).
 In this picture, both the gas content and the SFR of galaxies change to reflect the balance between accretion and outflows, 
and the ratio is determined by the assumed star formation law. 

{This interpretation is supported by the comparison of the reference model we use here
with model variations in \eagle\ presented in Appendix~\ref{ModelTests}. We show $4$ models in which the 
efficiency of AGN and stellar feedback is changed. We find that weakening the stellar feedback has the effect 
of changing the normalisation of the plane, but most importantly, increasing the scatter around it, while making 
feedback stronger tends to tighten the plane. The effect of AGN feedback is very mild due to most of the galaxies 
shown being on the main sequence of galaxies in the SFR-$M_{\rm stellar}$ plane, and therefore not affected by 
AGN feedback. A similar change in scatter is seen if we now look at models where the stellar feedback strength has a different 
scaling (i.e. depending on metallicity alone or on the velocity dispersion of the dark matter). 
Both model variations produce less feedback at higher redshift ($z>1$; see Fig.~$5$ in C15) compared to the reference model, which 
leads to both models producing a more scattered `fundamental plane of star formation' at high redshift. If feedback was not sufficient to 
balance the gas inflows, the scatter would increase even further, erasing the existence of the fundamental plane of star formation 
discussed here.}

We find that the curvature of the 2-dimensional surface is mainly driven by how the gas populates the 
probability distribution function (PDF) of densities in galaxies at different redshifts 
and how star formation depends on the density in \eagle\ (see $\S$~\ref{sub-gridsec}). 
Galaxies at high redshift tend to form stars at higher ISM pressures than galaxies at $z=0$, 
on average (see Fig.~$12$ in \citealt{Lagos15}), which 
together with the super-linear star formation law, lead to higher-redshift galaxies having higher 
star formation efficiencies (i.e. the ratio between the SFR and the gas content above the density threshold for star formation). 
{In Appendix~\ref{ModelTests} we show that changing the dependency of the SFR density on the 
gas density changes the slope of the plane significantly, supporting our interpretation.}

\subsubsection{Example galaxies residing in the fundamental plane of star formation}
\begin{figure*}
\begin{center}
\includegraphics[width=0.25\textwidth]{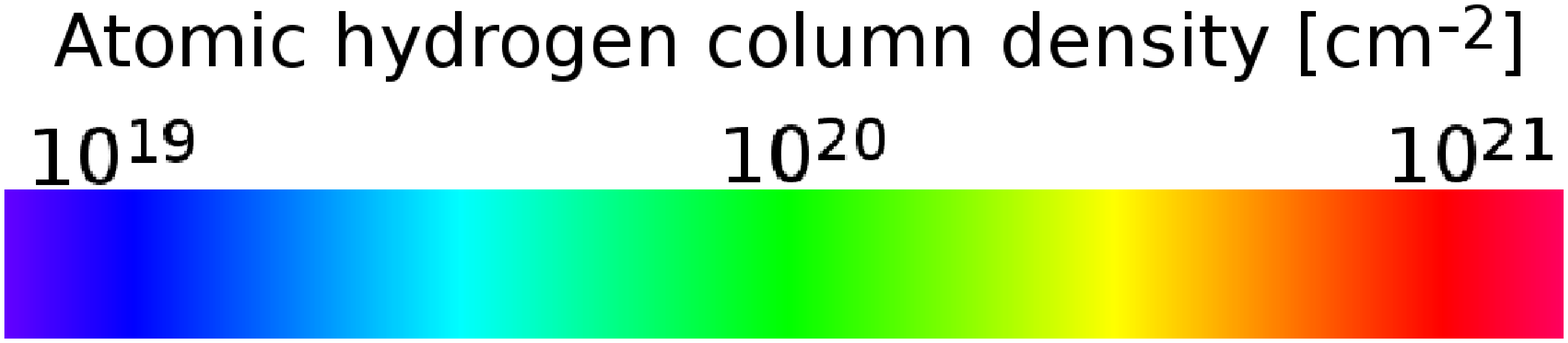}
\includegraphics[width=0.25\textwidth]{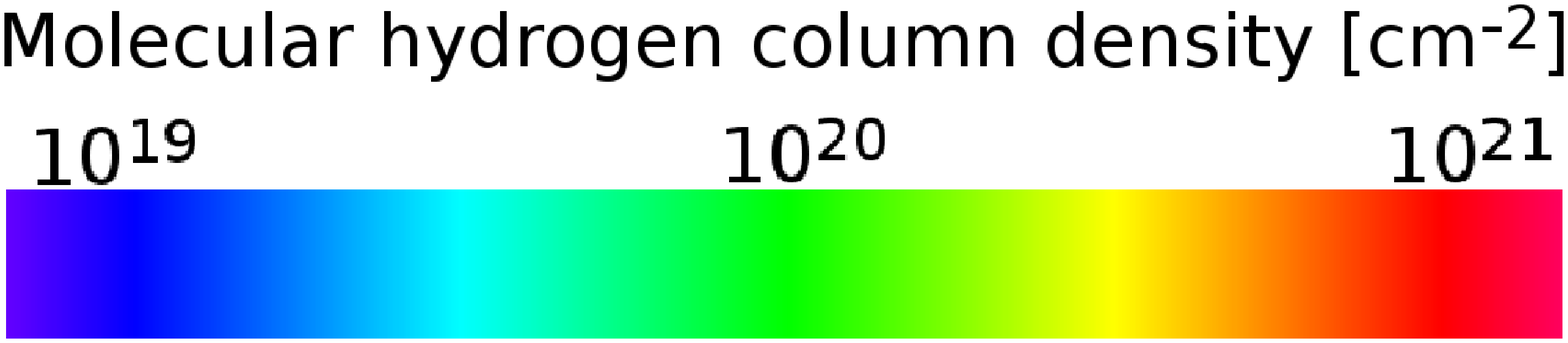}
\includegraphics[width=0.25\textwidth]{Figs/TableColoursrgb.eps}
\includegraphics[width=0.23\textwidth]{Figs/TableColoursrgb.eps}\\
\includegraphics[width=0.25\textwidth]{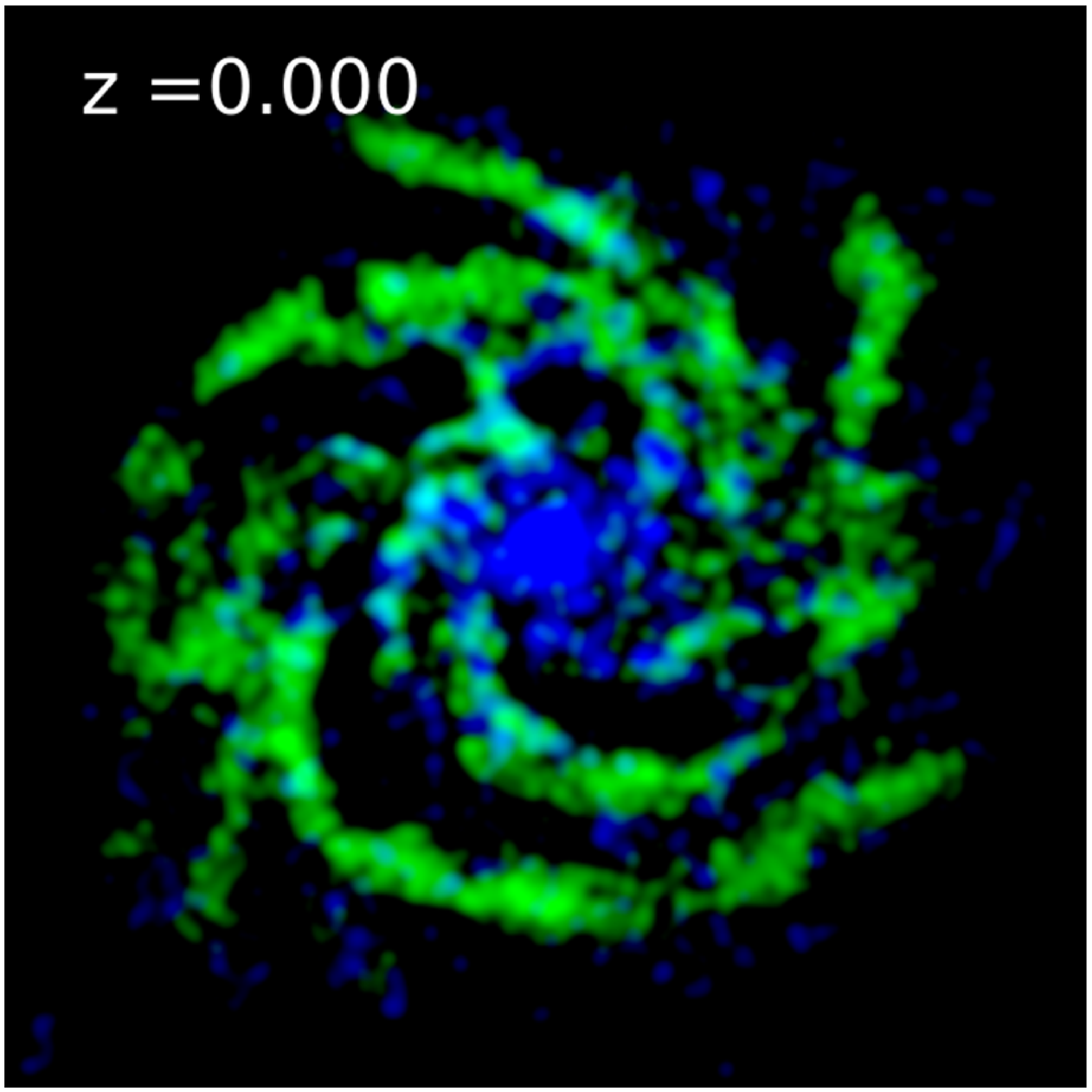}
\includegraphics[width=0.25\textwidth]{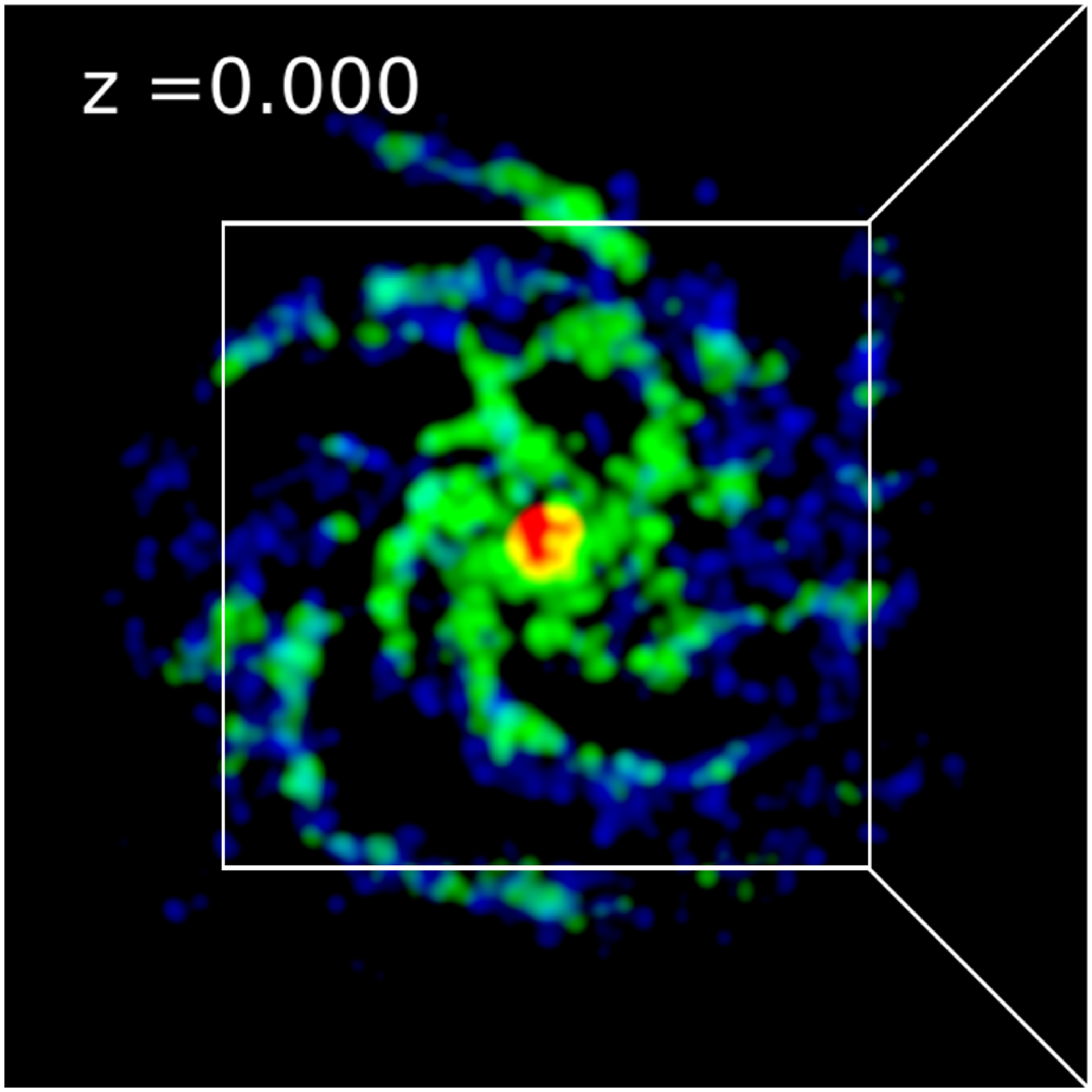}
\includegraphics[width=0.25\textwidth]{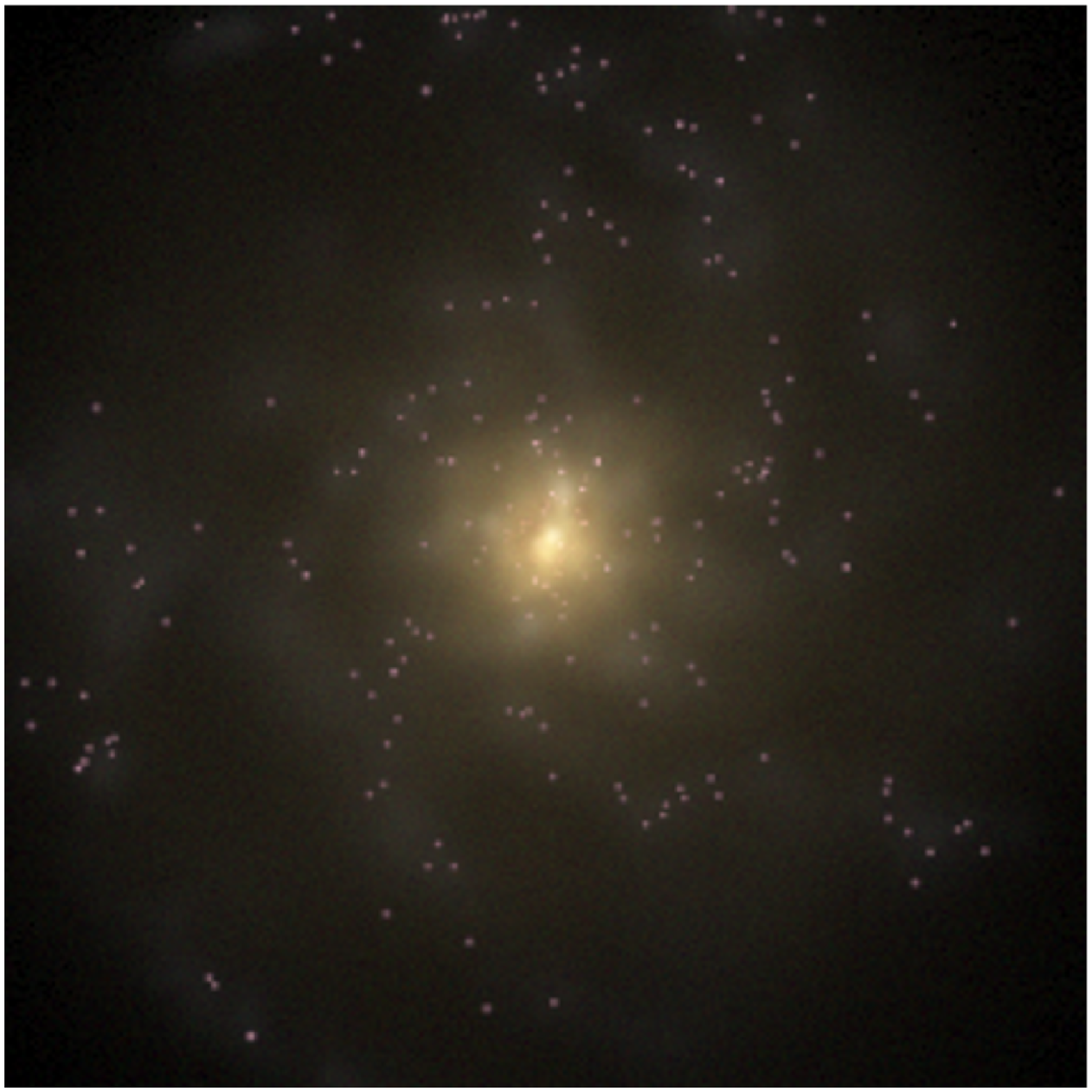}
\includegraphics[width=0.23\textwidth]{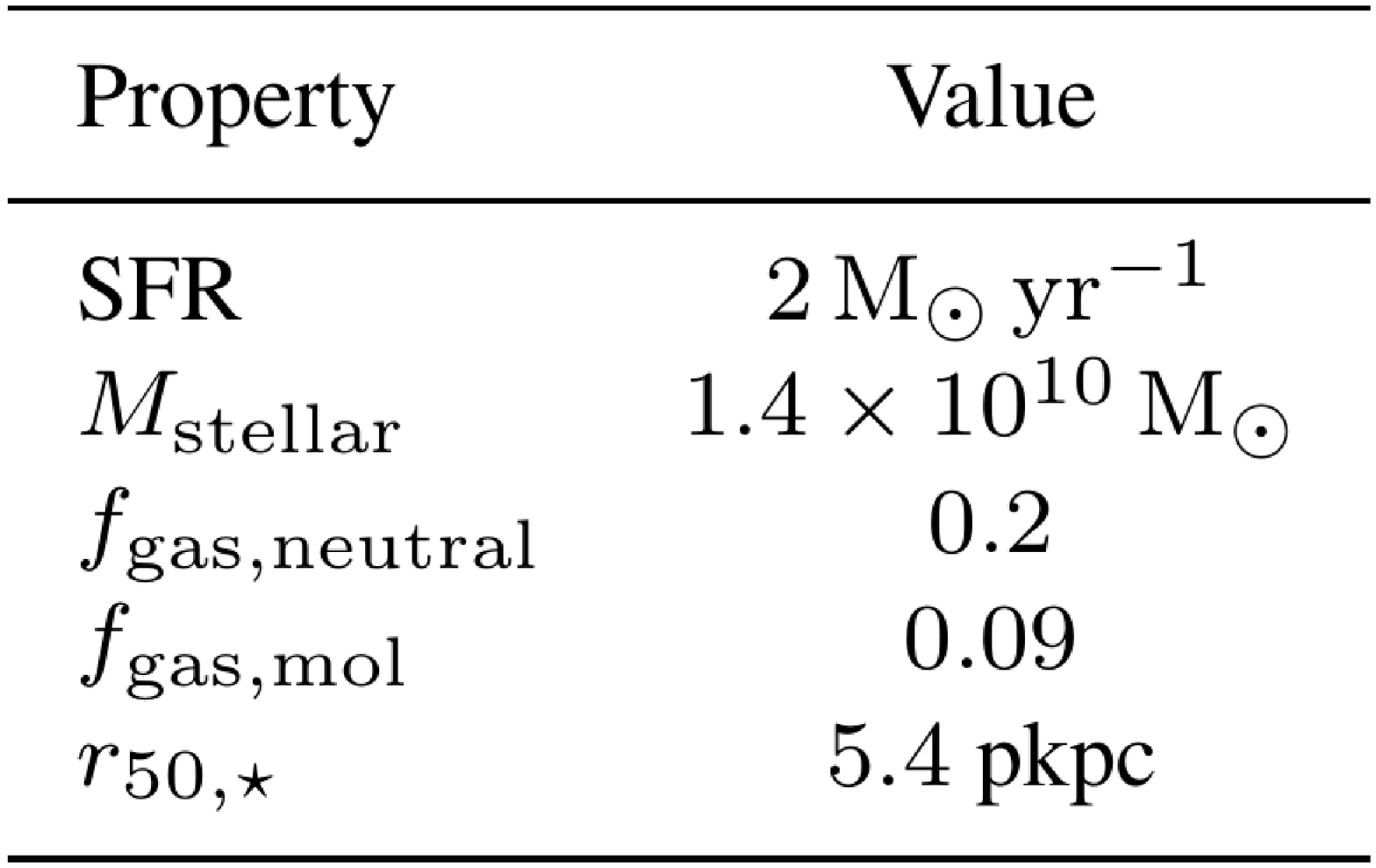}
\includegraphics[width=0.25\textwidth]{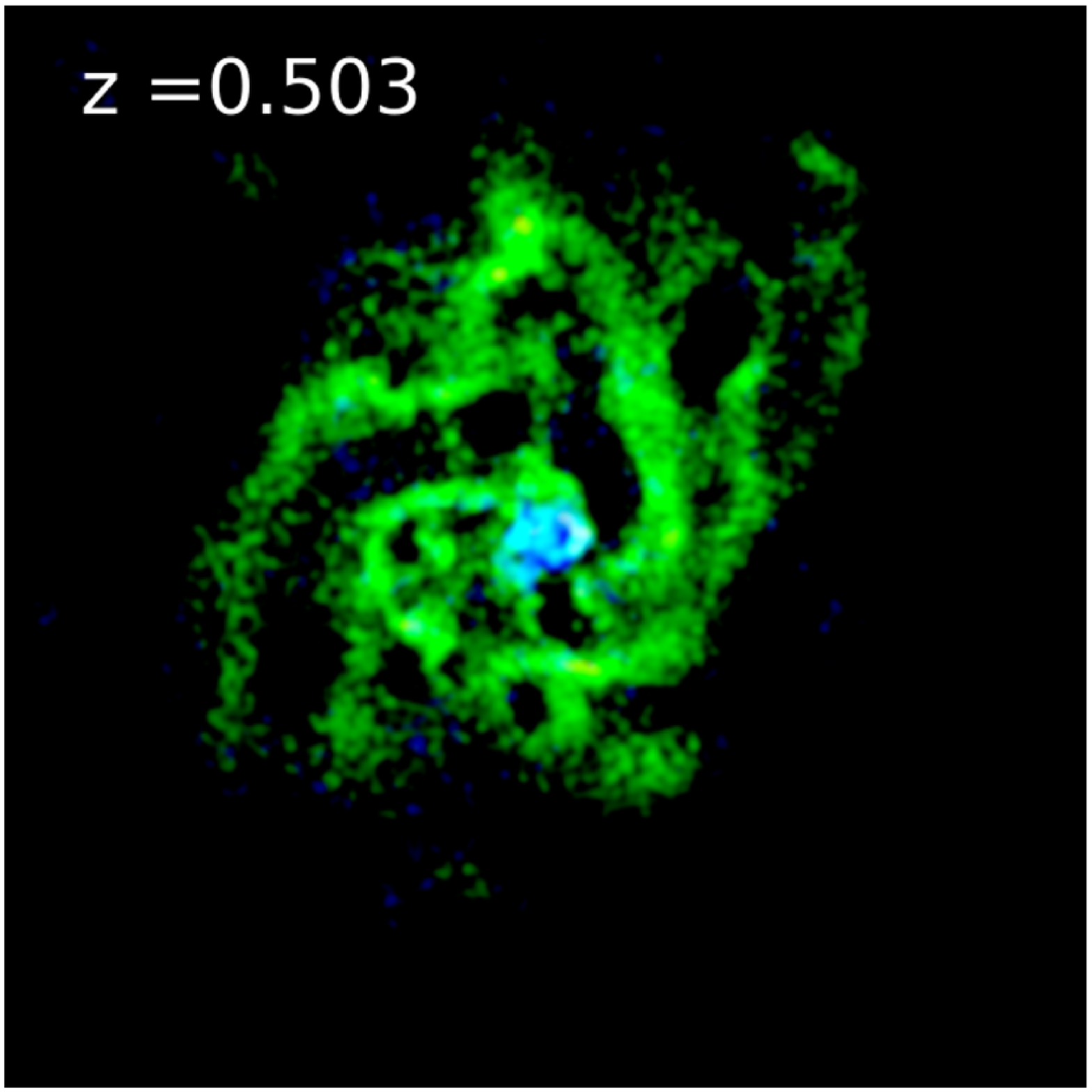}
\includegraphics[width=0.25\textwidth]{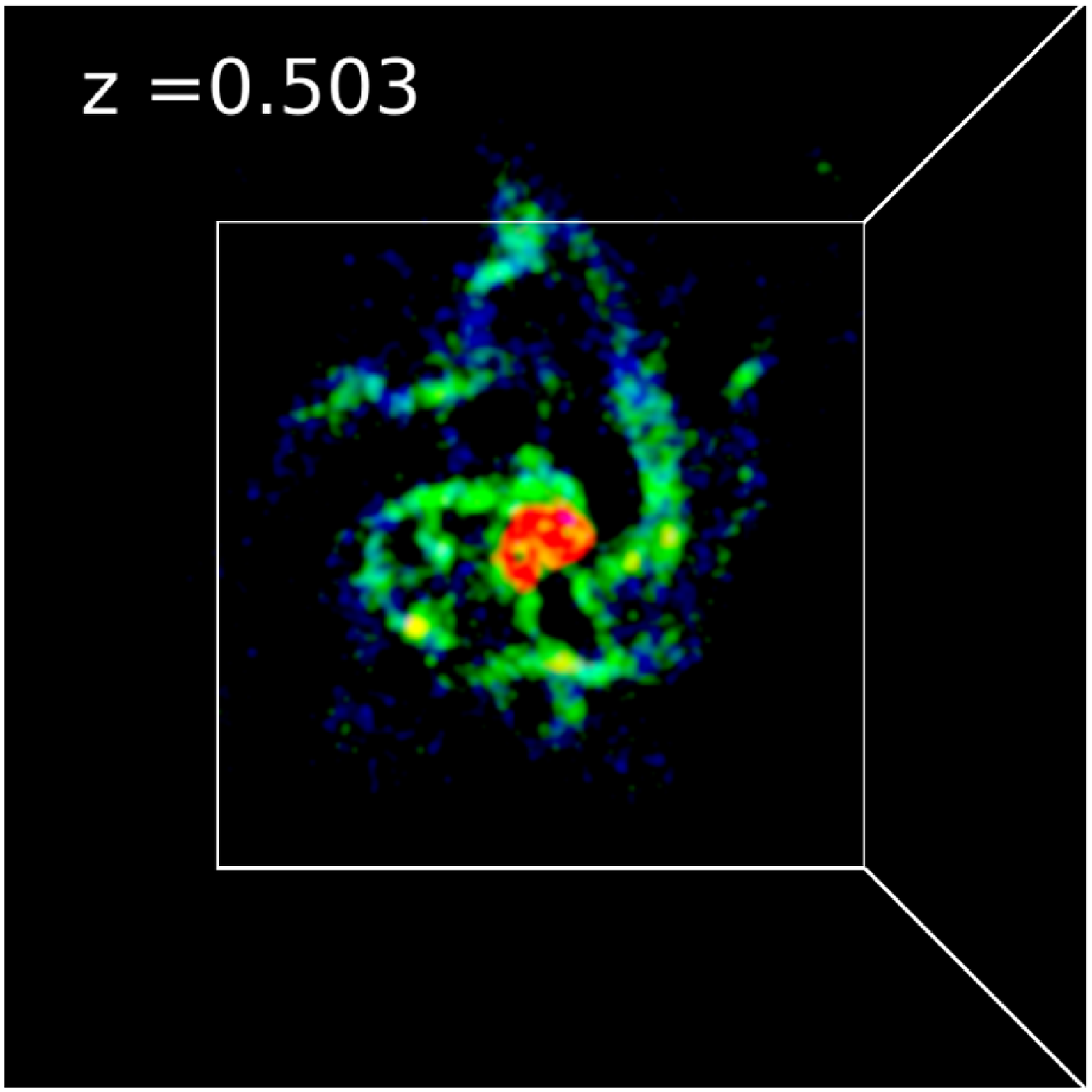}
\includegraphics[width=0.25\textwidth]{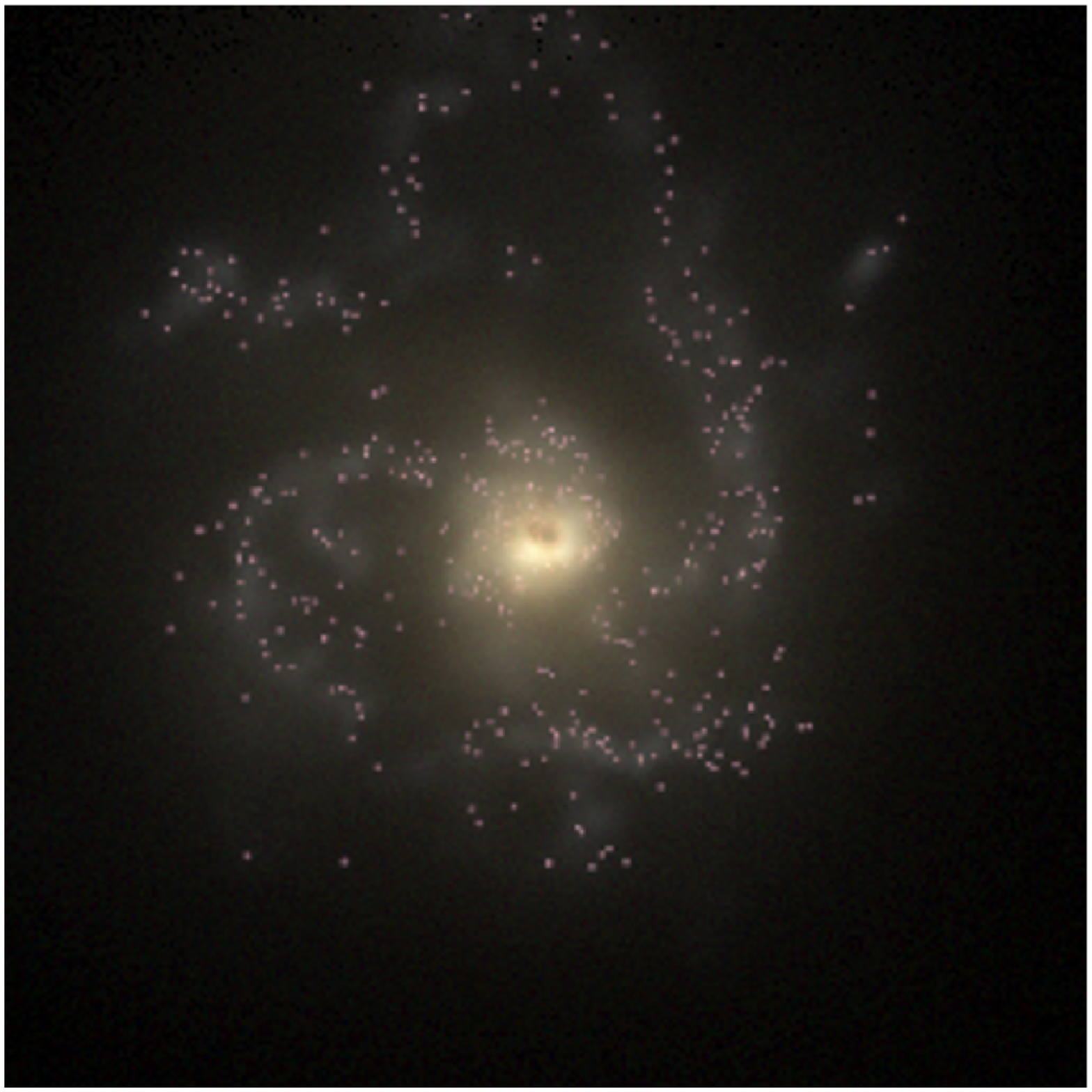}
\includegraphics[width=0.23\textwidth]{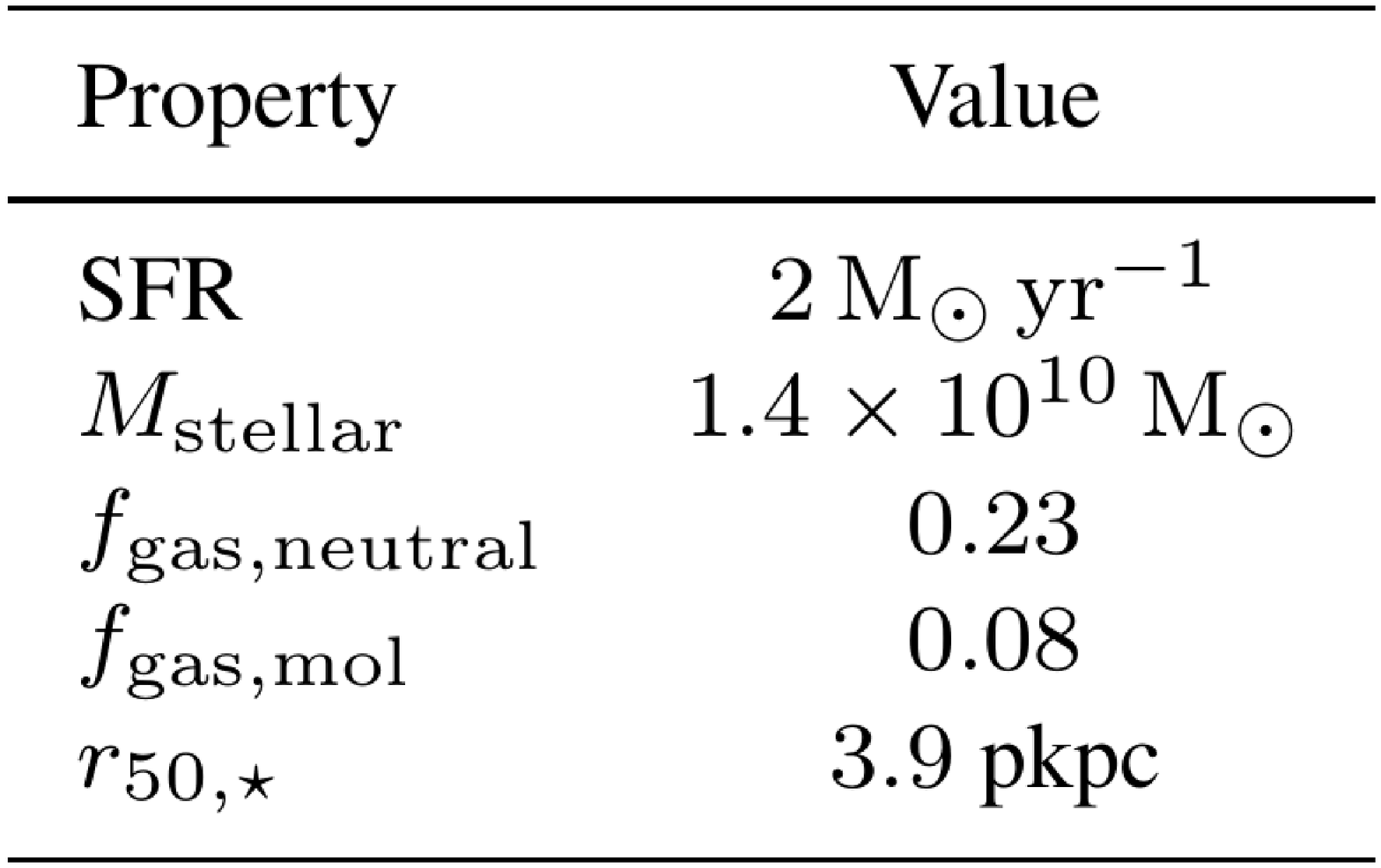}
\includegraphics[width=0.25\textwidth]{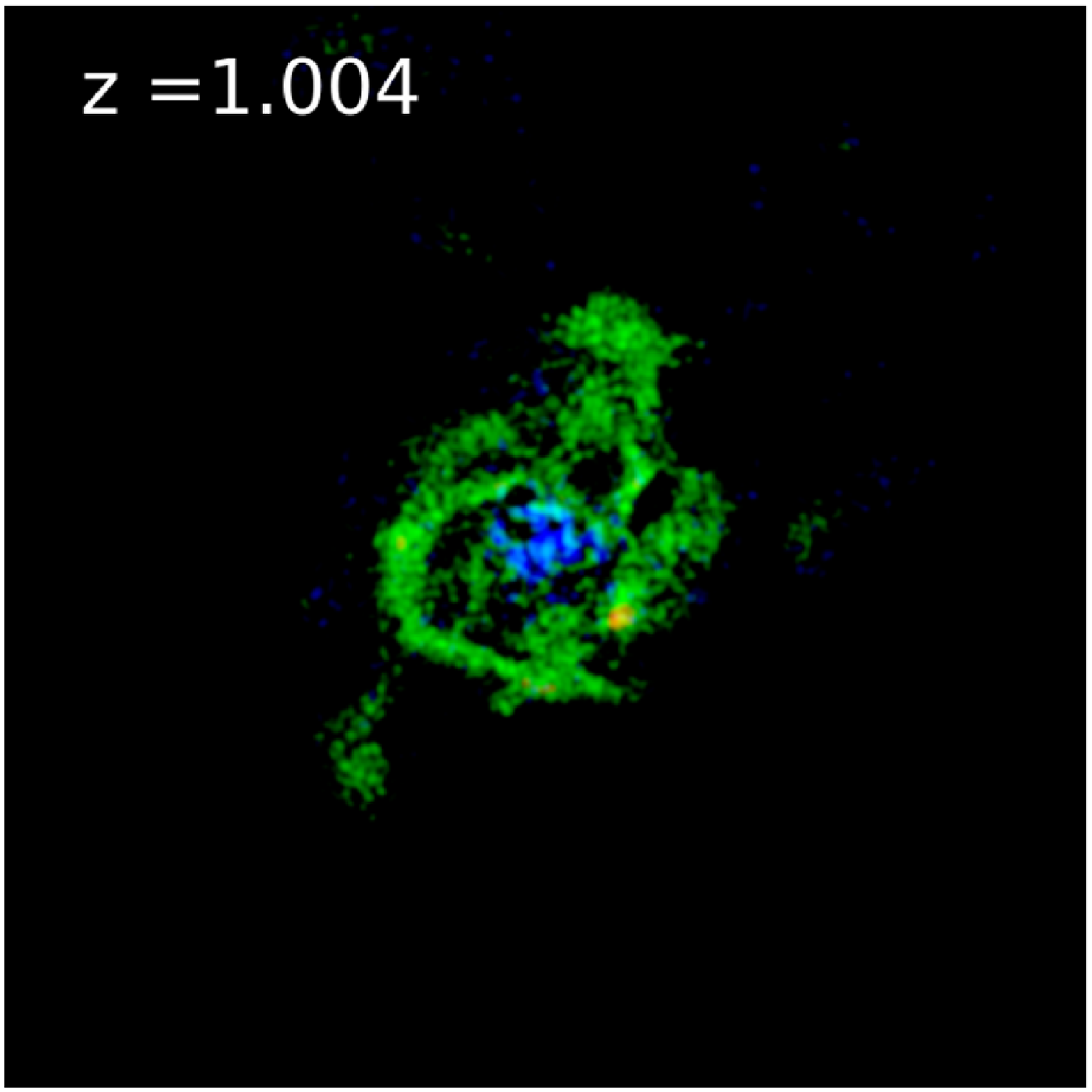}
\includegraphics[width=0.25\textwidth]{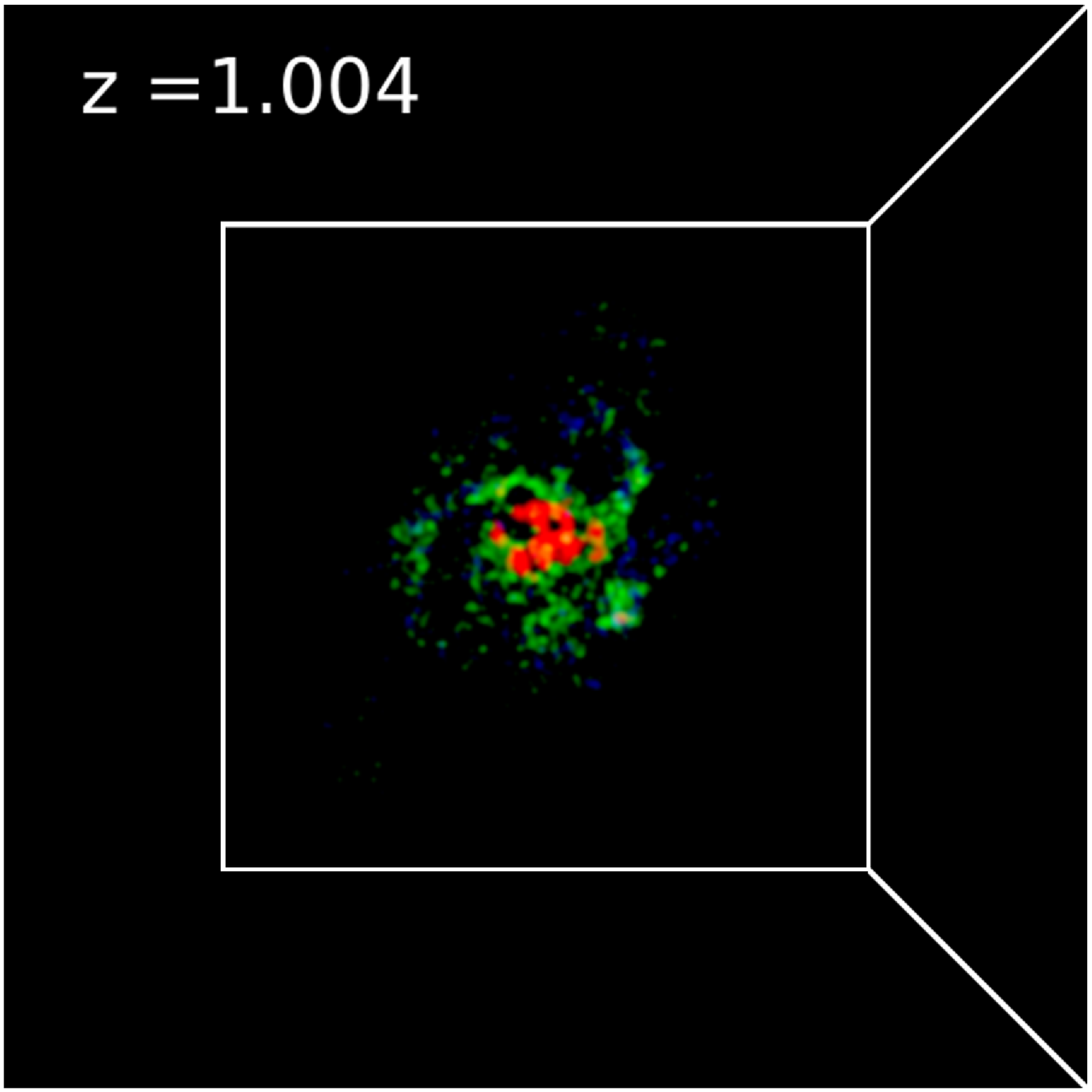}
\includegraphics[width=0.25\textwidth]{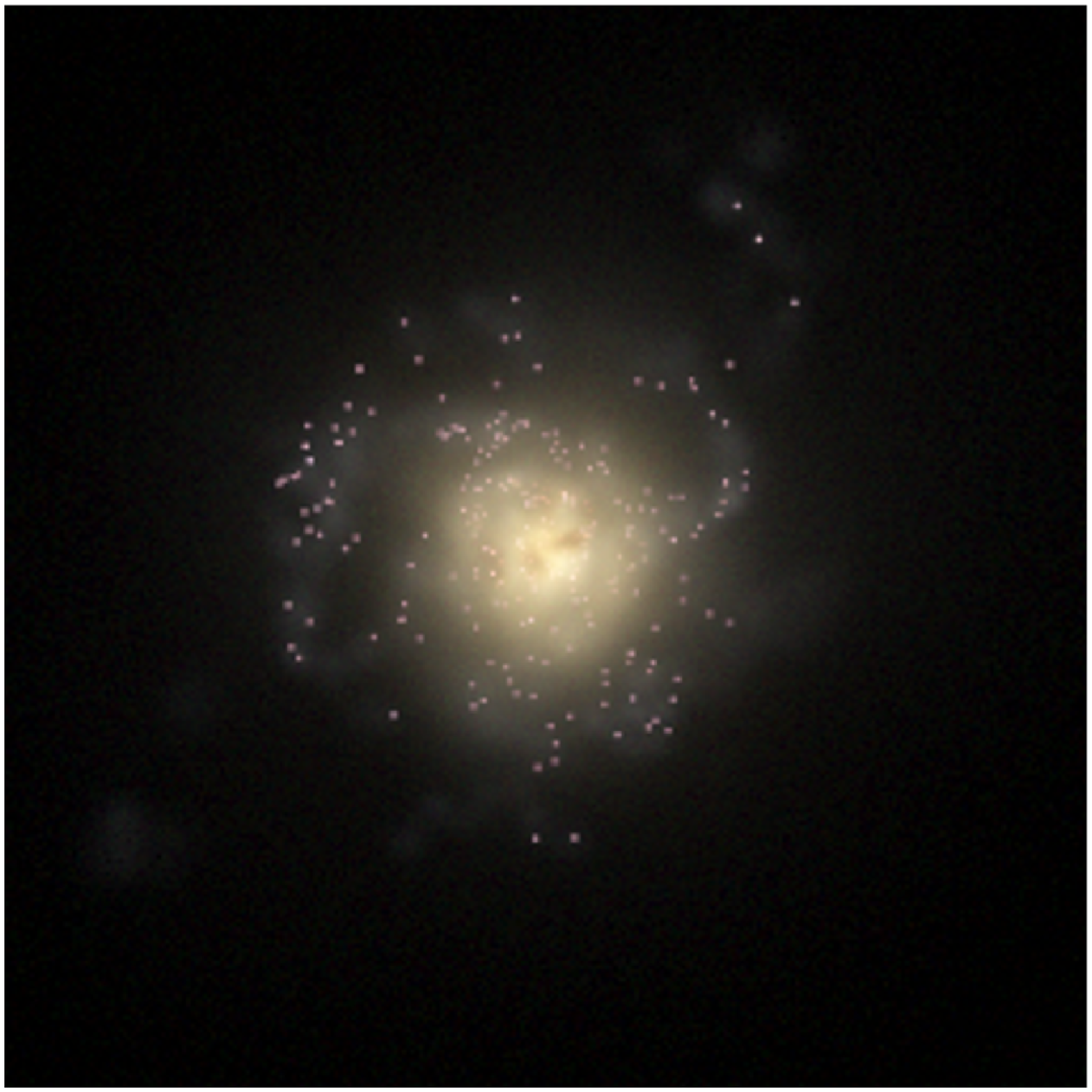}
\includegraphics[width=0.23\textwidth]{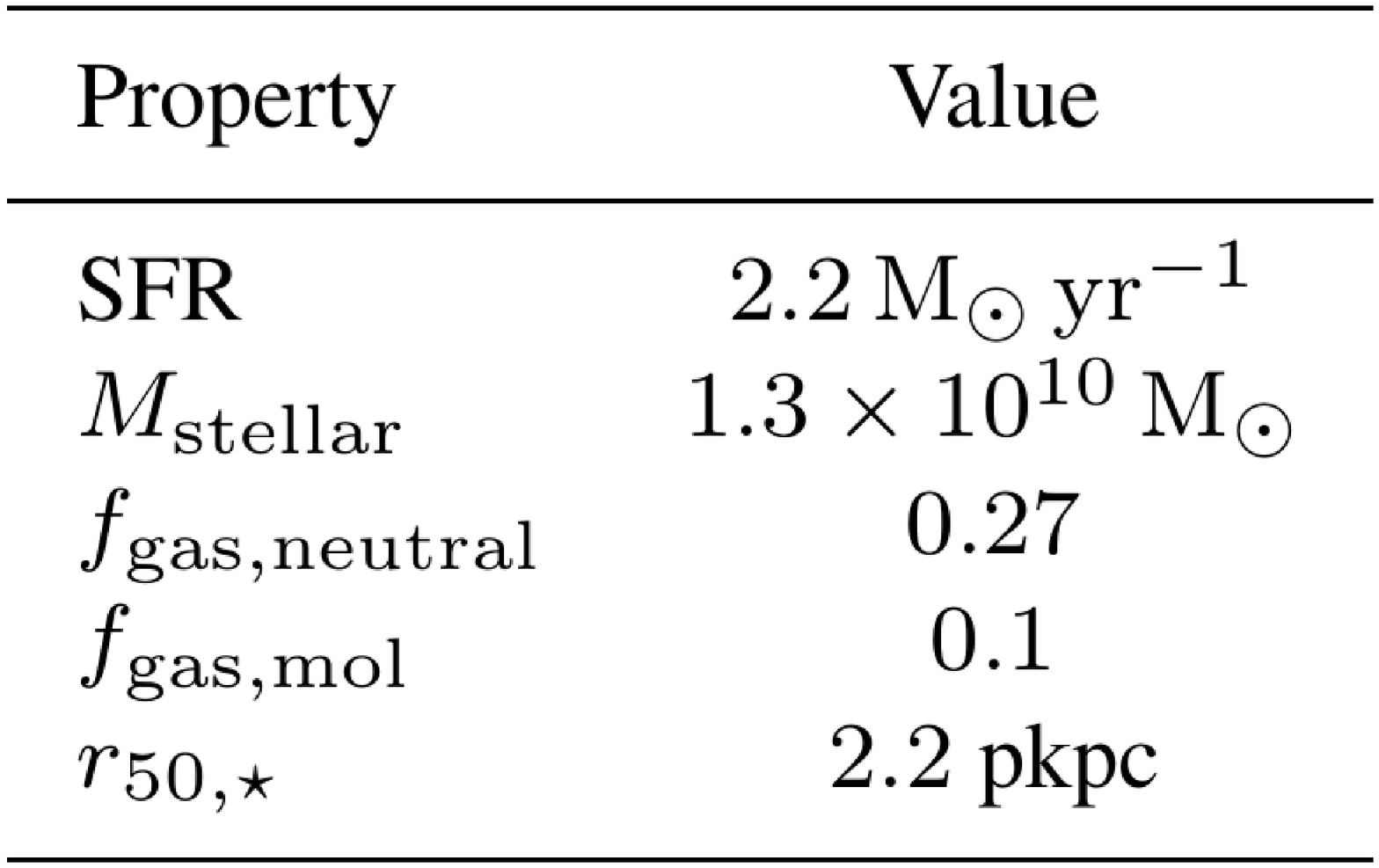}
\includegraphics[width=0.25\textwidth]{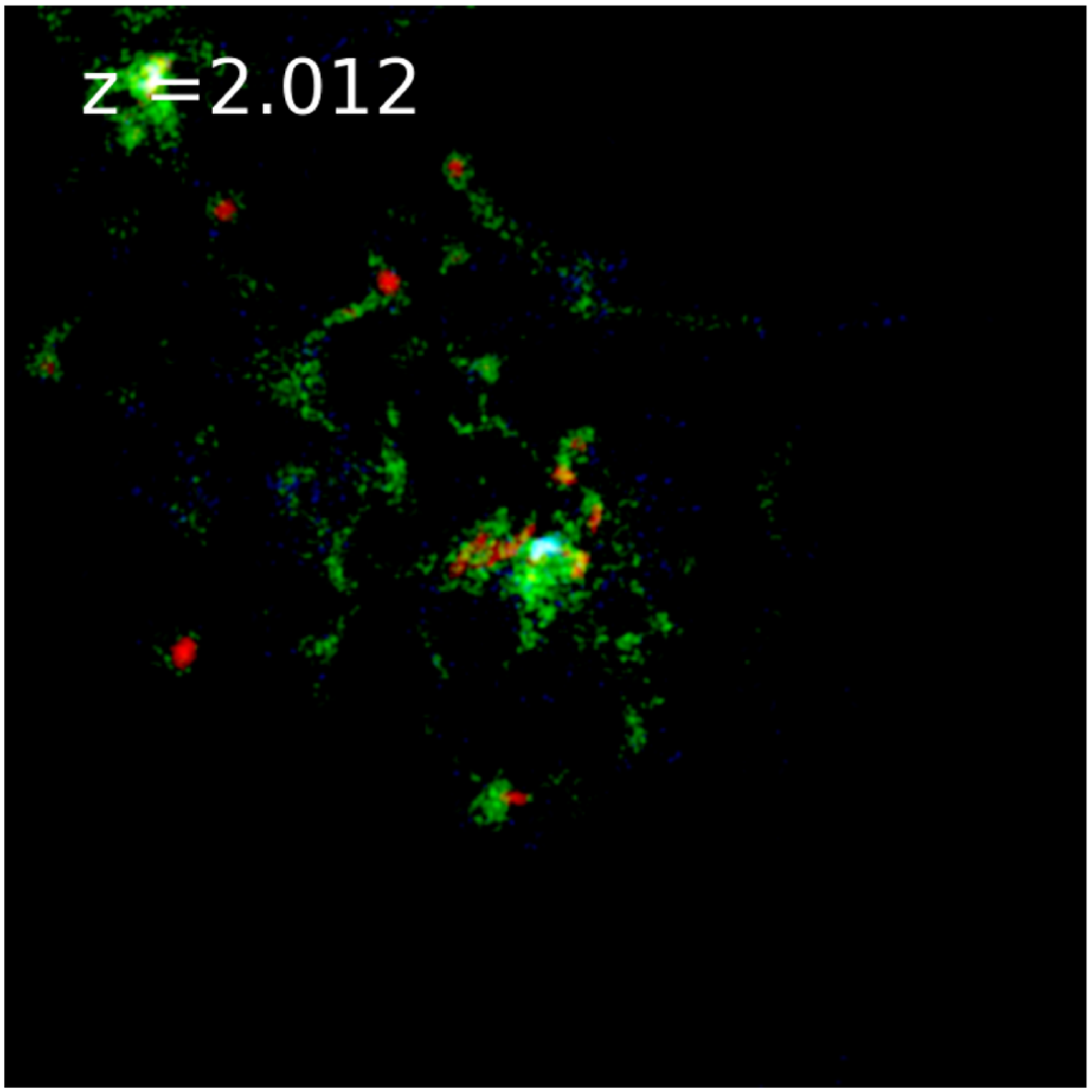}
\includegraphics[width=0.25\textwidth]{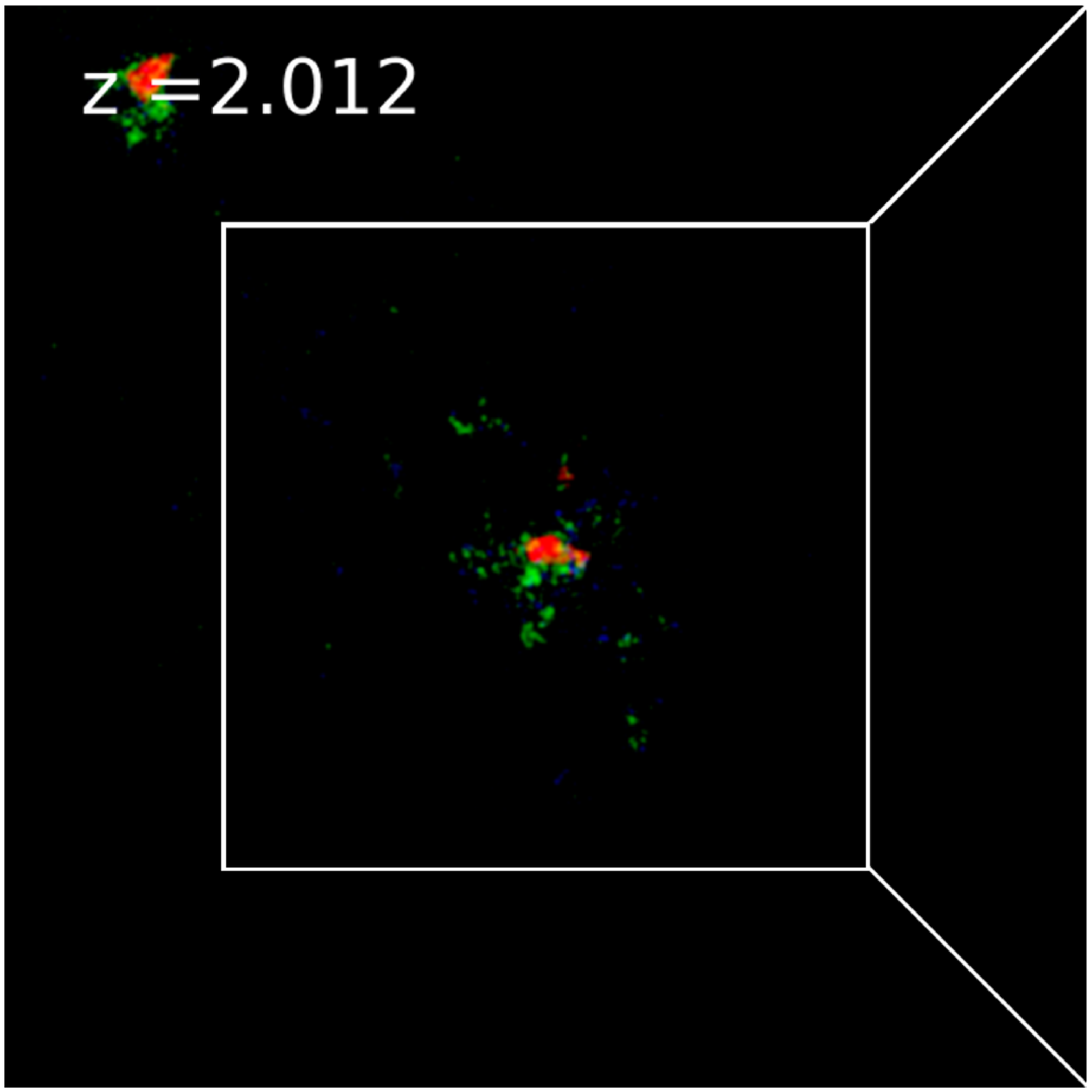}
\includegraphics[width=0.25\textwidth]{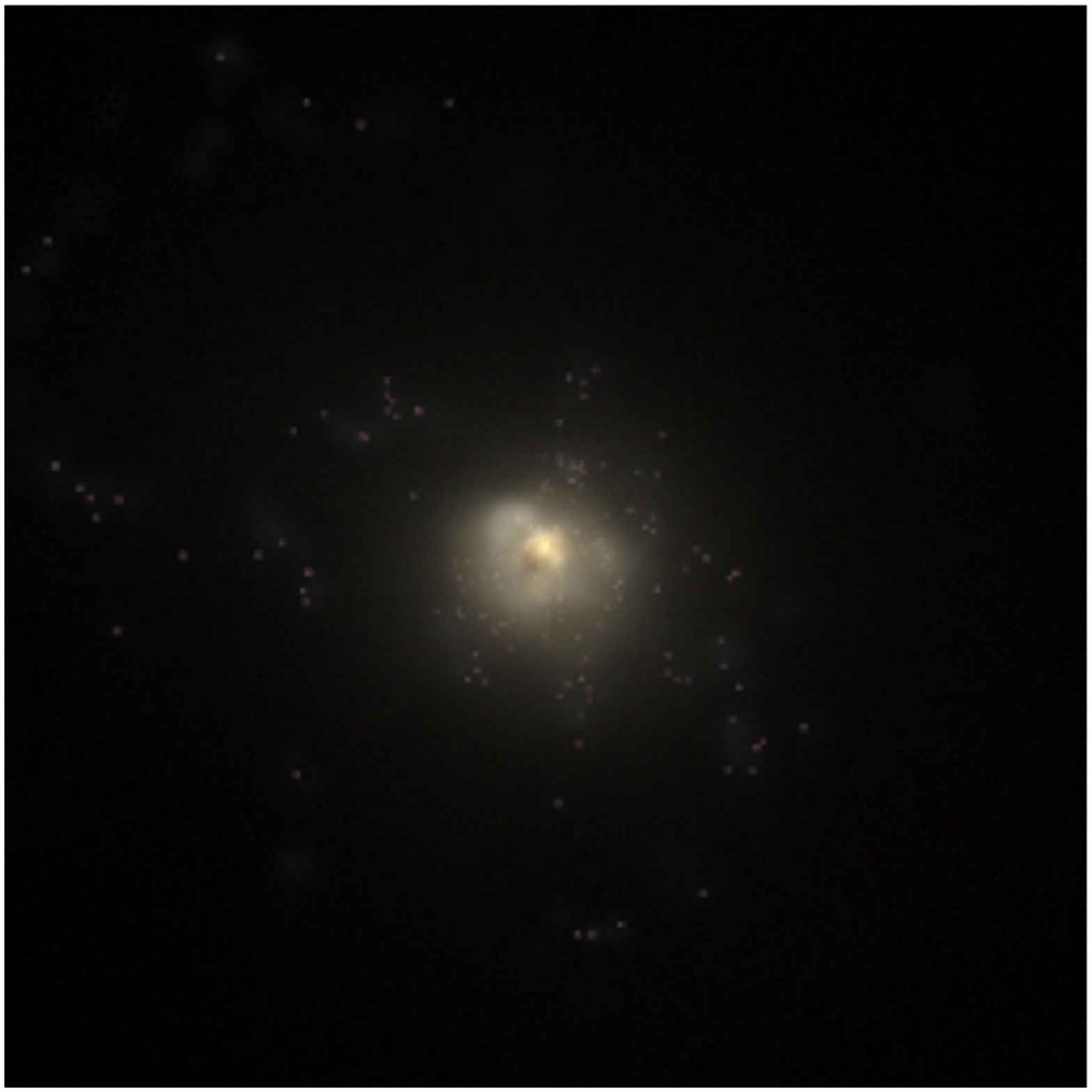}
\includegraphics[width=0.23\textwidth]{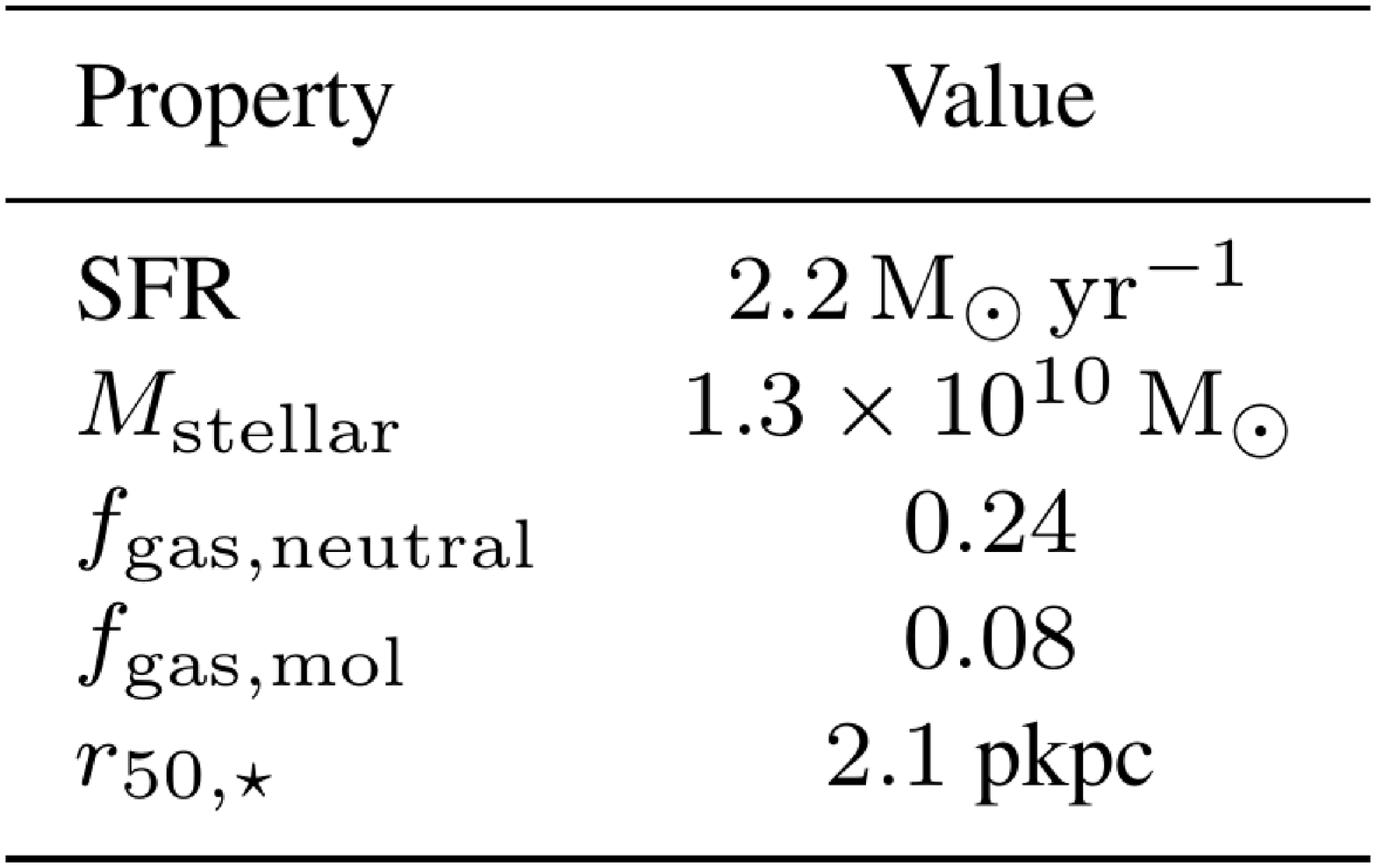}
\caption{Visualisation of $4$ galaxies in \eagle\ (at redshifts $z=0$, $z=0.5$, $z=1$ and $z=2$)
 which were chosen to have $M_{\rm stellar}\approx 10^{10}\rm M_{\odot}$, $\rm SFR\approx2\rm \, M_{\odot}\,yr^{-1}$ and 
$f_{\rm gas,neutral}\approx 0.25$.
The redshift of each galaxy is shown in the HI and H$_2$ maps.
The HI and H$_2$ maps are coloured by column density, according to the colour bars at the top, with 
column densities in units of $\rm cm^{-2}$. The right panels show SDSS gri images, which were constructed using the radiative transfer 
code {\tt SKIRT} \citep{Baes11} (see Trayford et al. in prep. for details).
Particles are smoothed by $1$~ckpc
in the $N_{\rm H_2}$ and $N_{\rm HI}$ maps. HI and H$_2$ maps 
have a size of $100\times100$~pkpc$^2$, while the gri images are of $60\times60$~pkpc$^2$ 
(scale that is shown in the middle panels as a white square frame). At the right of every row we show the integrated values
for the stellar mass, SFR, neutral gas fraction, molecular gas fraction and the {projected half-stellar mass radius. 
Masses and SFR were calculated in spherical apertures of $30$~pkpc, while the radius is calculated using a 2D circular aperture 
of $30$~pkpc (averaged over three orthogonal projections)}.}
\label{Images1}
\end{center}
\end{figure*}

We select examples of 
galaxies of a similar stellar mass, SFR and neutral gas fraction at different redshifts to examine their similarities and differences.
Fig.~\ref{Images1} shows the atomic and molecular column density maps and the optical gri images of $4$ galaxies at $z=0$, $0.5$, $1.0$ and $2$
 with $M_{\rm stellar}\approx 1.1\times 10^{10}\,\rm M_{\odot}$, $\rm SFR \approx 2\,\rm M_{\odot}\,yr^{-1}$
and $f_{\rm gas,neutral}\approx 0.2$. The optical images were created using radiative transfer simulations 
performed with the code {\tt SKIRT} \citep{Baes11} in the SDSS g,~r and i filters \citep{Doi10}. Dust extinction was implemented 
using the metal distribution of galaxies in the simulation, and assuming $40$\% of the metal mass is locked up is dust grains \citep{Dwek98}. The images were produced 
using particles in spherical apertures of $30$~pkpc around the centres of sub-halos (see \citealt{Trayford15}, and in prep. for more details).

At $z=0$,  $\rm SFR \approx 2\,\rm M_{\odot}\,yr^{-1}$ and $f_{\rm gas,neutral}\approx 0.2$ are typical values of 
galaxies with $M_{\rm stellar}\approx 10^{10}\,\rm M_{\odot}$ in the main sequence of star formation. However, at higher 
redshifts, the normalisation of the sequence increases, and therefore a galaxy 
with the stellar mass, SFR and neutral gas fraction 
 above lies below the main sequence of star formation and is thus considered an unusually passive galaxy. 
Nonetheless, it is illuminating to visually inspect galaxies of the same properties at different redshifts. 

We find that the $z=0$ galaxy in Fig.~\ref{Images1} is an ordered disk (which is a common feature of  galaxies
 with these properties at $z=0$), with most of the star formation proceeding
in the inner parts of the galaxy and in the disk 
(compare H$_2$ mass with stellar density maps). However, in the $z=0.5$ and $z=1$ galaxies we see striking differences:
the higher-redshift galaxies are smaller {(see the values of their half-mass radius listed in 
Fig.~\ref{Images1})}, have more disturbed disks, have steeper H$_2$ density profiles, and are more clumpy.
This is particularly evident when we compare the $z=0$ galaxy with its $z=1$ counterpart with the same integrated properties.
The picture at $z=2$ again changes completely: the neutral gas of the $z=2$ galaxy displays a very irregular morphology with filaments at $\approx 50-100$~pkpc from the 
galaxy centre, which is much more evident in HI than in H$_2$, but still present in the latter. 
In the $z=2$ galaxy, a significant fraction of the H$_2$ is locked up in big clumps, which is in contrast with the 
smooth distribution of H$_2$ in the $z=0$ galaxy.

Although galaxies 
follow a tight plane relating $f_{\rm gas,neutral}$, stellar mass and SFR with little redshift evolution, 
they can have strikingly different morphologies 
even at fixed $f_{\rm gas,neutral}$, stellar mass and SFR. We analyse this in detail in an upcoming paper 
(Lagos et la. in prep.).

\subsection{The Mass-Metallicity relation}\label{MMrelation} 

The PCA performed with \eagle\ galaxies 
shows that the mass-metallicity relation emerges mostly in the second principal 
component (that accounts for $24$\% of all the variance seen in the galaxy population of the simulation). However, the relation between 
stellar mass and gas metallicity is not so strong, and other variables are also relevant in the principal component, such as 
SFR and gas mass. We find that the neutral gas fraction again plays a more important role than the molecular gas fraction. 
Here we analyse in detail this multi-dimensional scaling.

\begin{figure}
\begin{center}
\includegraphics[width=0.49\textwidth]{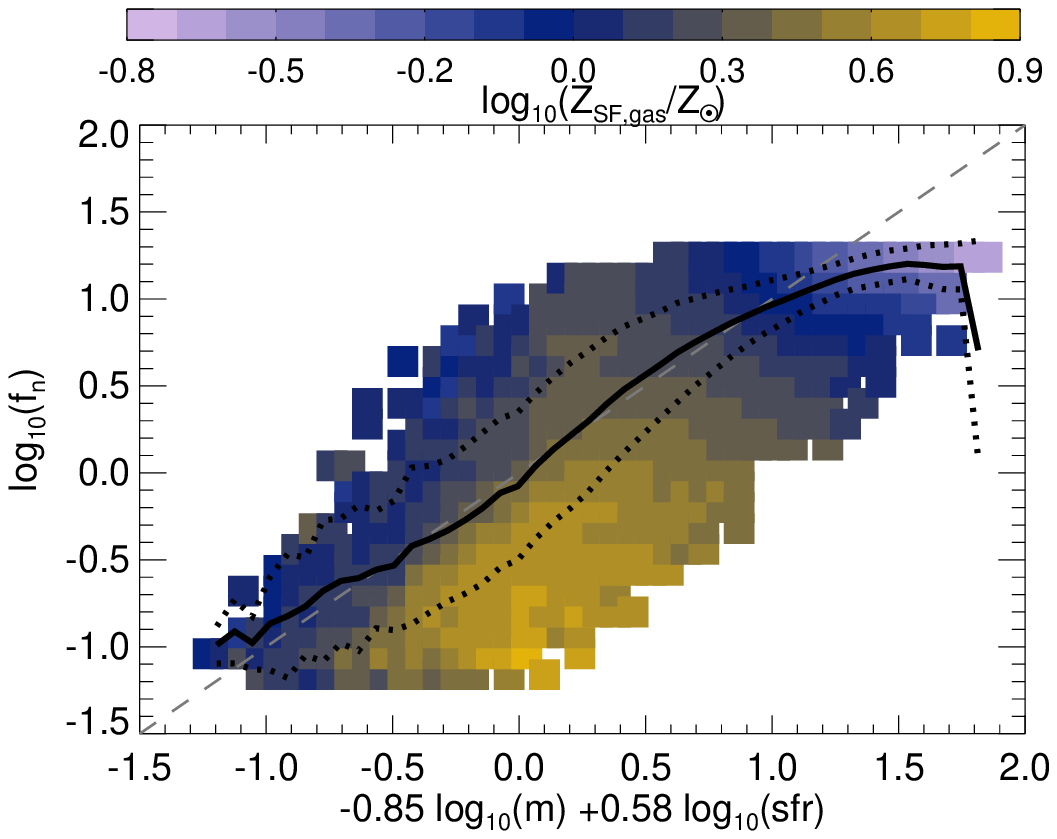}
\includegraphics[width=0.49\textwidth]{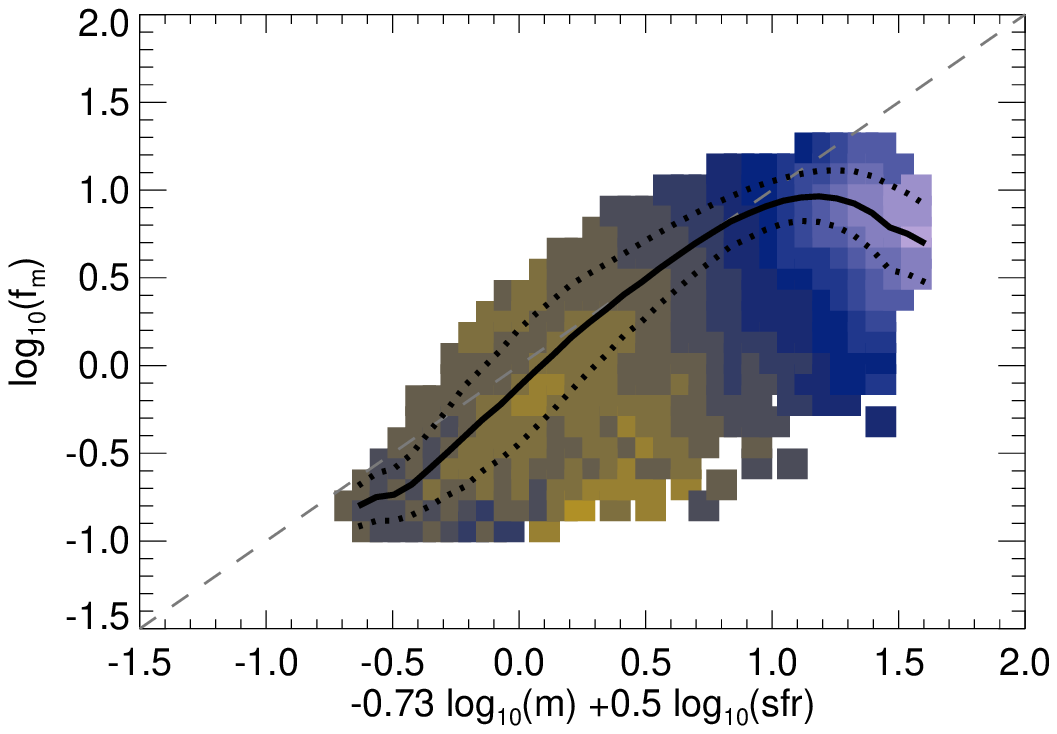}
\caption{{\it Top panel:} As in the top-left panel of 
Fig.~\ref{HIFP},  but here we colour pixels by the metallicity of the star-forming gas, as the colour bar at the top of the panel shows.
{\it Bottom panel:} As in the top-left panel of Fig.~\ref{H2FP}, but here we colour pixels by the metallicity of the star-forming gas.
For movies rotating over the 3-dimensional spaces see {\small \tt www.clagos.com/movies.php}.
{The dashed lines in both panels show the one-to-one relationship (i.e. best fits of Eqs.~\ref{Eqfneutral}~and~\ref{Eqfmol})}.}
\label{HIFPZ}
\end{center}
\end{figure}

Fig.~\ref{HIFPZ} shows edge-on views of the planes in the space comprising $f_{\rm gas,neutral}$ or $f_{\rm gas,mol}$ and stellar mass and SFR 
(defined as in Eqs.~\ref{Eqfneutral}~and~\ref{Eqfmol}, respectively).
 All galaxies in the redshift range $0\le z\le 4.5$ and with $M_{\rm stellar}>10^9\rm \,M_{\odot}$ 
were included in the figure. Pixels are coloured according to the median star-forming gas (i.e. ISM) metallicity, as indicated by the colour bar. 
Gas metallicity decreases as the neutral gas fraction increases at fixed x-axis value. 
Galaxies with high neutral gas fractions 
and high SFRs are almost exclusively metal poor. For example, galaxies with $\rm SFR>15\rm \,M_{\odot}\,yr^{-1}$ and 
$f_{\rm gas,neutral}>0.7$ have a median metallicity of the star-forming gas of $Z_{\rm SF,gas}\approx 0.3\,\rm Z_{\odot}$. 
For galaxies with slightly lower SFRs, $\rm 10\rm \,M_{\odot}\,yr^{-1}<SFR<15\rm \,M_{\odot}\,yr^{-1}$ and
$0.4<f_{\rm gas,neutral}<0.6$, the median metallicity of the star-forming gas is $Z_{\rm SF,gas}\approx 0.8\,\rm Z_{\odot}$.
The trend of decreasing metallicity with increasing gas fraction is not driven by how galaxies at different 
redshift populate the plane, given that the metallicity trend of Fig.~\ref{HIFPZ} 
is still seen at fixed redshift (this is not shown in Fig.~\ref{HIFPZ}).
 Note that the direction in which the metallicity of the star-forming gas changes 
is orthogonal to the plane defined by Eq.~\ref{Eqfneutral}. 
{Galaxies with $\rm log_{10}(f_n)\gtrsim 1$, that are among the most metal-poor galaxies 
in EAGLE, lie in the region where the relation in the top panel of 
Fig.~\ref{HIFPZ} flattens. These galaxies correspond to star-forming dwarf galaxies in EAGLE (which have 
SFR$\approx 2-3\,\rm M_{\odot}\,yr^{-1}$, $M_{\rm stellar}\approx 2\times 10^9\rm M_{\odot}$ and 
$f_{\rm gas, neutral}\gtrsim 0.5$).}

Galaxies with high molecular gas fractions also tend to be more metal 
poor than galaxies with lower $f_{\rm gas,mol}$. For example, galaxies with $\rm SFR>15\rm \,M_{\odot}\,yr^{-1}$ and
$f_{\rm gas,mol}>0.3$ have a median metallicity of the star-forming gas of $Z_{\rm SF,gas}\approx 0.6\,\rm Z_{\odot}$. 
For galaxies with slightly lower SFRs, with $\rm 10\rm \,M_{\odot}\,yr^{-1}<SFR<15\rm \,M_{\odot}\,yr^{-1}$ and
$0.1<f_{\rm gas,mol}<0.2$, the median metallicity of the star-forming gas is $Z_{\rm SF,gas}\approx 1.2\,\rm Z_{\odot}$.
However, the direction of the correlation here is different to the one found for $f_{\rm gas,neutral}$. 
The metallicity of the star-forming gas decreases parallel to the plane of Eq.~\ref{Eqfmol}, which means 
that little extra information is gained through adding gas metallicity as an extra dimension in the dependence  
$f_{\rm gas,mol}$-stellar mass-SFR.

\begin{figure}
\begin{center}
\includegraphics[width=0.49\textwidth]{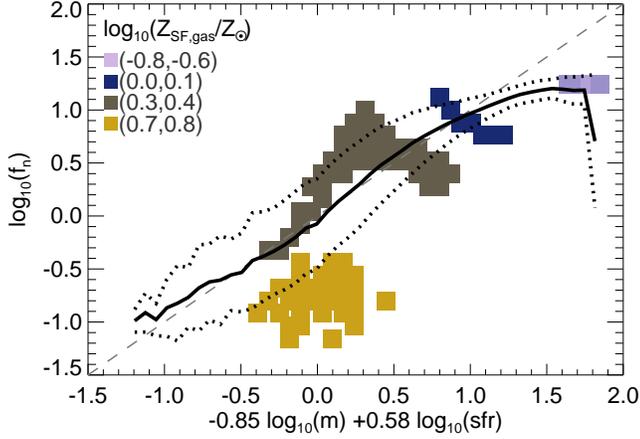}
\caption{As in the top panel of Fig.~\ref{HIFPZ} but here we show 4 discrete bins in metallicity, as labelled. 
{Galaxies in a narrow range of metallicity occupy well defined regions of the fundamental plane of star formation}.}
\label{HIFPZBins}
\end{center}
\end{figure}

In order to help visualise this more clearly, we show in Fig.~\ref{HIFPZBins} the same edge-on view of the top panel of Fig.~\ref{HIFPZ} 
but for four narrow bins of star-forming gas metallicity.
We see that narrow ranges in metallicity result in a very small portion of the 
plane being sampled. This implies that the position of the galaxy  
on the plane comprised of $f_{\rm gas,neutral}$, stellar mass and SFR is a 
good proxy for the star-forming gas metallicity.
 A consequence of this is that the scatter in the SFR 
or neutral gas fraction largely determines the scatter in the stellar mass-gas metallicity 
relation. {This agrees with recent claims by \citet{Zahid14}, which based on observations and a simple model of 
chemical enrichment, claim that gas metallicity is strongly correlated with the gas fraction, with the latter relation not 
evolving in time.} 

We find that the gas metallicity in \eagle\ is more strongly correlated with 
the neutral gas fraction than with the molecular gas fraction.
{Recently, \citet{Bothwell15}, using a sample comprising $221$ galaxies in the redshift range 
$0\le z\le 2$, claimed that the residuals of the MZ relation 
are more strongly correlated with H$_2$ than SFR. However, due to the lack of data, Bothwell et al. 
were not able to test whether atomic hydrogen or neutral hydrogen masses are better predictions of the scatter than 
the H$_2$ mass.}

We use the {\tt HYPER-FIT} R package of \citet{Robotham15} to fit the dependence of $Z_{\rm SF,gas}$ on stellar mass, SFR and 
$f_{\rm gas,neutral}$ and find that the least scatter $3$-dimensional surface has a very weak dependence 
on stellar mass and SFR, and a strong dependence on 
$f_{\rm gas,neutral}$. This means that the metallicity of the star-forming gas in galaxies can be predicted from the neutral gas fraction alone 
to within $40$\%. {We perform these fits independently of Eqs.~\ref{Eqfneutral}~and~\ref{Eqfmol}.} 
The best fit between $Z_{\rm SF,gas}$ and $f_{\rm gas,neutral}$ is:

\begin{eqnarray}
{\rm log_{\rm 10}}\left(\frac{Z_{\rm SF,gas}}{Z_{\odot}}\right)&=&-0.57\, {\rm log_{\rm 10}}\left(\frac{f_{\rm gas,neutral}}{0.09}\right).\label{EqMMFNeutral}
\end{eqnarray}

\noindent The standard deviation perpendicular to the fitted relation of 
Eq.~\ref{EqMMFNeutral} is $0.17$~dex, while the standard deviation parallel to the 
metallicity axis is $0.19$~dex. 
We find that the metallicity can also be predicted from a combination of the stellar mass and SFR, although with a slightly larger scatter:

\begin{eqnarray}
{\rm log_{\rm 10}}\left(\frac{Z_{\rm SF,gas}}{Z_{\odot}}\right)&=&0.2+0.45\,{\rm log_{10}(m)}\nonumber\\
                & & -0.37\,{\rm log_{10}(sfr)},\label{EqMMSFR}
\end{eqnarray}

\noindent where m and sfr are defined in Eq.~\ref{defvariables}.
The standard deviation perpendicular to the fitted relation of Eq.~\ref{EqMMSFR} is $0.19$~dex, while the 
scatter parallel to the metallicity axis is $0.2$~dex.
From the standard deviations above, we can say 
that Eqs.~\ref{EqMMFNeutral}~and~\ref{EqMMSFR} 
are similarly good representations of $Z_{\rm SF,gas}$ in \eagle\ galaxies.
 
\begin{figure}
\begin{center}
\includegraphics[width=0.49\textwidth]{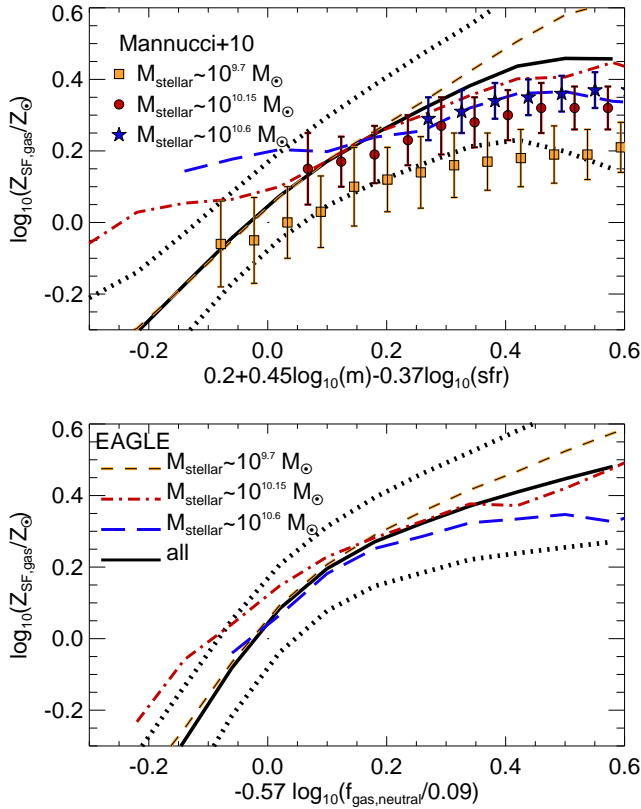}
\caption{{\it Top panel:} Edge-on view of the plane of Eq.~\ref{EqMMSFR}, 
comprised of star-forming gas metallicity, SFR and stellar mass 
 in \eagle. The thick solid line and the dotted lines show the median and 1$\sigma$ scatter, respectively, of all galaxies with 
$M_{\rm stellar}>10^{9}\,\rm M_{\odot}$ and SFR$>0.01\,\rm M_{\odot}\,yr^{-1}$. 
We also show the results for \eagle\ galaxies in narrow bins of stellar masses in lines as labelled in the bottom panel.
Observations at $z=0$ from \citet{Mannucci10} are shown for the same $3$ bins of stellar mass we used for \eagle. Observations are shown 
as symbols (labelled at the top-left corner). 
\eagle\ galaxies with $M_{\rm stellar}>10^{10}\,\rm M_{\odot}$ and the observations agree to within $0.15$~dex, while lower-mass galaxies 
show discrepancies with the observations (up to $\approx 0.4$~dex).
To convert the observations of Mannucci et al. from oxygen abundance to metallicity, we 
adopted a solar oxygen abundance of $12+\rm log_{10}(O/H)_{\odot}=8.69$ and $Z_{\odot}=0.0127$.
{\it Bottom panel:} As in the top panel, but here we show an 
edge-on view of the relation between star-forming gas metallicity and neutral gas fraction (Eq.~\ref{EqMMFNeutral}).}
\label{ZFits}
\end{center}
\end{figure}

We assess the performance of the fits of Eqs.~\ref{EqMMFNeutral}~and~\ref{EqMMSFR} 
and compare with the observations of \citet{Mannucci10} in Fig.~\ref{ZFits}. 
In \eagle, deviations from 
Eqs.~\ref{EqMMSFR}                 
and~\ref{EqMMFNeutral} are seen at $Z_{\rm SF,gas}\gtrsim 3\,Z_{\odot}$ and $Z_{\rm SF,gas}\lesssim 0.7\,Z_{\odot}$. However, 
$73$\% of the galaxies at $0\le z\le 4.5$ have $0.7\,Z_{\odot} \le Z_{\rm SF,gas}\le 3\,Z_{\odot}$, and thus 
the fits of Eqs.~\ref{EqMMFNeutral}~and~\ref{EqMMSFR} are good descriptions of the majority of the galaxies in \eagle.
We also show how observed galaxies populate the plane of Eq.~\ref{EqMMSFR}. For this we took the tabulated results for 
the dependence of gas metallicity on 
SFR and stellar mass from \citet{Mannucci10} and show here $4$ bins of stellar mass.
We find that observed galaxies follow a plane in the 3-dimensional space of metallicity, SFR and stellar mass that is very similar to the 
one that \eagle\ galaxies follow. The agreement between 
observations and \eagle\ galaxies with $M_{\rm stellar}\gtrsim 10^{10}\,\rm M_{\odot}$ is good (deviations are 
of $\lesssim 0.15$~dex). However, galaxies with $M_{\rm stellar}<10^{10}\,\rm M_{\odot}$ in the observations have metallicities that are 
$\approx 0.3-0.4$~dex lower than \eagle\ galaxies of the same stellar mass.
This is consistent with the discrepancies seen in the comparison presented in S15 between the predicted MZ relation in \eagle\ and the 
observations of \citet{Tremonti04}. S15 show that this is related to the resolution of the simulation, as the higher resolution run that 
is recalibrated to reach a similar level of agreement with the $z=0.1$ stellar mass function and size-stellar mass relation, displays a 
MZ relation in much better agreement than the simulation we use here.
The effect this discrepancy has on the results 
presented in Fig.~\ref{ZFits} is minimal because the fit of Eq.~\ref{EqMMSFR} was calculated using 
 the inverse of the number density as weight, and therefore low-mass galaxies, which display the 
largest discrepancies with the observed metallicity of galaxies, 
do not significantly skew the fit.

\section{Conclusions}\label{ConcluSec}

We have studied the evolution of the gas fraction and the 
multi-dimensional dependence between stellar mass, star formation rate, 
gas fraction and gas metallicity in the {EAGLE} suite of hydrodynamical simulations. 
We use the gas phase transitions from ionised to neutral, and from neutral to molecular, implemented on a particle-by-particle 
bases in post-processing by \citet{Lagos15}. 
The post-processing is done using the fitting functions of \citet{Rahmati13} for the transition from ionised to 
neutral gas, and of \citet{Gnedin11} for the transition from neutral to molecular gas.

We summarise our main results below:
\begin{itemize}
  \item We find that at fixed stellar mass, both the neutral and molecular gas fractions 
increase with redshift. In the case of the neutral gas fraction, this increase is a factor of $\approx 5$
between $z=0$ and $\approx 2.5$, while the same increase is seen in the molecular gas fraction over a shorter timescale, 
from $z=0$ to $z\approx 1.5$. The gas fractions at higher redshifts plateaus or even decreases. The specific SFR on the other hand 
increases by a factor of $\approx 15$ over the same redshift interval. The difference is due to high-$z$ galaxies having higher 
$\rm SFR/M_{\rm H_2}$ and $\rm SFR/M_{\rm neutral}$ than $z=0$ galaxies, which in turn is caused by the superlinear star formation 
law adopted in \eagle\ and the higher gas pressure at high redshift.
  \item The evolution of the gas fraction is related to that of the SFR and the stellar mass.
Galaxies show little evolution in their gas fraction at fixed stellar mass and SFR. 
This is a consequence of galaxies in \eagle\ following with little scatter a 2-dimensional 
surface in the 3-dimensional space of stellar mass, SFR and neutral (or molecular) gas fraction. 
We term the plane tangential to this surface at the mean location of galaxies the
``fundamental plane of star formation'', and provide fits derived from \eagle in Eqs.\ref{Eqfneutral}~and~\ref{Eqfmol}.  
These 2-dimensional surfaces are also seen in a 
compilation of observations of galaxies at $0\le z\le 3$ that we presented here. Observed and simulated 
galaxies populate the three-dimensional space of SFR, stellar mass and gas fractions in a very similar manner.
A PCA analysis reveals that the relation between the neutral gas fraction, stellar mass and SFR contains most of the variance ($55$\%)
seen in the galaxy population of \eagle, and therefore is one of the most fundamental correlations, which we term 
the ``fundamental plane of star formation''.
   \item We attribute the existence of the 2-dimensional surfaces 
above to the self-regulation of star formation in galaxies: SFR is set by the balance between the 
accretion and outflow of gas. 
We suggest that the curvature of the plane in \eagle is set by the model of star formation adopted, 
and affected by the relation between the ISM pressure and the SFR in \eagle.
We base these arguments on the analysis of how the plane changes 
when we change the SNe and AGN feedback strength, the equation of state imposed on the unresolved
ISM and the power-law index in the star formation law adopted in \eagle (Appendix~\ref{ModelTests}).
  \item The positions of galaxies in the 2-dimensional surface in the space of gas fraction, SFR and stellar mass are very well correlated with gas metallicity. 
 The metallicity of the star-forming gas can therefore be predicted from the stellar masses and SFRs of galaxies, 
or from the neutral gas fraction of galaxies alone, to within $\approx 40$\%. 
This relation between metallicity, stellar mass, SFR and neutral gas fraction appears in the PCA 
in the second component, and contributes $24$\% to the variance seen in the galaxy population in \eagle.
  \item The neutral gas fraction is more strongly correlated with the scatter in the stellar mass-metallicity (MZ) 
relation than the molecular gas fraction.
{Upcoming surveys will be able to test these predictions, as they will increase the number of galaxies} 
sampled in their HI content by a factor of $\approx 100$ (see for instance the accepted
proposals of the ASKAP HI All-Sky Survey \footnote{{\tt http://www.atnf.csiro.au/research/WALLABY/proposal.html}}, WALLABY, and
the Deep Investigation of Neutral Gas Origins survey\footnote{\tt http://askap.org/dingo}, DINGO, \citealt{Johnston08}). 
Similarly, the Atacama Large Millimeter Array (ALMA\footnote{\tt http://almaobservatory.org/}) 
and the NOrthern Extended Millimeter Array (NOEMA\footnote{\tt http://iram-institute.org/EN/noema-project.php}) will be able to 
carry out a similar task but for H$_2$ in galaxies.
\end{itemize}

This is the first time it has been shown in simulations and a large compilation of observations 
that the stellar mass, SFR and gas fraction (either neutral or molecular) of galaxies follow a well-defined surface in the 
3-dimensional space of stellar mass, SFR and neutral (or molecular) gas fraction.
The fidelity to which \eagle\ predictions describe the observations is remarkable, particularly since we focused on galaxy properties that 
were not used to constrain any of the free parameters in the sub-grid models.

\section*{Acknowledgements}

We thank Luca Cortese, Matt Bothwell, Paola Santini and Tim Davis for providing observational
datasets, and Aaron Robotham, Luca Cortese and Barbara Catinella for useful discussions.
CL is funded by a Discovery Early Career Researcher Award (DE150100618).
CL also thanks the MERAC Foundation for a Postdoctoral Research Award. 
This work used the DiRAC Data Centric system at Durham University, operated by the Institute for Computational Cosmology on behalf of the STFC DiRAC HPC Facility ({\tt www.dirac.ac.uk}). This equipment was funded by BIS National E-infrastructure capital grant ST/K00042X/1, STFC capital grant ST/H008519/1, and STFC DiRAC Operations grant ST/K003267/1 and Durham University. DiRAC is part of the National E-Infrastructure.
Support was also received via the Interuniversity Attraction Poles Programme initiated
by the Belgian Science Policy Office ([AP P7/08 CHARM]), the
National Science Foundation under Grant No. NSF PHY11-25915,
and the UK Science and Technology Facilities Council (grant numbers ST/F001166/1 and ST/I000976/1) via rolling and 
consolidating grants awarded to the ICC.
The research was supported in part by the
European Research Council under the European Union’s Seventh
Framework Programme (FP7/2007-2013)/ERC grant agreement
278594-GasAroundGalaxies.

%----------------------------------------------
\bibliographystyle{mn2e_trunc8}
\bibliography{EAGLE_Plane}
%---------------------------------------------------------------------

\label{lastpage}
\appendix
\section[]{Strong and weak convergence tests}\label{ConvTests}

\begin{table*}
\begin{center}
  \caption{\eagle\ simulations used in this Appendix.  The columns list:
    (1) the name of the simulation, (2) comoving box size, (3) number
    of particles, (4) initial particle masses of gas and (5) dark
    matter, (6) comoving gravitational
    softening length, and (7) maximum proper comoving Plummer-equivalent
    gravitational softening length. Units are indicated below the name of
    column. \eagle\
    adopts (6) as the softening length at $z\ge 2.8$, and (7) at $z<2.8$. The simulation Recal-L025N0752 
    has the same
    masses of particles and softening length values than the simulation Ref-L025N0752.}\label{TableSimus2}
\begin{tabular}{l c c c c c l}
\\[3pt]
\hline
(1) & (2) & (3) & (4) & (5) & (6) & (7) \\
\hline
Name & $L$ & \# particles & gas particle mass & DM particle mass & Softening length & max. gravitational softening \\
Units & $[\rm cMpc]$   &                &  $[\rm M_{\odot}]$ &  $[\rm M_{\odot}]$ & $[\rm ckpc]$ & $[\rm pkpc]$\\
\hline
Ref-L025N0376 & $25$ &   ~$2\times 376^3$  &$1.81\times 10^6$   &   ~~$9.7\times 10^6$ & $2.66$ & ~~$0.7$  \\
Ref-L025N0752 & $25$ &   ~$2\times 752^3$  &$2.26\times 10^5$   &   $1.21\times 10^6$  & $1.33$ &  $0.35$ \\
\hline
\end{tabular}
\end{center}
\end{table*}

S15 introduced the concept of `strong' and `weak' convergence
tests. Strong convergence refers to the case where a simulation is
re-run with higher resolution (i.e. better mass and spatial resolution)
adopting exactly the same subgrid physics and parameters. Weak
convergence refers to the case when a simulation is re-run with higher
resolution but the subgrid parameters are recalibrated to recover, as
far as possible, similar agreement with the adopted calibration
diagnostic (in the case of \eagle, the $z=0.1$ galaxy stellar mass
function and disk sizes of galaxies). 

S15 introduced two higher-resolution versions of \eagle, both in a box of
($25$~cMpc)$^{3}$ and with $2\times 752^3$ particles, Ref-L025N0752
and Recal-L025N0752 (Table~\ref{TableSimus2} shows some details of these simulations). 
These simulations have better spatial and mass
resolution than the intermediate-resolution simulations by factors of
$2$ and $8$, respectively. In the case of Ref-L025N0752, the parameters of the sub-grid physics are 
kept fixed (and therefore comparing with this simulation is a strong convergence test), while 
the simulation Recal-L025N0752 has $4$ parameters whose values have
been slightly modified with respect to the reference simulation (and therefore comparing with this simulation is a weak 
convergence test).

Here we compare the results presented throughout the paper obtained using the Ref-L100N1504 simulation with 
the results of the higher-resolution simulations Ref-L025N0752 and Recal-L025N0752. 

\begin{figure}
\begin{center}
\includegraphics[width=0.49\textwidth]{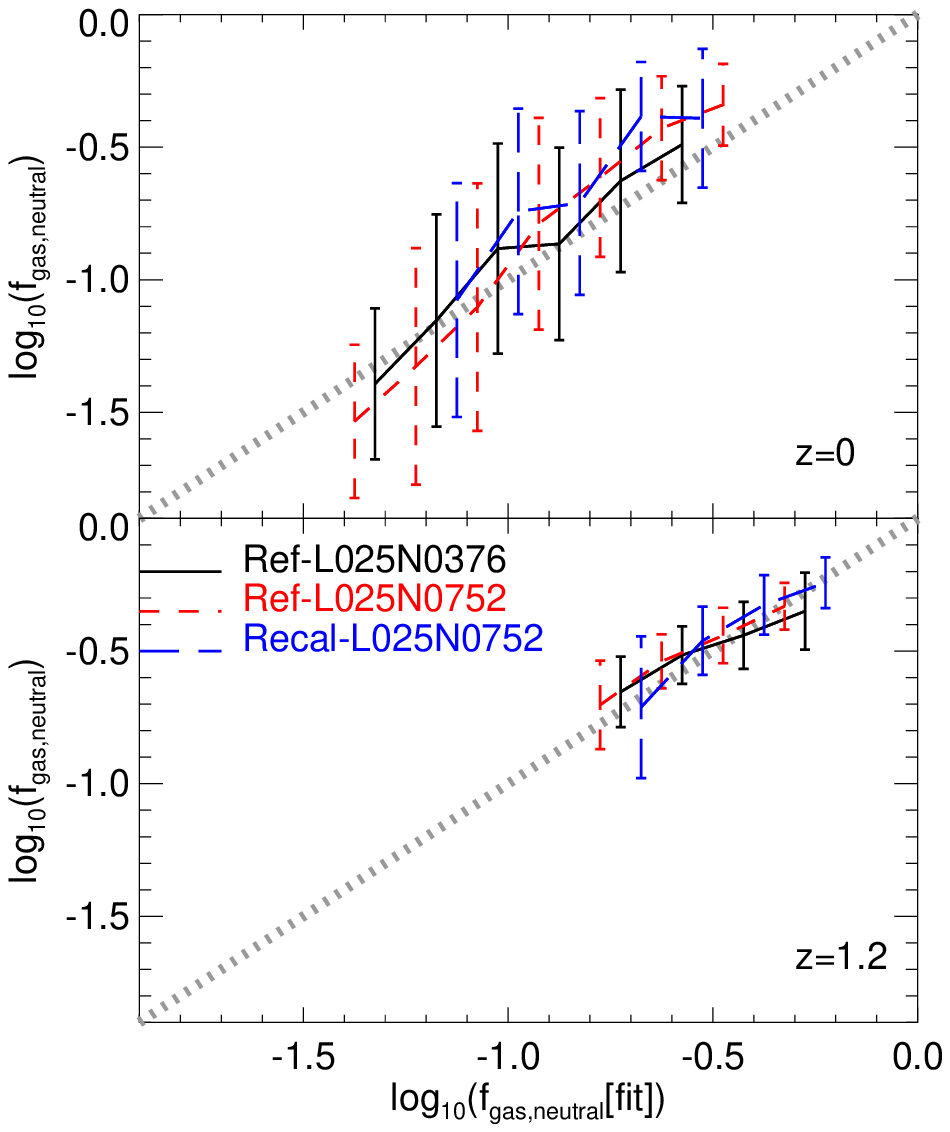}
\caption{Strong and weak convergence tests. {\it Top panel:} 
The neutral gas fraction of galaxies, defined as in Eq.~\ref{fgasneutral}, in the simulations
Ref-L025N0376, Ref-L025N0752 and Recal-L025N0752 at $z=0$, 
as a function of the neutral gas fraction calculated from the stellar masses and SFRs 
of galaxies by applying
Eq.~\ref{Eqfneutral}.
 We include all galaxies with 
$M_{\rm stellar}>10^9\,\rm M_{\odot}$. 
 Lines with error bars show the medians and the $16^{\rm th}$ and $84^{\rm th}$ percentiles, respectively.
The dotted line shows the relation
$f_{\rm gas,neutral}\equiv f_{\rm neutral,fit}$.
{\it Bottom panel:} As in the top panel but at $z=1.2$.}
\label{ConvergenceTestEagle1}
\end{center}
\end{figure}

\begin{figure}
\begin{center}
\includegraphics[width=0.49\textwidth]{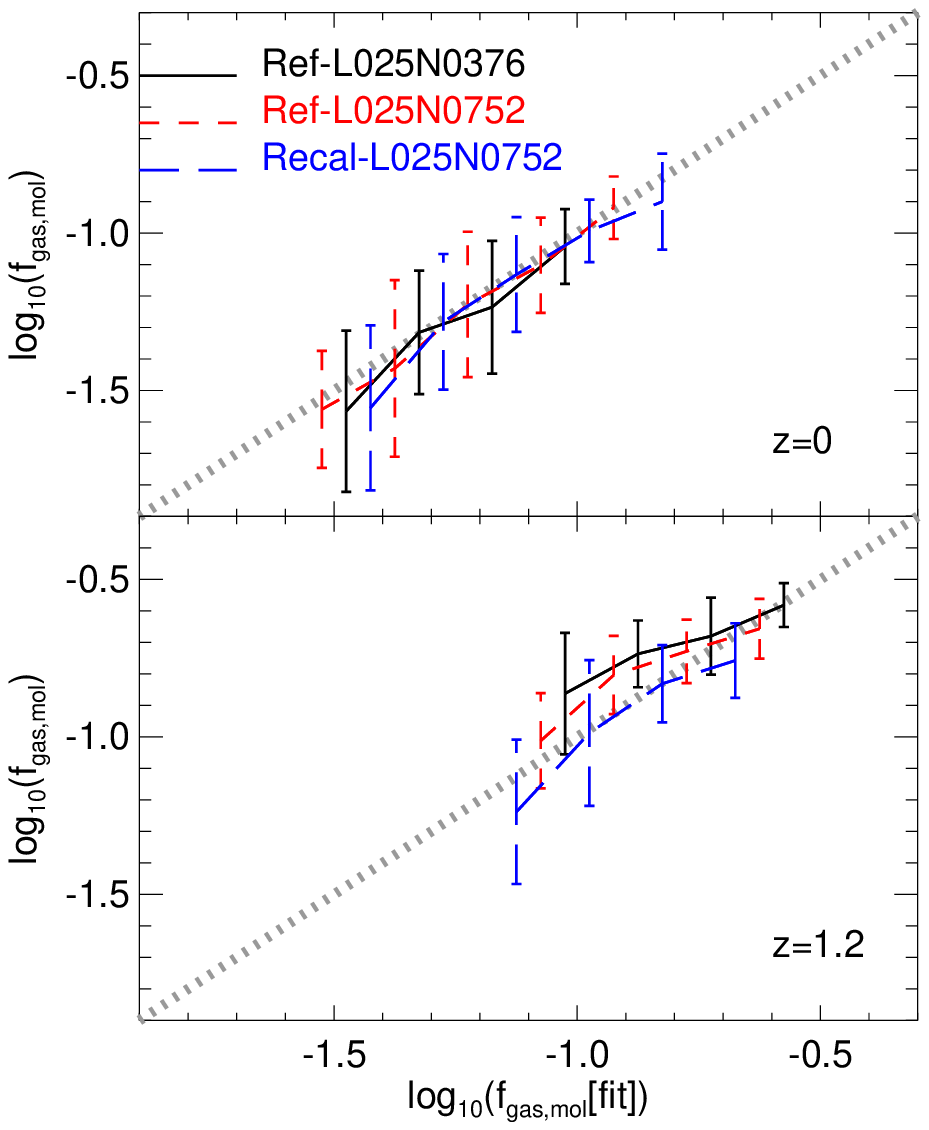}
\caption{As in Fig.~\ref{ConvergenceTestEagle1} but for the molecular 
gas fraction, defined as in Eq.~\ref{fgasmol}. We use Eq.~\ref{Eqfmol} to calculate the molecular gas fraction from the stellar masses and SFRs of galaxies.}
\label{ConvergenceTestEagle2}
\end{center}
\end{figure}

Fig.~\ref{ConvergenceTestEagle1} shows the neutral gas fraction as a function of the 
neutral gas fraction calculated from the stellar masses and SFRs of galaxies (i.e. applying Eq.~\ref{Eqfneutral}). We show the 
relation for the simulations Ref-L025N0376, Ref-L025N0752 and Recal-L025N0752 at $z=0$ and 
$z=1.2$. We find that Ref-L025N0376 follows the same relation 
as Ref-L100N1504, showing that the box size has no effect on the results. The higher-resolution runs 
are very similar, both in terms of the median and the scatter of the relation. Deviations of the high resolution runs from the 
Ref-L025N0376 simulation are of $\lesssim 0.2$~dex at $z=0$ and of $\lesssim 0.1$~dex at $z=1.2$, 
with the largest values corresponding to galaxies with 
$f_{\rm gas,neutral}\gtrsim 0.25$. 

In the case of the molecular gas fraction (Fig.~\ref{ConvergenceTestEagle2}), we find 
that deviations of the high resolution runs from the
Ref-L025N0376 simulations are $\lesssim 0.1$~dex at $z=0$ and $\lesssim 0.25$~dex at $z=1.2$. The scatter of the relations does 
not vary with resolution. Galaxies showing the largest variations with resolution are those 
that have $f_{\rm gas,mol}\lesssim 0.1$ at $z\gtrsim 1$ that populate the turnover of the relation between 
molecular fraction, stellar mass and SFR. Interestingly, we find that Recal-L025N0752 deviates more 
from Ref-L025N0376 than Ref-L025N0752 in the metrics used here. This is due to gas metallicities in  
Recal-L025N0752 being on average a factor of $\approx 2$ lower than the  
gas metallicities in Ref-L025N0376 and Ref-L025N0752 (see S15 for details). 
This influences the density threshold for star formation and explicitly enters in the second principal component of $\S$~\ref{PCASec}. 

We conclude that there is good convergence of the results for the gas fractions 
with the numerical resolution.

\section[]{Principal Component Analysis}\label{PCATest}

{Here we perform three principal component analyses over stellar mass, SFR, metallicity of the star-forming gas and 
HI, H$_2$ or total neutral gas mass. 
We do this with the aim of studying how much the variance contained in the 
first and second principal components change if only one gas phase mass is included. If the 
``fundamental plane of star formation of galaxies'' is indeed responsible for most of the variance seen in the galaxy population, 
we should find stellar mass, SFR and neutral gas mass in the first principal component, and being responsible for the largest variance 
compared to principal component vectors we obtain if we were to use atomic or molecular gas masses instead.}

\begin{table}
\begin{center}
\caption{PCA of galaxies in the Ref-L100N1504 simulation. Galaxies 
with $M_{\rm stellar}>10^9\,\rm M_{\odot}$, $\rm SFR<0.01\,M_{\odot}\,yr^{-1}$ and 
$M_{\rm H_2}/(M_{\rm H_2}+M_{\rm stellar})>0.01$ and in the redshift range $0\le z \le 4.5$ were included in the analysis.
The PCA was conducted with the variables stellar mass, SFR, metallicity of the 
star-forming gas and HI, H$_2$ or total neutral gas mass. Here we show the first two principal components for each 
PCA together with the variance they are responsible for.}
\label{TablePCA2}
\begin{tabular}{l c c c c c}
\\[3pt]
\hline
(1) & (2) & (3) & (4) & (5) & (6) \\
\hline
vector & & $\hat{\rm x}_1$ & $\hat{\rm x}_2$ &$\hat{\rm x}_3$ &$\hat{\rm x}_4$ \\
\hline
Property & variance & $\frac{M_{\rm stellar}}{\rm M_{\odot}}$ & $\rm \frac{SFR}{M_{\odot}\,yr^{-1}}$ & $\rm \frac{Z_{\rm SF,gas}}{Z_{\odot}}$ & $\frac{M_{\rm H_2}}{\rm M_{\odot}}$\\
\hline
PC1 & $59$\% & $0.49$ & $0.5$ & $0.58$ & $0.4$\\
PC2 & $35$\% & $0.59$ & $-0.24$ & $0.21$ & $-0.74$\\
\hline
Property & & $\frac{M_{\rm stellar}}{\rm M_{\odot}}$ & $\rm \frac{SFR}{M_{\odot}\,yr^{-1}}$ & $\rm \frac{Z_{\rm SF,gas}}{Z_{\odot}}$ & $\frac{M_{\rm HI}}{\rm M_{\odot}}$\\
\hline
PC1 & $58$\% & $0.28$ & $0.76$ & $0.06$ & $-0.59$\\
PC2 & $34$\% & $0.6$ & $0.18$ & $-0.63$ & $0.45$\\
\hline
Property & & $\frac{M_{\rm stellar}}{\rm M_{\odot}}$ & $\rm \frac{SFR}{M_{\odot}\,yr^{-1}}$ & $\rm \frac{Z_{\rm SF,gas}}{Z_{\odot}}$ & $\frac{M_{\rm neutral}}{\rm M_{\odot}}$\\
\hline
PC1 & $61$\% & $0.36$ & $0.69$ & $0.05$ & $-0.62$\\
PC2 & $34$\% & $0.62$ & $0.03$ & $-0.71$ & $0.34$\\
\hline
\end{tabular}
\end{center}
\end{table}

{The three PCA are shown in Table~\ref{TablePCA2}.
In the first PCA (which uses $M_{\rm H_2}$),   
the first principal component (PC1) has a strong contribution from the gas metallicity, which we do not 
see in PC1 when $M_{\rm HI}$ or $M_{\rm neutral}$ are used instead of $M_{\rm H_2}$. 
We see that PC1 in the case $M_{\rm neutral}$ is used has the largest variance, which is what we expected if 
the fundamental plane of star formation introduced in $\S$~\ref{FPGas} was indeed responsible for most of the variance 
seen in the galaxy population. This confirms the analysis presented in $\S$~\ref{localU}. We again see that 
the metallcity of the star-forming gas has a large weight in the second principal component in the three case we analyse here.}

\section[]{Effects of subgrid modelling on the fundamental plane of star formation}\label{ModelTests}

One of the main advantages of the \eagle\ hydrodynamical simulation suite is that it includes a series of runs in which 
the subgrid models and model parameters are varied, in addition to the reference run that we analysed in $\S$~$3$~and~$4$. 
C15 introduced $9$ runs in which the parameters of the reference model
 are altered (see Table~$1$ in C15). Here we show how the fundamental plane of star formation in \eagle\ is affected 
by changes in the stellar and AGN feedback strength and model, in the polytropic index of the equation of state 
applied to gas particles that have temperatures below an imposed temperature floor, and 
in the power-law index adopted in the star formation law (see $\S$~\ref{sub-gridsec} for details). 
We use the models of Table~\ref{Models} to study how the relation between neutral gas 
fraction, SFR and stellar mass changes under changes in the parameters of the subgrid physics.

\begin{table}
\begin{center}
  \caption{Variations of the reference model studied here. These variations were run in the 
L025N0376 setup (a simulation in a cubic volume of length $25$~cMpc on a side,
using $376^3$ particles of dark matter and an equal number of
baryonic particles; {see Table~\ref{TableSimus2} for more details}).
The reference model 
  has the following parameter values: $f_{\rm th,max}=3.0$, $f_{\rm th,min}=0.3$ (govern the stellar feedback strength), 
$\rm log_{10}(\Delta T_{\rm AGN}/K)=8.5$ (governs the AGN feedback strength), 
$\gamma_{\rm eos}=4/3$ (governs the relation between pressure and density for gas particles 
with temperatures below $T_{\rm eos}=8\times 10^3$~K), and a power-law dependence in the star formation law of 
$n=1.4$. 
{In the bottom two models, the scaling of $f_{\rm th}$ is changed to depend on the velocity dispersion of the DM (FB$\sigma$) and 
on the gas metallicity alone (FBZ). In the reference model, $f_{\rm th}$ depends on gas metallicity and density.}
  See $\S$~\ref{sub-gridsec} for a description of the subgrid models and more details on the parameters included in \eagle.}\label{Models}
\begin{tabular}{l l}
\\[3pt]
\hline
(1) & (2) \\
\hline
Model & Parameters changed\\
\hline
WeakFB & $f_{\rm th,max}=1.5$, $f_{\rm th,min}=0.15$\\
StrongFB & $f_{\rm th,max}=6$,$\,\,$ $f_{\rm th,min}=0.6$\\
AGNTd8 &  $\rm log_{10}(\Delta T_{\rm AGN}/K)=8$\\
AGNTd9 &  $\rm log_{10}(\Delta T_{\rm AGN}/K)=9$\\
eos1 & $\gamma_{\rm eos}=1$\\
eos5$/$3& $\gamma_{\rm eos}=5/3$\\
KSNormHi &  $n=4.8$\\
KSNormLow & $n=0.48$\\
\hline
Model & $f_{\rm th}$-scaling \\
\hline
FB$\sigma$ & $\sigma^2_{\rm DM}$\\
FBZ & $Z_{\rm SF,gas}$\\
\hline
\end{tabular}
\end{center}
\end{table}

\begin{figure}
\begin{center}
\includegraphics[width=0.49\textwidth]{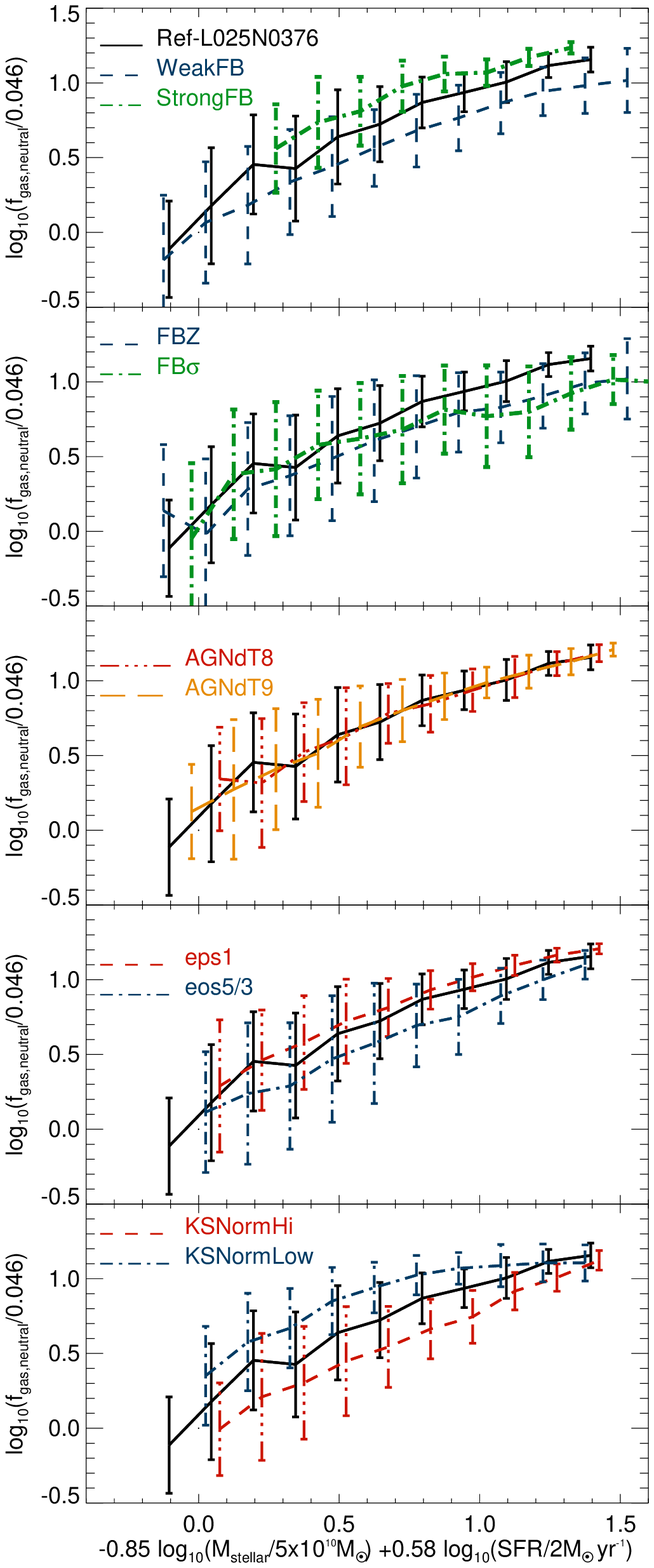}
\caption{An edge-on view of the plane of neutral gas fraction, stellar mass and SFR, as defined in Eq.~\ref{Eqfneutral}. Here we include galaxies 
in all the simulations that have $M_{\rm stellar}>10^9\,\rm M_{\odot}$ and that are in the redshift range $0\le z\le 4.5$. The results 
for the simulation Ref-L025N0376 are shown in each panel, while the results of the simulations WeakFB, StrongFB are shown in the top panel, 
FBZ and FB$\sigma$ in the next-to-top panel, 
AGNTd8 and AGNTd9 in the middle panel, eos1 and eos5$/$3 in the next-to-bottom panel, and 
KSNormHi and KSNormLow in the bottom panel, as labelled in each panel. Lines with error bars show the 
medians and $16^{\rm th}$ to $84^{\rm th}$ percentiles for each model, respectively. 
All variations of the reference model were run in the L025N0376 setup.}
\label{ModelsTests}
\end{center}
\end{figure}

We find that changing the strength of stellar feedback (WeakFB and StrongFB; top panel in Fig.~\ref{ModelsTests}) 
has a mild effect on the normalisation 
of the fundamental plane of star formation.
Weaker feedback produces higher neutral gas fractions at fixed stellar mass and SFR with respect to the 
reference model (and the opposite is true for stronger feedback). 
%This is because less gas is expelled out of the galaxy if we impose weaker feedback. 
Under self-regulation of star formation, weaker feedback would produce lower outflow rates, which 
in turn produces higher gas fraction and SFRs compared to stronger feedback.  
Note that the median relations in the models with weak/strong stellar feedback are almost parallel to each other 
which supports our interpretation that the curvature of the fundamental plane of 
star formation is set by the physics of star formation (i.e. how the gas gets converted into stars), 
while the normalisation is set by the self-regulation of star formation. 
{The effect of the self-regulation is suggested by the change in the scatter of the fundamental plane when different feedback 
strength are adopted.
A weaker feedback produces larger scatter, consistent with the less efficient self-regulation. 
A change in the scatter is also seen if we change the model of stellar feedback (FBZ and FB$\sigma$ models), where both lead 
to an increase in the scatter, due to the change in the timescale of self-regulation.}
Regarding the curvature, we see that changing the index in the equation of state imposed on the unresolved 
ISM, $\gamma_{\rm eos}$ (next-to-bottom panel in Fig.~\ref{ModelsTests}), 
does not change the normalisation of the fundamental plane of star formation, but changes the slope. 
For example, model eos5$/$3 produces a slightly steeper relation compared to the reference model and 
the model eos1 at $\rm log_{10}(f_{\rm gas,neutral}/0.046)\gtrsim 0.5$. 
Effectively, changing the index in the equation of state 
has the effect of slightly changing the curvature of the fundamental plane of star formation 
while not changing the normalisation significantly.  
The effect on the curvature of the fundamental plane of star formation by how star formation is modelled in \eagle\ becomes 
more apparent in the bottom panel of Fig.~\ref{ModelsTests}) where we varied the power-law index in the star formation law, $n$. 
 A lower $n$ produces a much flatter plane than in reference model, while increasing $n$ steepens the plane. 

Changing the AGN feedback strength has little effect on the 
fundamental plane of star formation, pointing to stellar feedback being the main mechanism driving outflows in 
star-forming galaxies.
 
In short, feedback from star formation (or AGN) produces an effective amount of feedback to balance the cosmological 
gas accretion rate, and the star formation recipe determines the rate at which gas gets converted into stars. These two `rules' help
 us understanding what we see in Fig.~\ref{ModelsTests} (see \citealt{Haas13a} and \citealt{Haas13b} for detailed 
studied of the self-regulation of star formation in galaxies).

\end{document}